\documentclass[reqno,12pt]{book}
\usepackage{amsmath} 
\usepackage{amsfonts} 
\usepackage{amssymb}
\usepackage{makeidx}

\usepackage{bm}
\usepackage{graphicx}
\usepackage[dvips]{color}
\usepackage{amssymb,amsmath,amsthm,amssymb}

\newtheorem{theorem}{Theorem}
\newtheorem{lemma}{Lemma}
\newtheorem{corollary}{Corollary}
\newtheorem{definition}{Definition}
\def\d{{\mathrm{d}}}
\def\implies{\Rightarrow}

\makeindex

\begin{document}
\bibliographystyle{plain}
\begin{titlepage}

\center

\vspace*{2mm}

{\Huge \bf Cosmological milestones and Gravastars --- topics in General Relativity}\\[0.3cm]


\vspace*{5mm}

{\sf \large

by\\[2mm]

\Large C\'eline CATTO\"EN\\[4cm]

Supervisor:
Dr. Matt Visser\\[2mm]
{\normalsize  A thesis
submitted to the Victoria University of Wellington
in fulfilment of the
requirements for the degree of
Master of Science
in Mathematics.}\\
\large Victoria University of Wellington\\[2mm]

New Zealand\\[0.5cm]
}
\begin{minipage}[t][4cm][c]{10cm}
\begin{center}
\includegraphics{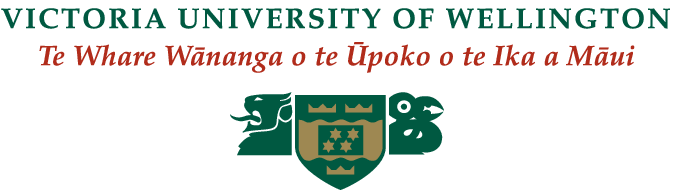} 
\end{center}
\end{minipage}
%

\end{titlepage}

\thispagestyle{empty}
\frontmatter

\chapter{Abstract}
In this thesis, we consider two different problems relevant to general relativity. Over the last few years, opinions on physically relevant singularities occurring in FRW cosmologies have considerably changed. We present an extensive catalogue of such cosmological milestones using generalized power series both at the kinematical and dynamical level. We define the notion of ``scale factor singularity" and explore its relation to polynomial and differential curvature singularities. We also extract dynamical information using the Friedmann equations and derive necessary and sufficient conditions for the existence of cosmological milestones such as big bangs, big crunches, big rips, sudden singularities and extremality events. Specifically, we provide a complete characterization of cosmological milestones for which the dominant energy condition is satisfied. The second problem looks at one of the very small number of serious alternatives to the usual concept of an astrophysical black hole, that is, the gravastar model developed by Mazur and Mottola. 
 By considering a generalized class of similar models with continuous pressure (no infinitesimally thin shells) and negative central pressure, we demonstrate that gravastars cannot be perfect fluid spheres: anisotropc pressures are unavoidable. We provide bounds on the necessary anisotropic pressure and show that these transverse stresses that support a gravastar permit a higher compactness than is given by the Buchdahl--Bondi bound for perfect fluid stars. We also comment on the qualitative features of the equation of state that such gravastar-like objects without any horizon must have.

\chapter{Acknowledgment}

I would like to thank my supervisor, Dr.~Matt Visser for his assistance, guidance and support. I would also  like to give some special thanks to Petarpa Boonserm, Tristan Faber and Silke Weinfurtner for numerous interesting discussions relating to topics discussed in this thesis. \\

I wish to acknowledge the School of Mathematics, Statistics, and Computer Science for providing me with office space and all the facilities.\\

I am also very grateful to the Marsden fund that has indirectly supported me financially during my work at Victoria University of Wellington.\\

Finally, I would like to thank my family and Ian for being there for me.

\tableofcontents
\listoffigures

\mainmatter

\chapter{Introduction}
This thesis is composed of three main chapters, and explores two separate problems in classical general relativity. 
\begin{itemize}
\item The first chapter assembles some basic notions in general relativity that will be exploited in the two following research chapters. 
\item The second chapter  is relevant to cosmology. It presents general definitions as a function of the scale factor of generic cosmological milestones (big bangs, sudden singularities...), and investigates their properties both at the kinematical and dynamical level. 
\item The last chapter explores the properties of a gravastar like object --- that would be an alternative to black holes without any horizon.
\end{itemize}

The first problem deals with cosmology and singularities. In Friedmann--Robertson--Walker
cosmologies, the physically relevant singularities had traditionally been thought to be restricted to the ``big bang" or ``big crunch". Even the most important ``singularity theorems" (e.g. Hawking and Penrose~\cite[p 240]{Wald:1984rg}) do not specify which type of singularities are dealt with. However, in the last few years, additional different types of singularities have appeared in the literature, making the list of cosmological singularities more extensive --- with Òbig ripsÓ and Òsudden singularitiesÓ added to the mix, as well as renewed interest in non-singular cosmological events such as ÒbouncesÓ and ÒturnaroundsÓ.
We define the notion of ``scale factor singularity" and define generic cosmological milestones using generalized power series of the scale factor of the universe $a(t)$. Once this classification is in place, we can investigate properties of these cosmological milestones on kinematical and dynamical levels.  
\begin{itemize}
\item Can we find interesting relations between this notion of ``scale factor singularity", and polynomial and differential curvature singularities?
\item Introducing the Friedmann equations, is it possible to place constraints on whether or not the classical energy conditions are satisfied at the cosmological milestones?
\end{itemize}
Since the classification of the cosmological milestones as generalized power series of the scale factor is extremely general, the corresponding results are to a high degree model-independent.
We will demonstrate that there exists a particular class of singularities that do not imply a polynomial curvature singularity near the milestone and only an even smaller sub-class of them do not imply a differential curvature singularity.
 We will also determine, with a minimum of technical assumptions, the necessary and sufficient conditions for the occurrences of all these cosmological milestones.\\

The second problem looks at an alternative model to black holes, the ``gravastar model", developed by Mazur and Mottola. Instead of having a star contracting and collapsing until matter arrives at a singularity at the centre, the gravastar model suggests that a gravitationally collapsing star would force spacetime itself to undergo a phase transition that would prevent further collapse. Thus the star would be transformed into a spherical Òquantum vacuumÓ surrounded by a form of super-dense matter. In the Mazur and Mottola model, the gravastar-like objects are composed of three layers (interior de Sitter space, outer region consisting of a finite-thickness shell of stiff matter, exterior Schwarzschild vacuum geometry) and two infinitesimally-thin shells with surface densities $\sigma_{\pm}$, and surface tensions $\vartheta_{\pm}$.
\begin{itemize}
\item Is it possible to replace the thin shell completely with a continuous layer of finite thickness?
\item Is the pressure anisotropy (implicit in the Mazur--Mottola infinitesimally thin shell) a necessity for any gravastar-like objects? 
\item Is it really possible to build a gravastar-like objects using only perfect fluid with a continuous layer of finite thickness? Could a horizon or naked singularity form and why?
\end{itemize}
Assuming pressure is continuous and differentiable, we will first analyze the resulting
static geometry with the resulting isotropic TOV equation, and point out all the fatal problems arising with this situation. Secondly, we will turn to the resulting anisotropic TOV equation, and discuss when the anisotropy is necessary for the TOV equation to hold and what the resulting bounds are on transverse pressures in the anisotropic region. Finally, we will discuss what specific features the equation of state must have for an horizon-avoiding gravastar-like object to be sustained. Note that we derive these properties from an agnostic point of view as to the existence or non-existence of gravastars.\\

To analyze these two problems, we have made use of the theory of classical general relativity, but we have not used a quantum field theory point of view. \\

This thesis has three chapters but also four appendices. The first appendix exhibits all of the most important spacetime metrics used in this thesis, the last three appendices are papers published on work relating to this thesis, they were produced as a collaboration with Dr.~Matt Visser, Petarpa Booserm, Tristan Faber and Silke Weinfurtner. At the time of writing, two papers (``Effective refractive index tensor for weak-field gravity")~\cite{Boonserm:2004wp} and (Gravastars must have anisotropic pressures)~\cite{Cattoen:2005he} have been published in Classical Quantum Gravity, and one paper (Necessary and sufficient conditions for big bangs,
bounces, crunches, rips, sudden singularities, and extremality events)~\cite{Cattoen:2005dx} has been accepted for publication in Classical Quantum Gravity.

\chapter{Some essential notions in General Relativity}   \label{intro}

This chapter introduces some basic notions in general relativity and cosmology. This will serve as the starting point for the specific topics investigated in this thesis. The notions of general relativity presented in this chapter are non-exhaustive.\\

In 1915, Einstein\index{Einstein} formulated the theory of general relativity, a theory of space, time, and gravitation. The new viewpoint it introduced on the nature of space and time  often appears to be abstruse as it goes against some deeply ingrained, intuitive notions and requires some specific and sophisticated mathematic tools such as differential geometry.

Later developments, and renewed interest in general relativity began in the 1960s, when the theory was related to other areas of physics and astronomy. The modern theory of gravitational collapse, singularities, and black\index{black hole} holes was developed at this time.

A deeper understanding of general relativity is another factor for renewed interest. Indeed, further understanding of the laws of nature are necessary to make progress toward the goal of developing a quantum theory of gravitation. Quantum gravity is the field of theoretical physics attempting to unify the theory of quantum mechanics, which three of the fundamental forces of nature, with general relativity, the theory of the fourth fundamental force: gravity. The theory of relativity in its own right makes many remarkable statements concerning the structure of space and time and the structure of the gravitational field.

One of the conceptual cores of general relativity is the \textit{Equivalence principle}\index{equivalence principle} which comes in three different versions:
\begin{enumerate}
\item \textit{Weak Equivalence Principle:} ``All bodies (subject to no force other than gravity) will follow the same paths given the same initial positions and velocities." This is also often called the Principle of Uniqueness of Free Fall.
\item \textit{Einstein's (1907) Principle of Equivalence} of gravitation and inertia: ``All motions in an external static homogeneous gravitational field are identical to those in no gravitational field if referred to a uniformly accelerated coordinate system."
\item \textit{Strong Equivalence Principle:} ``At any event, always and everywhere, it is possible to choose a local inertial frame (L.I.F) such that in a sufficiently small spacetime neighbourhood all (non-gravitational) laws of nature take on their familiar forms appropriate to the absence of gravity, namely the laws of special relativity."
\end{enumerate}

All the three versions are correct, but it is the Strong Equivalence Principle (now often called the Einstein Equivalence Principle) which is now viewed as the most fundamental.
 \newline
 
Another conceptual core of general relativity is the field equations.

\section{Notions of metric, geodesics, affine connexion, Killing vector field}  \label{notions} \index{geodesic}\index{metric}\index{Killing!vector field}
The bending of space and time nearby an object is called a \textit{gravitational field}, and its potential can be described by a \textit{spacetime metric} commonly written as:
\begin{equation}
ds^2=g_{ab}dx^{a} dx^{b}.
\end{equation}
A \textit{spacetime} consists of a four-dimensional \textit{manifold} \index{manifold}$M$ and a Lorentzian metric\index{metric!Lorentzian} $g_{ab}$ defined everywhere on $M$. The metric has a Lorentzian signature $(-,+,+,+)$.
 Indices $a$ and $b$ run from $0$ (usually denoted as the time direction) to $3$. 
With the components of the metric $g_{ab}$, we can construct a connexion known as the affine connexion \index{affine connexion} or Christoffel \index{Christoffel symbol}symbol:
\begin{equation}
\Gamma^a{}_{bc} = \frac{1}{2}g^{ad} \left( g_{db,c} +g_{dc ,b} +g_{bc ,d} \right)
\end{equation}
where $g^{ad}$ is the inverse \index{metric!inverse}metric. The connection indirectly represents the coordinate \textit{acceleration} of a free falling particle in a gravitational field. 

A geodesic\index{geodesic} is the curved-space generalization of the notion of a ``straight line" in Euclidean
space. By definition, a straight line is the path of the shortest distance between two points. More technically, a straight line is a path which parallel transports its own tangent vector.,i.e a curve\index{curve} whose tangent, $T^a$, satisfies the equation:
 \begin{equation}
T^a \nabla_aT^b =0 
\end{equation}
If we consider a parametric equation of the curve\index{curve}  $x^a=x^a(\lambda)$,  we obtain the geodesic\index{geodesic} equation:
\begin{equation}
\frac{d^2x^a}{d\lambda^2}+\Gamma^a{}_{bc} \frac{dx^b}{d\lambda}\frac{dx^c}{d\lambda}=0
\end{equation}

If we take two events $X_1$ and $X_2$, define $\triangle X= X_2-X_1$,  and define the quadratic form 
 \begin{equation}
g(\triangle X,\triangle X)=(\triangle X)^T g (\triangle X),
\end{equation}
we have the following definitions:
 \begin{itemize}
\item If $g(\triangle X,\triangle X)<0$, the two events are \textit{timelike} separated. It means that normal particles can successfully travel from one event to the other.
\item  If $g(\triangle X,\triangle X)=0$, the two events are \textit{lightlike} separated. It means that one has to travel exactly at the speed of light to travel from one event to the other.
\item  If $g(\triangle X,\triangle X)>0$, the two events are \textit{spacelike} separated. It means that one would have to travel faster than light to get from one event to the other.    
\end{itemize}

In general relativity, timelike geodesics\index{geodesic} represent spacetime motions of free falling particles and null geodesics\index{geodesic} represent the paths of light rays.\\

Another important mathematical concept is the one of \emph{Killing vector field}\index{Killing!vector field}. Some propositions on Killing vector fields are typically used when investigating properties related to stars (e.g. static and stationary spacetimes).

A Killing vector field is a vector field on a Riemannian\index{Riemann!Riemannian manifold} manifold \index{manifold}that preserves the metric. \index{Killing!vector field}Killing fields are the infinitesimal generators of isometries; that is, flows generated by Killing fields are continuous isometries of the \index{manifold}manifold.
Specifically, a vector field $\xi^a$ is a Killing field if the Lie derivative $\mathcal{L}$ with respect to $\xi^a$ of the metric\index{metric} $g_{ab}$ vanishes:
\begin{equation}
\mathcal{L}_{\xi} \; g_{ab}=0.
\end{equation}
In terms of the covariant derivative, this is
\begin{equation}
\nabla_a \xi_b + \nabla_b \xi_a =0. \label{killing_eq}
\end{equation}
Equation (\ref{killing_eq}) is also known as \emph{Killing's equation}\index{Killing!'s equation}, thus, the necessary and sufficient condition that $\xi^a$ be a \index{Killing!vector field}Killing vector field is that it satisfy the Killing's equation.

If a spacetime has a\index{Killing!vector field} Killing vector, then we know we can find a coordinate system in which the \index{metric}metric is independent of one of the coordinates.
By far the most useful fact about Killing vectors is the following lemma:
\begin{lemma} \label{lemma_killing}
Let $\xi^a$ be a Killing vector field and let $\lambda$ be an affinely parametrized geodesic\index{geodesic} with tangent $u^a$. Then $\xi_a u^a$ is constant along $\lambda$.
\end{lemma}
Lemma \ref{lemma_killing} can be interpreted as saying that every \index{Killing!vector field}Killing vector field gives rise to a conserved quantity for particles and light rays. This conserved quantity enables one to determine the gravitational redshift in stationary spacetimes and is extremely useful for integrating the geodesic \index{geodesic}equation when symmetries are present.

 \section{Tensors}
According to \index{Riemann}Riemann's theory of curved \index{manifold}manifolds, the geometry of space-time is completely described by the metric \index{metric!tensor}tensor $g_{ab}$, which has $10$ algebraically independent components at each event. 
\subsection{Riemann tensor} 
 The Riemann \index{Riemann!tensor}tensor, and its derived tensors (Ricci\index{Ricci!tensor}, Weyl\index{Weyl tensor}), are the only tensors that can be constructed from the metric tensor\index{metric!tensor} and its first and second derivatives:
 \begin{equation}
R^a{}_{bcd}=\partial_c  \Gamma^a{} _{bd}-\partial_d  \Gamma^a {}_{bc}+ \Gamma^e {}_{bd} \Gamma^a {}_{ec}-\Gamma^e{}_{bc}\Gamma^a{} _{ed}
\end{equation}

The Riemann\index{Riemann!tensor} curvature tensor has $20$ algebraically independent components at each event. The components of the Riemann tensor identically satisfy a differential equation (the Bianchi identity\index{Bianchi identity} (\ref{Bianchi})):
 \begin{equation}
 \nabla_a R_{debc}+ \nabla_c R_{deab}+ \nabla_b R_{deca}=0 \label{Bianchi}
 \end{equation}

The Riemann curvature tensor $R_{abcd}$ is used to define other important tensors. 

\subsection{Ricci tensor, Ricci scalar, \index{Weyl tensor}Weyl tensor}\index{Ricci!tensor}
The Ricci tensor $R_{ab}$ (\ref{Ricci_tensor}), the Ricci\index{Ricci!scalar} scalar $R$ (\ref{Ricci_scalar}) and the Weyl tensor $C_{abcd}$ for $n\geqslant 3$ dimensions (\ref{Weyl_tensor}), are defined by contractions, in a manner analogous to decomposing a matrix into trace and tracefree parts. 
\begin{equation}
R_{ab}=g^{cd}R_{dacb}=R^c{}_{acb} \label{Ricci_tensor}
\end{equation}
\begin{equation}
R=g^{ab}R_{ab}  \label{Ricci_scalar}
\end{equation}
\begin{eqnarray}
C_{abcd}&=&R_{abcd}\;+\; \frac{1}{n-2} \left( g_{ad} R_{cb} \; + \;  g_{bc} R_{da} - g_{ac} R_{db} - g_{bd} R_{ca} \right) \; \nonumber \\
& & + \; \frac{1}{(n-1)(n-2)} \left( g_{ac}g_{bd}-g_{ad}g_{cb} \right) R          \label{Weyl_tensor}
\end{eqnarray}

If the Riemann \index{Riemann!tensor}tensor vanishes on a neighborhood of space-time, this neighborhood is locally isometric to  Minkowski space-time, it is \textit{locally flat}. Otherwise, if the Weyl tensor\index{Weyl tensor} vanishes on a neighborhood of space-time, the neighborhood is locally conformally equivalent to Minkowski space-time. Thus, the Riemann, Ricci\index{Ricci!tensor}, and Weyl tensors all have geometric meaning independent of any physical interpretation.

\subsection{Einstein tensor}\index{Einstein!tensor}
Finally, the Ricci tensor\index{Ricci!tensor} and the Ricci scalar\index{Ricci!scalar} can be used to define the Einstein tensor \index{Einstein!tensor}(\ref{Einstein_tensor}):
\begin{equation}
G_{ab}=R_{ab}-\frac{1}{2}g_{ab} \; R \label{Einstein_tensor}
\end{equation}

\subsection{Energy-momentum tensor}
Now let's introduce the energy-momentum tensor $T^{ab}$ (also called the stress-energy tensor), this is a symmetric $T^2_0$ tensor which formulates the energy-like aspects of a system: energy density, pressure, stress, and so on.

The stress-energy tensor is defined by:
\begin{equation}
T^{ab}=
\left[
\begin{array}{c|c}
\rho & F^i \\\hline
 F^j & \Pi^{ij}
 \end{array}
 \right],
\end{equation}
where $\rho$ is the energy density (at rest in this coordinate system), $F$ is the energy flux (generalization of the Poynting vector), and $\Pi$ is the stress tensor (pressure, shear etc...).

The three most important energy-momentum tensors in general relativity are, namely, \textit{matter or dust, perfect fluid} \index{perfect fluid}and \textit{the electromagnetic field}.

\section{Einstein's field equations}  \label{Einsteins_equations}
\index{Einstein!field equations}
We can now write the Einstein's  field equations (\ref{einstein_field_eq}), which govern how the\index{metric} metric responds to energy and momentum.
\begin{equation}
G_{ab}=\frac{8 \pi G}{c^2} \; T_{ab}   \label{einstein_field_eq}
\end{equation}
Here $G$ is Newton's constant of universal gravitation, $G=6.67257\times 10^{-11} \; m^3 \; kg^{-1}\; s^{-2}$.

In other words, Einstein's field \index{Einstein!field equations}equation states that
\begin{equation}
R_{ab}-\frac{1}{2} R \; g_{ab}   =\frac{ 8 \pi G}{c^2}T_{ab},  
\end{equation}
or equivalently 
\begin{equation}
R_{ab} =  \frac{ 8 \pi G}{c^2} \left( T_{ab} - \frac{1}{2} g_{ab} T  \right) ,
\end{equation}
where $T$ is the trace of the stress energy tensor. Note that often, the gravitational constant $G$ and the speed of light $c$ are set equal to one, this is called using ``geometrized units". 

Thus, the \index{Ricci!curvature}Ricci curvature is directly coupled to the immediate presence of matter at a given event. If there is no mass-energy at a given event, the Ricci\index{Ricci!tensor} tensor vanishes. If it were not for the \index{Weyl tensor}Weyl tensor, this would mean that matter here could not have a gravitational influence on distant matter separated by a void. Thus, the Weyl tensor represents that part of space-time curvature which can propagate across and curve up a void. 

There are two particularly physically important solutions of Einstein's field equation\index{Einstein!field equations} that represent two extremes of curvature. The Kerr vacuum solution, which models space-time outside a rotating body such as a star, has zero Ricci \index{Ricci!curvature}curvature but nonzero Weyl curvature at each event. The FRW spacetimes\index{Friedmann}, which model the universe on a very large scale, has zero Weyl curvature but nonzero Ricci curvature at each event. 
 
\section{Weak field theory}
When the gravitational field is ``weak", general relativity can be reduced to a \textit{linearized theory}.
The weakness of the gravitational  field allows us to  decompose the metric \index{metric!Minkowski}into the flat Minkowski spacetime (\ref{Minkowski}) of special relativity plus a small perturbation,
\begin{equation}
g_{ab}  = \eta_{ab}   + h_{ab} \; , \qquad    |h_{ab}| \ll 1 ,
\end{equation}
where
\begin{equation}
\eta_{ab}=     
\left[
\begin{array}{cccc}
-1 & 0 & 0 & 0 \\
0 & 1 & 0 & 0 \\
0 & 0 & 1 & 0 \\
0 & 0 & 0 & 1 \\ 
 \end{array}
 \right].              \label{Minkowski}
\end{equation}

The assumption that $h_{ab}$ is small allows us to ignore anything that is higher than first order in this quantity, from which we immediately obtain:
\begin{equation}
g^{ab}=\eta^{ab}   + h^{ab} + O(h^2)  ; \qquad h^{ab}=\eta^{ac}\eta^{bd}h_{cd}
\end{equation}
The Christoffel \index{Christoffel symbol}symbol can now be rewritten as :
\begin{equation}
\Gamma^a{}_{bc}=\frac{1}{2} \eta^{ad} \left( h_{db,c} +h_{dc,b}-h_{bc,d} \right) +O(h^2).
\end{equation}
The Riemann\index{Riemann!tensor} tensor now becomes:
\begin{equation}
R_{abcd}=-\frac{1}{2} \left(   h_{ac,bd}+ h_{bd,ac} - h_{ad,bc} - h_{bc,ad}   \right) +O(h^2).
\end{equation}
The Ricci\index{Ricci!tensor} tensor is:
\begin{equation}
R_{ab}=\frac{1}{2} \left(   h^c_{a,bc} +h^c_{b,ac} -\nabla^2 h_{ab} -h_{,ab}   \right) +O(h^2).
\end{equation}
By making coordinate changes, we can adopt the Einstein gauge (also called de Donder gauge, or Hilbert gauge, or Fock gauge) which provides the simple result,
\begin{equation}
\left[  h^c{}_b-\frac{1}{2} h \delta^c{}_b \right]_{,c}=0,
\end{equation}
and therefore, in this gauge, we have the simple results,
\begin{equation}
R_{ab}=-\frac{1}{2}\nabla^2h_{ab}+O(h^2),
\end{equation}
and 
\begin{equation} 
G_{ab}=-\frac{1}{2}\nabla^2 \left( h_{ab}-\frac{1}{2}h \eta_{ab} \right) +O(h^2).
\end{equation}

Thus, using the weak field limit, we can write
\begin{equation}
\nabla^2h_{ab}=-16 \pi G \left(   T_{ab} -\frac{1}{2}T\eta_{ab} \right)+O(h^2),
\end{equation}
which gives  a lot more information.

\section{Strong field theory}
Let's move from the domain of the weak field limit to solutions of the full nonlinear Einstein's equations\index{Einstein!field equations}. 

\subsection{Black\index{black hole} holes, event horizon}  \label{black_holes_event_horizon}\index{horizon}
\textit{Horizons} \index{horizon}are typically associated with strong gravitational fields (it is still possible to have a strong gravitational field without forming a horizon\index{horizon}). By definition an \textit{event horizon} is the boundary of the region from which future-directed null curves\index{curve!null} cannot escape to infinity. In other words, event horizons are ``one-way membranes" permitting the passage of light and matter in only one direction and at which time slows to a stop. 

Some stars (whose masses are of the order of the sun's mass), can reach a final equilibrium state, e.g. white dwarf or a neutron star. However, for stars with much larger masses, such an equilibrium state is not possible. In this case, stars will contract to such an extent that the gravitational effects will overcome the internal pressure and stresses which will not oppose the contraction. The theory of general relativity predicts that such spherically symmetric stars will necessarily contract until all matter contained in the star arrives at a singularity at the centre of the symmetry. Those specific stars are surrounded by an event horizon\index{horizon} and are referred to as  \textit{black holes}\index{black hole}.

\subsection{Schwarzschild's solution to the Einstein equations}  \label{Schwarzschild_solution}
\index{Einstein!field equations} \index{Schwarzschild!exterior solution}
One of the most important solution is the one discovered by Schwarzschild, which describes spherically
symmetric vacuum spacetimes. Since it is in vacuum, Einstein's equations\index{Einstein!field equations} become:
\begin{equation}
R_{ab}=0
\end{equation}
Thus, to obtain a solution, we need to find all four-dimensional Lorentz signature \index{metric}metrics whose Ricci \index{Ricci!tensor}tensor vanishes and which are static and possess spherical symmetry. But first, we need to define more precisely the meaning of the terms ``stationary",  ``static" and ``spherically symmetric" and to choose a convenient coordinate system for analyzing this class of spacetimes.

\begin{itemize}
\item A \textit{stationary} spacetime is a spacetime that possess a timelike\index{Killing!vector field} Killing vector field\footnote{See section \ref{notions} on Killing vector fields.}, $\xi^a$, (more precisely, a Killing vector field that is timelike near spatial infinity).

\item A spacetime is \textit{static} if it is \textit{stationary} and, if in addition, there exists a spacelike hypersurface $\Sigma$ orthogonal to the orbits of the isometry. It is equivalent to the requirement that the timelike Killing \index{Killing!vector field}vector field $\xi^a$ satisfy:
\begin{equation}
 \xi_{[a} \nabla_{b} \xi_{c]}=0.
\end{equation}

This implies that the spacetime \index{metric}metric $ds^2$ can be chosen to be invariant under a time revearsal about the origin of time, e.g. all cross terms $dx^t dx^a \; \forall \; a \neq t$ vanish in the chosen coordinate system with arbitrary $\{ x^a \}$ on $\Sigma$. Thus, in these coordinates, the metric components are of the form:
\begin{equation}
ds^2= -V^2 (x^1, x^2, x^3)dt^2 +\sum_{a,b=1}^{3} g_{ab}(x^1,x^2,x^3)dx^a dx^b,  \label{static}
\end{equation}
where $V^2=-\xi_a \xi^a$. The absence of $dt dx^a$ cross terms expresses the orthogonality of $\xi^a$ with $\Sigma$. From the explicit form of a static metric\index{metric} (\ref{static}) it is clear that the diffeomorphism defined by $t \rightarrow -t$ is an isometry. And hence, in addition to the time translation symmetry, $t \rightarrow t+\mathrm{constant}$, possessed by all stationary spacetimes, the static spacetimes also posses a time reflection symmetry. \\
The static spherically symmetric metric\index{metric!static}\index{metric!symmetric}\index{metric!spherical} will describe non-rotating stars or black\index{black hole} holes,
while rotating systems (which keep rotating in the same way at all times) will be described
by stationary metrics\index{metric!stationary}.

\item A \textit{spherically symmetric} solution means that there exists a privileged point, e.g. the origin, such that the system is invariant under spatial rotations about the origin.
\end{itemize}
If the spacetime is \textit{stationary and spherically symmetric}, it automatically implies that it is static. An example of a stationary but non-static case is the Kerr black hole\index{black hole}, as the rotation rate is not changing but the surfaces orthogonal to the timelike \index{Killing!vector field}Killing field are not spacelike. For stars we will always be looking at stationary cases (and therefore static by spherical symmetry).\\

The Schwarzschild\index{metric!Schwarzschild} \index{Schwarzschild!exterior solution}solution, describing static spherical symmetric vacuum spacetimes, is in standard coordinates $(t,r,\theta,\phi)$:
\begin{equation}
ds^2=- \left(1-\frac{2GM}{r}   \right)dt^2+\frac{1}{ \left(1-\frac{2GM}{r}   \right)} dr^2 +r^2d\Omega^2     \label{Schwarzschild_metric}
\end{equation}
where $d\Omega^2=d\theta^2+\sin^2 \theta d\phi^2$.
This is true for any spherically symmetric vacuum solution to Einstein's equations\index{Einstein!field equations}.
 $M$ is a parameter that can be interpreted as the conventional Newtonian mass that  would be measured by studying orbits at large distances from the gravitating source. 

As $M \longrightarrow0$ we recover the expected Minkowski space. The metric also has the property of \textit{asymptotic flatness}, as $r\longrightarrow \infty $, the metric becomes progressively \index{Minkowski}Minkowskian.

\begin{theorem} \label{Birkhoff_th}
(Birkhoff's theorem) In a spherically symmetric spacetime the only solution to the Einstein \index{Einstein!field equations}equations (without cosmological constant) in vacuum is a piece of the Schwarzschild\index{Schwarzschild!exterior solution} solution which is more properly called the "Schwarzschild exterior solution". Furthermore, a spherically symmetric vacuum solution in the exterior region is necessarily static.
\end{theorem}
There is a generalization of Theorem \ref{Birkhoff_th} with a cosmological constant included which leads to the Kottler solution (Schwarzschild-de Sitter\index{de Sitter} spacetime).
 It is interesting to note that the result is a static metric \index{metric!static}but that the source was not specified except that it be spherically symmetric. Specifically, the source itself does not need to be static, it could be a collapsing star, as long as the collapse were symmetric. Practically, the metric \index{metric!Schwarzschild}of the Schwarzschild \index{Schwarzschild!exterior solution}exterior solution corresponds to (\ref{Schwarzschild_metric}) restricted to the region $r>2GM$.

From the form of the Schwarzschild \index{metric!Schwarzschild}metric (\ref{Schwarzschild_metric}), we can see that the metric coefficients become infinite at $r = 0$ and $r = 2GM$.
The metric coefficients are coordinate-dependent quantities, but it is certainly possible to have a \textit{coordinate singularity} \index{coordinate singularity}which results from a breakdown of a specific  coordinate system rather than the underlying \index{manifold}manifold. The metric appears to have a singularity at $r=2GM$, but there is actually no \textit{physical singularity} at that point, the coordinate system is just breaking down there. The value $r=2m$ is known as the \index{Schwarzschild!radius}\textit{Schwarzschild radius}\footnote{The value $m$ is actually interpreted as the relativistic mass with $m=GM/c^2$ in metres.} , this is a removable coordinate singularity\index{coordinate singularity} as indicated by the Riemann\index{Riemann!tensor} tensor scalar invariant 
\begin{equation}
R_{abcd}R^{abcd}=48m^2r^{-6},
\end{equation}
which is finite at $r=2m$.

 However there is an interesting  physics phenomenon associated with this surface: it is the \textit{event horizon}\index{horizon} of the \index{Schwarzschild!black hole}Schwarzschild black hole\index{black hole}.

\subsection{Singularities}
 The term \textit{singularity} \index{singularity}  is used in many ways and in many theorems without being specifically, physically and fully explained.  In this section, we will discuss some main definitions of what a \textit{singularity} stand for and what it involves qualitatively.\\

In a manifold \index{manifold}$M$, there can be two points which are not connected by any causal curve\index{curve!timelike} \index{curve!lightlike} (timelike or lightlike) (in a Euclidean signature metric this is impossible, but not in a Lorentzian spacetime). In such a case, it is said  that the geodesic\index{geodesic}  runs into a singularity, which can be thought of as the \textit{edge of the manifold}\index{manifold}. Manifolds which have such singularities are known as geodesically incomplete. In fact the ``singularity theorems" of Hawking and Penrose (\cite[p 240]{Wald:1984rg}) state that, for reasonable matter content (no negative energies), spacetimes in general relativity are almost guaranteed to be geodesically incomplete. \\

The Big bang \index{singularity!big bang}or the \textit{initial singularity} \index{singularity!big bang}  is defined by the fact that the\index{scale factor} scale factor of the universe $a(t)\to 0$ in a finite time. This particular singularity results from a homogeneous contraction of space down to ``zero size", but does not represent an explosion of matter concentrated at a point of preexisting nonsingular spacetime.   A big rip \index{singularity!big rip} is a singularity for which  the scale factor of the universe $a(t)\to \infty$ in a finite time\footnote{For more definitions, see Chapter \ref{Chapter_bangs}.}.\\

What kind of coordinate-independent feature is a warning that something goes wrong in the geometry, e.g a \textit{curvature singularity} occurs? \index{singularity!curvature singularity} One simple criterion is when the curvature becomes infinite. The curvature is measured by the \index{Riemann!tensor}Riemann tensor from which  can be constructed various scalar quantities, such as the \index{Ricci!scalar}Ricci scalar $R$, or higher order scalars $R^{ab}R_{ab}$, $R^{abcd}R_{abcd}$, etc...  If any of those scalars go to infinity when approaching some point $x$, there is a curvature singularity at this point $x$. \\

A \textit{naked singularity} \index{naked singularity}\index{singularity!naked singularity} is any curvature singularity not surrounded by an event horizon\index{horizon}.
Since there is no event horizon, there is no obstruction to an observer travelling to the singularity and returning to report on what was observed. 

\section{Energy conditions} \label{section_energy_cond}
\index{energy conditions}
In classical general relativity, there are several types of energy conditions~\cite{Visser:1995cc}:
\begin{itemize}
\item the null energy condition (NEC);\index{energy conditions!NEC} \index{NEC}
\item the weak energy condition (WEC);\index{energy conditions!WEC}\index{WEC}
\item the strong energy condition (SEC);\index{energy conditions!SEC}\index{SEC}
\item the dominant energy condition (DEC). \index{DEC}\index{energy conditions!DEC}
\end{itemize}

The energy conditions\index{energy conditions} of general relativity permit one to deduce very powerful and general theorems about the behaviour of strong gravitational fields and cosmological geometries.
In an orthonormal frame, the components of the stress energy tensor are given by:
\begin{equation}
T ^{ \hat{a} \hat{b}} =
\left[ 
\begin{array}{cccc}
 \rho & 0        & 0        &0       \\
      0     & p_{1} & 0        & 0       \\
      0     & 0        & p_{2} & 0        \\
      0     & 0        & 0        & p_{3} \\
\end{array}
\right] .
\end{equation}

The components of $T ^{ \hat{a} \hat{b}}$ are the energy density and the three principal pressures.

\subsection{Null Energy condition (NEC)}  \label{NEC}
\index{energy conditions!NEC}\index{NEC}
For all future pointing null vectors $k^a$, we ask that:
\begin{equation}
T_{ab}k^a k^b \geqslant 0
\end{equation}
In terms of pressures and density, we have:
\[
\forall \; i  \qquad \rho +p_i \geqslant 0.
\]

HawkingÕs area theorem for black hole\index{black hole} \index{horizon}horizon relies on the NEC\index{energy conditions!NEC}\index{NEC}, and hence evaporation of a black hole\index{black hole} must violate the NEC\index{energy conditions!NEC}\index{NEC}. 

\subsection{Weak Energy condition (WEC)} \label{WEC}
\index{energy conditions!WEC}\index{WEC}
Sometimes it is useful to think about Einstein's equations \index{Einstein!field equations}without specifying the theory
of matter from which $T ^{ \hat{a} \hat{b}}$  is derived. This leaves us with a great deal of arbitrariness, in the absence of some constraints on   $T ^{ \hat{a} \hat{b}}$, any metric\index{metric} can satisfy the Einstein\index{Einstein!field equations} equations. The real concern is  the existence of solutions to Einstein's equations\index{Einstein!field equations} with  ``realistic" sources of energy
and momentum. The most common property that is demanded of $T ^{ \hat{a} \hat{b}}$ is that it represent positive energy densities --- no negative masses are allowed. In a
locally inertial frame this requirement can be stated as $\rho  = T_{00} >   0$. To turn this into a
coordinate-independent statement, we ask that:
\[
T_{ab}V^a V^b \geqslant 0 \; \forall \; \text{timelike vector } V  \label{WEC_eq}
\]
In terms of pressures and density, we have:
\[
\rho \geqslant 0 \qquad \text{and}  \qquad  \forall \; i  \; \rho +p_i \geqslant 0.
\]
Any  timelike vector can be a tangent to an observerÕs world
line. The WEC\index{energy conditions!WEC}\index{WEC} condition states that the energy density measured by
any timelike observer is non-negative. 
It seems like a fairly reasonable requirement, and many of the important theorems about solutions to general relativity (such as the singularity theorems of Hawking and Penrose  (\cite[p 240]{Wald:1984rg}) ) rely on this condition or something very close to it. Unfortunately it is not set in stone; indeed, it is straightforward to invent
otherwise respectable classical  field theories which violate the WEC\index{energy conditions!WEC}\index{WEC}, and almost impossible
to invent a quantum  field theory which obeys it. Nevertheless, it is legitimate to assume
that the WEC\index{energy conditions!WEC}\index{WEC} holds in all but the most extreme conditions. 

\subsection{Strong Energy Condition (SEC)}
\index{energy conditions!SEC}\index{SEC}
For any timelike vectors $V^a$, we ask that:
\[
\left( T_{ab}-\frac{T}{2}g_{ab} \right)V^a V^b \geqslant 0   \label{SEC_eq}
\]
where $T$ is the trace of the stress-energy tensor: $T=T_{ab}g^{ab}$.

In terms of pressures and density, we have:
\[
T=-\rho+\sum_i p_i
\]
\[
\forall i \; \rho+p_i  \geqslant 0 \; \text{and}  \;   \rho +\sum_i p_i \geqslant 0.
\]
Note that the SEC\index{energy conditions!SEC}\index{SEC} implies the NEC\index{energy conditions!NEC}\index{NEC}, it does not imply the WEC\index{energy conditions!WEC}\index{WEC}. For example, matter with a negative energy density but sufficiently high pressures could satisfy the SEC \index{energy conditions!SEC}\index{SEC}but would violate the WEC\index{energy conditions!WEC}\index{WEC}.

The Penrose--Hawking singularity theorem relevant to the cosmological singularity uses the SEC\index{energy conditions!SEC}\index{SEC}.

\subsection{Dominant Energy Condition (DEC)}\index{DEC}
\index{energy conditions!DEC}\index{DEC}
For any timelike vectors $V^a$, we ask that:
\[
T_{ab}V^a V^b \geqslant 0 \qquad  \text{and that}  \qquad T_{ab}V^b \; \; \text{is a future directed non-spacelike vector.}
\]

 The DEC\index{energy conditions!DEC}\index{DEC} assumes that the WEC\index{energy conditions!WEC}\index{WEC} holds, and that for all future directed timelike vectors $V^a$  that
$T_{ab}V^b $ is a future directed non-spacelike vector. This ensures that the
net energy flow does not exceed the speed of light. The dominant energy condition\index{energy conditions} implies the weak energy condition\index{energy conditions} and also the null energy condition\index{energy conditions}, but does not necessarily imply the strong energy condition\index{energy conditions}. 

In terms of pressures and density, we have:
\[
\rho \geqslant 0 \qquad \text{and}  \qquad \forall \; i  \;  -\rho \leqslant  p_i \leqslant \rho.
\]
The dominant energy condition \index{energy conditions!DEC}\index{DEC}can be interpreted as saying that the speed of energy flow of matter is always less than the speed of light.\\

\subsection{Comments}\index{energy conditions}
Note that the null energy condition implies the weak energy condition, but otherwise the NEC\index{energy conditions!NEC}\index{NEC}, the WEC\index{energy conditions!WEC}\index{WEC} and the SEC\index{energy conditions!SEC}\index{SEC} are mathematically independent assumptions. In particular, the SEC\index{energy conditions!SEC}\index{SEC} does not imply the WEC\index{energy conditions!WEC}\index{WEC}. It is stronger only in the sense that it appears to be a stronger physical requirement to assume equation (\ref{SEC_eq}) rather than equation (\ref{WEC_eq}). Violating the NEC\index{energy conditions!NEC}\index{NEC} implies violating the DEC\index{DEC}, SEC and WEC as well.\\

The energy conditions\index{energy conditions} are looking a lot less secure than they once seemed:
\begin{itemize}
\item  There are quantum effects that violate all of the energy conditions.
\item There are even relatively benign looking classical systems that violate all the energy conditions ~\cite{Visser:1995cc}.
\end{itemize}

HawkingÕs area theorem for black hole\index{black hole} horizon \index{horizon}relies on the NEC\index{energy conditions!NEC}\index{NEC}, and hence evaporation of a black hole\index{black hole} must violate the NEC.

\section{Cosmology}  \label{cosmology}
\subsection{Introduction}
  
Cosmology is the study of the dynamical structure of the universe considered as a whole. Contemporary cosmological models are based on the idea that the universe is, on average, the same overall. This is based on a very simple principle, called ``the cosmological principle", which is a generalization of the Copernican principle:

The cosmological principle: at each epoch, the universe presents the same aspect from every point, except for local irregularities~\cite{D'inverno}.

When averaged over sufficiently large volumes the universe and the matter in the universe should be \textit{isotropic} and \textit{homogeneous}.
\begin{itemize}

\item \textit{Isotropy}\index{isotropy} states that space looks the same no matter what direction one looks at (direction independent). 

\item \textit{Homogeneity} is the statement that the \index{metric}metric is the same throughout the space (position independent). \index{homogeneity}

\end{itemize}

Astronomical observations reveal that the universe is homogeneous and isotropic on the largest scales. Traditionally this homogeneity has been assumed up to``small" fluctuations that are large enough to include clusters of galaxies. The scale at which homogeneity sets in is still not completely certain. Voids with diameters of order $10^8$ light years are ubiquitous, forming at least $40\%$ of the volume of the universe \cite{Hoyle:2001kn}, \cite{Hoyle:2003hc}, and are typically surrounded by bubble walls containing galaxy clusters. The largest feature observed - the Sloan Great Wall \cite{Gott:2003pf} -  is $1.47\times 10^9$ light years long. We simply assume homogeneity for some suitable defined cell size.

When looking at distant galaxies, they seem to be receding from our galaxy. It appears that the universe is not static, but changing with time. Thus most cosmological models are built on the fact that the universe is homogeneous and isotropic in space, but not in time.  Observationally, the universe today is significantly different from the universe of $10^{10}$ years ago, and radically different from the universe of $1.5 \times 10^{10}$ years ago.

\subsection{``Cosmography"}
Simply by using the assumptions of isotropy\index{isotropy} and homogeneity\index{homogeneity}, a cosmological model can be derived without yet using Einstein equations\index{Einstein!field equations}.
This homogeneous and isotropic cosmological model is called the \index{Friedmann}Friedmann-Robertson-Walker geometry (\ref{FLRW}), (FLRW Friedmann-Lema\^itre-Robertson-Walker geometry) and is given by:\index{universe!FRW} \index{universe!FLRW}
\begin{equation}
ds^2=-dt^2+a(t)^2 \left\{   \frac{dr^2}{1-kr^2} +r^2 \left[  d\theta^2+\sin^2 \theta  \;  d\phi^2 \right]   \right\}   \label{FLRW}
\end{equation}
where, $a(t)$ is the\index{scale factor} scale factor of the universe.
There are only three values of interest for the parameter $k$:
\begin{itemize}
\item $k=-1$, this corresponds to a negative curvature (for the hyperboloid)
\item $k=0$, this corresponds to no curvature (flat space)
\item $k=+1$, this corresponds to a positive curvature (for the $3$-sphere)
\end{itemize}
Therefore, the assumptions of homogeneity\index{homogeneity} and isotropy\index{isotropy} alone have determined the spacetime metric \index{metric}up to three discrete possibilities of spatial geometry $k$ and the arbitrary positive function of the\index{scale factor} scale factor $a(t)$.\\

Observational evidence strongly suggests that our universe (or part of our universe within our causal past), is well described by a Friedmann\index{Friedmann}--Robertson--Walker model, and  indeed a $k=0$ model, at least as far as the decoupling time of matter and radiation.

\subsection{``Cosmodynamics"}  \label{Cosmodynamics}
Now, by substituting  the spacetime metric  (\ref{FLRW}) into Einstein's equations\index{Einstein!field equations} (\ref{einstein_field_eq}), some predictions for the dynamical evolution of the system can be obtained. But first, we need to describe the matter content of the universe in terms of the stress-energy tensor. Using 
 the assumptions of isotropy\index{isotropy} and \index{homogeneity}homogeneity, the stress-energy tensor of matter in the present universe is approximated in an orthonormal frame by:
 \begin{equation}
T ^{ \hat{a} \hat{b}} =
\left[ 
\begin{array}{cccc}
 \rho & 0        & 0        &0       \\
      0     & p    & 0         & 0       \\
      0     & 0        & p    & 0        \\
      0     & 0        & 0    & p\\
\end{array}
\right] .
\end{equation}
Here $\rho$ and $p$ are the average density and pressure due to the galaxies, stars, clouds of dusts etc...

Applying the Einstein \index{Einstein!field equations}equations imply the two following Friedmann\index{Friedmann} equations:
\begin{eqnarray}
8\pi G_N \rho & = & 3 \left[    \frac{\dot{a}^2}{a^2}+\frac{k}{a^2}  \right]   \label{friedmann_eq1} \\
8\pi G_N p & = & - \left[    \frac{\dot{a}^2}{a^2}+\frac{k}{a^2}  +2\frac{\ddot{a}}{a} \right].  \label{friedmann_eq2}
\end{eqnarray}
And consequently, equation (\ref{friedmann_eq1}) and equation (\ref{friedmann_eq2}) imply:
\begin{equation}
8\pi G_N \left[ \rho+3p \right] = -6 \frac{\ddot{a}}{a} . \label{friedmann_eq3}
  \end{equation}
The Friedmann\index{Friedmann} equations completely specify the evolution of the universe as a function of time. The difficulty remains to detemermine a suitable matter model for $\rho$ and $p$, that is to make even more progress, it is necessary to choose an \textit{equation of state} between $\rho$ and $p$.

\subsection{Cosmological parameters}  \label{cosmo_parameters}
This section introduces some of the basic terminology associated with the cosmological parameters.

The \textit{rate of expansion} is characterized by the \textit{Hubble parameter}: \index{Hubble!parameter}
\begin{equation}
H=\frac{\dot{a}}{a}
\end{equation}

The Hubble \index{Hubble!parameter} parameter quantifies the ``speed" with which the size of the universe is increasing.
The value of the Hubble parameter at the present epoch is the Hubble  \index{Hubble!constant}constant,$H_0$. There
is currently a great deal of controversy about what its actual value is, currently the measurements\footnote{See S. Eidelman \emph{et al.} from the Particle Data Group~\cite{PDBook} for recent measurement values.} give:
\begin{equation}
H_0=100 \; h\; \; \;  \mathrm{km/s/Mpc},
\end{equation}
with a present day normalized Hubble \index{Hubble!expansion rate} expansion rate $h$,
\begin{equation}
h=0.71^{+0.04}_{-0.03}
\end{equation}
``Mpc" stands for ``megaparsec", $1\; \mathrm{Mpc} \cong 3\times10^{24} \mathrm{cm}$.

The universe is expanding, therefore we know that $\dot{a}>0$. From equation (\ref{friedmann_eq3}) we also know that $\ddot{a}<0$ when assuming that the pressure $p$ and the density $\rho$ are both positive. The universe must have been expanding at a faster and faster rate when going back in time. If we consider that the universe have always been expanding at the present rate, then at the time $T=H^{-1}=a/\dot{a}$ ago, the\index{scale factor} scale factor $a$ would be null, $a=0$. However, the expansion rate was actually faster, therefore, the time at which $a=0$ was even closer to the present. By assuming homogeneity \index{homogeneity}and isotropy\index{isotropy}, general relativity makes the prediction that at a time less than $H^{-1}$ ago, the universe was in a singular state. This singular point referred to as the big bang\index{singularity!big bang} had an infinite density of matter and an infinite curvature of spacetime. \\

The value of the Hubble  \index{Hubble!parameter}parameter changes over time either increasing or decreasing depending on the sign of the \textit{deceleration parameter}:\index{deceleration parameter}
\begin{equation}
q=-\frac{a\ddot{a}}{\dot{a}^2}.
\end{equation}
The \textit{deceleration parameter}\index{deceleration parameter} measures the rate of change of the rate of expansion. Different values, or ranges of values, of $q_0$ correspond to different cosmological models. 
In principle, it should be possible to determine the value of $q_0$ observationally. For example, for a set of identical supernovae within remote galaxies, the relationship between apparent brightness and redshift is dependent on the value of the deceleration\index{deceleration parameter} parameter. Although measurements of this kind are notoriously difficult to make and to interpret, recent observations tend to favor accelerating universe models. In~\cite{Virey:2005ih}, it was found that
 \begin{equation}
q_0 = -0.55^{+0.26}_{-0.13}
\end{equation}

Another useful quantity is the energy density parameter,
 \begin{equation}
\Omega =\frac{8\pi G}{3 H^2}\rho = \frac{\rho}{\rho_{\mathrm{crit}}},
\end{equation}
where the critical density (Hubble density) is
\begin{equation}
\rho_{\mathrm{crit}} =\frac{3 H^2}{8\pi G}.
\end{equation}
This quantity (which will generally change with time) is called the ``critical" density and current measurements\footnote{See S. Eidelman \emph{et al.} from the Particle Data Group~\cite{PDBook} for recent measurement values.}  give:
\begin{equation}
\rho_{\mathrm{crit}} =2.775\times 10^{11} \; h^2 \; M_{\odot} \; Mpc^{-3},
\end{equation}
where $M_{\odot} $ is the solar mass and $h$ is the present day normalized Hubble \index{Hubble!expansion rate} expansion rate.
Using the Friedmann\index{Friedmann} equation (\ref{friedmann_eq1}), we can then write:
\begin{equation}
\Omega -1=\frac{k}{H^2a^2}.
\end{equation}
The sign of $k$ is therefore determined by whether the energy density parameter $\Omega$ is greater than, equal to, or less than
one. Indeed,
\begin{equation}
\left.\begin{array}{ccccccc}
\rho < \rho_{\mathrm{crit}} &\ \leftrightarrow & \Omega <1 &  \leftrightarrow  & k=-1 &  \leftrightarrow & \text{open} \\
\rho = \rho_{\mathrm{crit}} &\ \leftrightarrow & \Omega =1 &  \leftrightarrow  & k=0 &  \leftrightarrow & \text{flat} \\
\rho > \rho_{\mathrm{crit}} &\ \leftrightarrow & \Omega >1 &  \leftrightarrow  & k=+1 &  \leftrightarrow & \text{closed} 
\end{array}\right.
\end{equation}
The density parameter, then, indicates which of the three Robertson-Walker geometries describes
our universe. Determining it observationally is an area of intense investigation, however, presently, it is thought to be\footnote{See S. Eidelman \emph{et al.} from the Particle Data Group~\cite{PDBook} for recent measurement values.}:
\begin{equation}
\Omega = 1.02\pm 0.02.
\end{equation}

\section{Conclusion}
This chapter recalls the main important features of general relativity that makes remarkable statements concerning the structure of space and time and the structure of the gravitational field (e.g. Einstein's equations\index{Einstein!field equations}). \\
The energy conditions\index{energy conditions} of general relativity permit one to deduce very powerful and general theorems about the behaviour of strong gravitational fields and cosmological geometries.\\
A description of our universe has been outlined in section \ref{cosmology}. Many important issues in cosmology remain to be solved, but it is already clear that general relativity has provided a successful picture of our universe.

\chapter[Necessary and sufficient conditions for big bangs, bounces, rips...]{Necessary and sufficient conditions for big bangs\index{singularity!big bang}, bounces, crunches, rips, 
sudden singularities, and extremality events}  
\label{Chapter_bangs}

\def\bang{{\ast}}
\def\crunch{{\circledast}}
\def\bounce{{\bullet}}
\def\inflexion{{\circledcirc}}
\def\turnaround{{\circ}}
\def\rip{{\divideontimes}}
\def\sudden{{\circleddash}}
\def\generic{{\odot}}


\def\d{{\mathrm{d}}}
\def\union{\cup}
\def\implies{\Rightarrow}
\def\endu{\mathrm{end}}

\section{Introduction}
In section \ref{cosmology}, we have discussed that when averaged over sufficiently large volumes the universe and the matter in the universe should be \textit{isotropic} and \textit{homogeneous}.

As a consequence, the cosmological principle, which is ultimately a distillation of our knowledge of observational cosmology, leads one to consider cosmological spacetimes of the idealized FRW form~\cite{Carroll:2004st, Hartle:2003yu, Gravitation, Wald:1984rg}:
\begin{equation}  \label{metric_FRW}
\d s^{2}=-\d t^{2}+a(t)^{2}\left\{\frac{\d r^{2}}{1-kr^{2}}+r^{2}\;[\d\theta^{2}+\sin^2\theta\; \d\phi^{2}]\right\},
\end{equation}
\index{universe!FRW}\index{universe!FLRW}
where, $a(t)$ is the\index{scale factor} scale factor of the universe and should be positive, and, where, the parameter $k$ takes only three values:
\begin{itemize}
\item $k=-1$, negative curvature
\item $k=0$, flat space
\item $k=+1$, positive curvature. 
\end{itemize}

The central question in cosmology is now the prediction of the history of the scale factor $a(t)$~\cite{Visser:2004bf,Visser:2003vq}. \\

As discussed in  \ref{Cosmodynamics}, and assuming we can apply the Einstein\index{Einstein!field equations} equations of classical general relativity to cosmology in the large, we get some key dynamical equations: the Friedmann \index{Friedmann}equations (\ref{Fried_eq1}, \ref{Fried_eq2}, \ref{Fried_eq3}, \ref{Fried_eq4}).

Using units where $8\pi G_{N}=1$ and $c=1$, we have:
\begin{eqnarray}
\rho(t) & = & 3 \left(   \frac{\dot{a}^2}{a^2}+\frac{k}{a^2}  \right),   \label{Fried_eq1}\\
p(t)      & = & -2\frac{\ddot{a}}{a}-\frac{\dot{a}^2}{a^2}-\frac{k}{a^2},    \label{Fried_eq2}\\
\rho(t)+3p(t) & =& -6\frac{\ddot{a}}{a},                                                      \label{Fried_eq3}
\end{eqnarray}
and the related conservation equation
\begin{equation}
\dot{\rho}(t) a^3+3 \left[   \rho(t)+ p(t)    \right] a^2\dot{a}=0    \label{Fried_eq4}
\end{equation}

\begin{theorem}
 Any two of equations (\ref{Fried_eq1}, \ref{Fried_eq2}, \ref{Fried_eq3}) imply the remaining one, and imply equation (\ref{Fried_eq4}). Equation (\ref{Fried_eq4}) and any one of equations (\ref{Fried_eq1}, \ref{Fried_eq2}, \ref{Fried_eq3}) imply the two other equations but with a specific choice of integration constant.
\end{theorem}

\begin{lemma} \label{lemma_1}
Equations (\ref{Fried_eq1}) and (\ref{Fried_eq2}) imply equations (\ref{Fried_eq3}) and (\ref{Fried_eq4}).
\end{lemma}
\begin{proof}
\begin{eqnarray}
\rho+3p   & = & 3 \left(   \frac{\dot{a}^2}{a^2}+\frac{k}{a^2}  \right) -2\frac{\ddot{a}}{a}-\frac{\dot{a}^2}{a^2}-\frac{k}{a^2}  \\
                & = & \frac{1}{a^2}\left(   3\dot{a}^2+3k-6\ddot{a}a-3\dot{a}^2-3k    \right) \\
                 & =& -6\frac{\ddot{a}}{a}      \label{eq3}  
\end{eqnarray}
Equation (\ref{eq3}) is exactly equation (\ref{Fried_eq3}).
Now, using equation (\ref{Fried_eq1}), we derive:
\begin{eqnarray}
\dot{\rho}(t)   &=  & \frac{3}{a^4} \left(   2a^2 \dot{a}\ddot{a}-2a \dot{a}^3-2ka\dot{a}  \right)\\
\dot{\rho}(t)a^3    &=  & 6a \dot{a}\ddot{a}-6 \dot{a}^3-6k\dot{a}  \label{eq_cons1}
\end{eqnarray}
We also have:
\begin{eqnarray}
3 \left( \rho +p  \right) a^2 \dot{a} & = & 3\dot{a} \left(   -2a\ddot{a}+ 2\dot{a}+2k \right) \\
                                                          & = &  -6a \dot{a}\ddot{a}+6 \dot{a}^3+6k\dot{a}  \label{eq_cons2}
\end{eqnarray}
Now adding equations ( \ref{eq_cons1}) and (\ref{eq_cons2}): 
\begin{eqnarray}
3 \left( \rho +p  \right) a^2 \dot{a} +\dot{\rho}(t)a^3 &=& -6a \dot{a}\ddot{a}+6 \dot{a}^3+6k\dot{a} + 6a \dot{a}\ddot{a}-6 \dot{a}^3-6k\dot{a}  \\
                                                                                        &=&0
\end{eqnarray}
Therefore we obtain equation (\ref{Fried_eq4}) from equations  (\ref{Fried_eq1}) and ( \ref{Fried_eq2}).
\end{proof}

\begin{lemma}     \label{lemma_2}
Equations (\ref{Fried_eq1}) and (\ref{Fried_eq3}) imply equations (\ref{Fried_eq2}) and (\ref{Fried_eq4}).
\end{lemma}
\begin{proof}
Using equation (\ref{Fried_eq3}), we can isolate $p$ and replace $\rho$ by equation (\ref{Fried_eq1}):
\begin{eqnarray}
 p     & = &    -\frac{6\ddot{a}}{3a}-\frac{\rho}{3}, \\
        & =&  -2\frac{\ddot{a}}{a}-\frac{\dot{a}^2}{a^2}-\frac{k}{a^2}   \label{eq_p}
\end{eqnarray}
Thus, equations (\ref{Fried_eq1}) and (\ref{Fried_eq3}) imply equations (\ref{Fried_eq2}).

And as proven for Lemma \ref{lemma_1}., equations (\ref{Fried_eq1}) and (\ref{Fried_eq2}) imply equations (\ref{Fried_eq3}).
\end{proof}

\begin{lemma}
Equations (\ref{Fried_eq1}) and (\ref{Fried_eq4}) imply equations (\ref{Fried_eq2}) and (\ref{Fried_eq3}).
\end{lemma}
\begin{proof}
Using equation (\ref{Fried_eq1}), we replace $\rho$ and $\dot{\rho}$ in equation (\ref{Fried_eq4}) to obtain an equation for the pressure $p$.
\begin{eqnarray}
  \dot{\rho}(t) a^3+3 \left(   \rho(t)+ p(t)    \right)a^2\dot{a} &=& 0 \\
\Longrightarrow   \; \; \; \; \; \;   \; \; \; \; \; \;     \; \; \; \; \; \;   \; \; \; \; \; \;   \; \;  
3pa^2\dot{a} &=& -9\dot{a}^3-9k\dot{a}-6a\dot{a}\ddot{a}+6\dot{a}^3+6k \dot{a} \\
\Longrightarrow   \; \; \; \; \; \;    \; \; \; \; \; \;     \; \; \; \; \; \;   \; \; \; \; \; \;  \; \; \; \; \; \;   \; \; \; 
  p &=& -2\frac{\ddot{a}}{a}-\frac{\dot{a}^2}{a^2}-\frac{k}{a^2}
\end{eqnarray}
Thus, equations (\ref{Fried_eq1}) and (\ref{Fried_eq4}) imply equations (\ref{Fried_eq2}).

And as proven for Lemma \ref{lemma_1}. , equations (\ref{Fried_eq1}) and (\ref{Fried_eq2}) imply equations (\ref{Fried_eq3}).
\end{proof}

\begin{lemma}
Equations (\ref{Fried_eq2}) and (\ref{Fried_eq4}) imply equations (\ref{Fried_eq1}) and (\ref{Fried_eq3}) up to a term proportional to an arbitrary integration constant $K$: $\rho \sim \frac{K}{a^3}$. This integration constant has to be eliminated "by hand"\footnote{Physically, this extra term  $\frac{K}{a^3}$ corresponds to ``dust".}.
\end{lemma}

\begin{proof}
Replacing the pressure $p$ in equation  (\ref{Fried_eq4}), by its value given in equation (\ref{Fried_eq2}), we obtain the following differential equation:
\begin{eqnarray}
3\rho a^2\dot{a}+\dot{\rho}a^3  &=&-3pa^2\dot{a}    \\
3\rho \dot{a}+\dot{\rho}a  &=&  6\frac{\dot{a}\ddot{a}}{a}+3\frac{\dot{a}^3}{a^2}+3\frac{k\dot{a}}{a^2}\\
\dot{\rho} +3\frac{\dot{a}}{a}\rho  &=&   6\frac{\dot{a}\ddot{a}}{a^2}+3\frac{\dot{a}^3}{a^3}+3\frac{k\dot{a}}{a^3}
\end{eqnarray}
By integrating this differential equation in $\rho$, we get:
\begin{equation}
\rho = \frac{\int   \left\{ \exp \left( \int \frac{3\dot{a}}{a}dt   \right) \left(  6\frac{\dot{a}\ddot{a}}{a^2}+3\frac{\dot{a}^3}{a^3}+3\frac{k\dot{a}}{a^3}  \right)  \right\} dt +K}{\exp \left( \int \frac{3\dot{a}}{a}dt   \right)},
\end{equation}
where $K$ is an integration constant. 
Furthermore, 
\begin{equation}
\exp \left( \int \frac{3\dot{a}}{a}dt \right) = \exp \left( 3\ln(a) \right) = a^3,
\end{equation}
and therefore,
\begin{eqnarray}
\int   \left\{ \exp \left( \int \frac{3\dot{a}}{a}dt   \right) \left(  \frac{3\rho \dot{a}+\dot{\rho}a }{a}\right)  \right\} dt &= &\int  \left\{ a^3  \left(  \frac{3\rho \dot{a}+\dot{\rho}a }{a}  \right)  \right\} dt   \\
                                                     & = & \int   \left(  6a\dot{a}\ddot{a}+3\dot{a}^3+3k\dot{a}  \right) dt \\
                                                     & = &  3 a\dot{a}^2 +3 ka
\end{eqnarray}

We finally have:
\begin{equation}
\rho = 3 \left(  \frac{\dot{a}^2}{a^2} + \frac{k}{a^2} +\frac{K}{3a^3} \right)  \label{eq_constant}
\end{equation}
The last term $\dfrac{K}{3a^3}$  of equation (\ref{eq_constant}) has to be eliminated "by hand" to obtain exactly equation (\ref{Fried_eq1}). 

And as proven for Lemma \ref{lemma_1}., equations (\ref{Fried_eq1}) and (\ref{Fried_eq2}) imply equations (\ref{Fried_eq3}).
\end{proof}

\begin{lemma}
Equations (\ref{Fried_eq3}) and (\ref{Fried_eq4}) imply equations (\ref{Fried_eq1}) and (\ref{Fried_eq2}) with $k$ arising as an integration constant and being set to $k=-1/0/+1$ by suitably rescaling $a(t)$.
\end{lemma}

\begin{proof}
Using equations (\ref{Fried_eq3}) and  (\ref{Fried_eq4}), we get:
\begin{eqnarray}
\dot{\rho}(t) a^3+3 \left(   \rho(t)+   \left(-2\frac{\ddot{a}}{a}-\frac{\dot{a}^2}{a^2}-\frac{k}{a^2} \right)  \right)a^2\dot{a}  & = & 0 \\
\Longrightarrow \; \; \; \;  \dot{\rho} +2\frac{\dot{a}}{a}\rho & = & 6\frac{\dot{a}\ddot{a}}{a^2} 
\end{eqnarray}
 By integrating the differential equation, we obtain:
 \begin{equation}
\rho = \frac{\int  \left\{   \exp \left( \int 2\frac{\dot{a}}{a}dt  \right) 6\frac{\dot{a}\ddot{a}}{a^2} \right\}  dt +K }{ \exp \left( \int 2\frac{\dot{a}}{a}dt  \right) },
\end{equation}
where $K$ is an integration constant. Furthermore,
\begin{equation}
 \exp \left( \int 2\frac{\dot{a}}{a}dt  \right)  = \exp  \left(  2 \ln(a) \right)=a^2, 
\end{equation}
and therefore,
\begin{eqnarray}
\rho & = & \frac{\int 6 \dot{a}\ddot{a}dt +K}{a^2}  \\
        & = & \frac{\int 3 \frac{d\dot{a}^2}{dt} +K}{a^2}\\
        & =& \frac{ 3 \dot{a}^2+K}{a^2}
\end{eqnarray}

By suitably rescaling $a(t)$, we can, without loss of generality set this integration constant to $k=-1/0/+1$. That is \begin{itemize}
\item if $K=0$, then we do not need to rescale $a(t)$.
\item if $K\neq 0$, then we can define $a(t)_{\mathrm{new}}=a(t)/\sqrt{| K |}$. After this substitution, we get 
 \begin{eqnarray}
\rho & = & \frac{ 3\left( \dot{a}_{\mathrm{new}}\sqrt{| K |} \right)^2+K}{(a_{\mathrm{new}}\sqrt{| K |})^2}     \\
        & = &     3 \left(   \frac{\dot{a}_{\mathrm{new}}^2}{a_{\mathrm{new}}^2}+\frac{K}{|K|a_{\mathrm{new}}^2}  \right)       \\
        & =&  3 \left(   \frac{\dot{a}_{\mathrm{new}}^2}{a_{\mathrm{new}}^2}+\frac{k}{a_{\mathrm{new}}^2}  \right),
\end{eqnarray}
with the only possible values of $k=-1/0/+1$.

\end{itemize}

Finally, we get the expected equation:
\begin{equation}
\rho = 3 \left(   \frac{\dot{a}^2}{a^2}+\frac{k}{a^2}  \right)
\end{equation}

And as proven for Lemma \ref{lemma_2}., equations (\ref{Fried_eq1}) and (\ref{Fried_eq3}) imply equations (\ref{Fried_eq2}).
 \end{proof}

The physically relevant  singularities occurring in the \index{Friedmann}Friedmann--Robertson--Walker cosmologies had traditionally been thought to be limited to the ``big bang''\index{singularity!big bang}, and possibly a ``big crunch''\index{singularity!big crunch}\index{big crunch}. However, over the last few years, the zoo of cosmological singularities considered in the literature has become  considerably more extensive, with ``big rips'' and\index{singularity!sudden singularity} ``sudden singularities'' added to the mix, as well as renewed interest in non-singular cosmological events such as\index{singularity!extremality events!bounce} ``bounces'' and \index{singularity!extremality events!turnaround}``turnarounds''\footnote{See references ~\cite{Barrow:2004hk, Barrow:2004xh, Barrow:2004he, Bouhmadi-Lopez:2004me, Calderon:2004bi, Caldwell:1999ew, Chimento:2004ps, Curbelo:2005dh, Fernandez-Jambrina:2004yy, Frampton:2004tn, Frampton:2004xn, Lake:2004fu, Nojiri:2004ip,Stefancic:2004kb}. Examples of sudden singularities have been given earlier, for example in references \cite{Matzner:1986us}, \cite{Matzner:1986nt}, in which they are referred to as ``Crack of Doom" singularities.}. \\

Typically, those singularities are classified in relation to what happens to the\index{scale factor} scale factor $a(t)$ of the universe at the time when the singularity occurs.
\begin{itemize}
\item  For a big bang\index{singularity!big bang} and a big \index{singularity!big crunch}\index{big crunch}crunch, $a(t) \longrightarrow 0$.
\item For a big rip, $a(t) \longrightarrow \infty$.
\item For a \index{singularity!sudden singularity}sudden singularity, the definition varies according to different authors: $a(t)$ is finite at the time of the occurrence of the singularity (everyone agrees on that), however, the first derivative of the scale factor is infinite $\dot{a}(t) \longrightarrow \infty$, or it could be the second derivative $\ddot{a}(t) \longrightarrow \infty$, etc..., depending on how ``sudden" one is willing to describe a \index{singularity!sudden singularity}sudden singularity.
\item For \index{singularity!extremality events!bounce}bounces\index{singularity!extremality events!bounce} and \index{singularity!extremality events!turnaround}turnarounds, $a(t)$ is finite and reaches either a minimum or a maximum.
\item For points of \index{singularity!extremality events!inflexion}inflexion, $a(t)$ is finite and reaches a stationary point.
\end{itemize}

\subsubsection{Problem:}
Is it possible to come up with generic definitions that would apply to a complete catalogue of such cosmological milestones, both at the kinematical and dynamical level?
\begin{itemize}
\item Kinematics:

 Can we define a notion of \index{scale factor!scale factor singularity}``scale-factor singularity'' and find interesting relations between this notion and curvature singularities?
  
 \item Dynamics:
 
  Using the Friedmann\index{Friedmann} equations (without assuming even the existence of any equation of state) is it possible to place constraints on whether or not the classical energy conditions\index{energy conditions} are satisfied at the cosmological milestones?

\end{itemize}

In this chapter, we use these considerations to derive necessary and sufficient conditions for the existence of cosmological milestones such as bangs\index{singularity!big bang}, bounces\index{singularity!extremality events!bounce}, \index{singularity!big crunch}\index{big crunch}crunches, rips, sudden \index{singularity!sudden singularity}singularities, and \index{singularity!extremality events}extremality events. Since the classification is extremely general,  the corresponding results are to a high degree model-independent: in particular, we determine, with a minimum of technical assumptions, the necessary and sufficient conditions for the occurrences of all theses cosmological milestones. Note that we only use classical general relativity to derive those results.

\section{Cosmological milestones and kinematics.}
\subsection{Definitions}
In this section, we develop precise definitions of a hopefully complete catalogue of cosmological milestones.\\

It is clear that it would be really convenient to have some unspecified generic cosmological milestone defined in terms of the behaviour of the\index{scale factor} scale factor $a(t)$, and which occurs at some finite time $t_\generic$. 

Generalized Frobenius series are commonly used when expanding solutions of differential equations around their singular points. Motivated by this property, we will assume that in the vicinity of the milestone the scale factor has a generalized power series expansion that generalizes the notions of Taylor series, meromorphic Laurent series,  Frobenius series, and Liapunov expansions~\cite{Liapunov} and are even more general than the generalized Frobenius series adopted in~\cite{Visser:2002ww}.
Indeed, in the present context, if the scale factor $a(t)$ is representable by such a generalized power series, then by the Friedmann equations both $\rho(t)$ and $p(t)$ are representable by such power series.  Formal reversion of the power series then implies that the equation of state $\rho(p)$, and thence also the function $\rho(a)$ possess such generalized power series. Conversely, if $\rho(a)$ is representable by such a generalized power series, then by the first Friedmann equation, $\dot a(t)$ has a power series of this type, which upon integration implies that $a(t)$ itself possesses such a power series.  Similarly, if the equation of state $p(\rho)$ is representable by such a generalized power series then by integrating the conservation equation we have
\begin{equation}
a(\rho) =   a_* \; \exp\left\{ {1\over3} \int_{\rho_*}^\rho {\d\bar\rho\over\bar\rho+p(\bar\rho)}\right\},
\end{equation}
which will now also possess such a generalized power series. That an extension of the usual concept of a Frobenius series is likely to be useful is already clear from the analysis of~\cite{Visser:2002ww}. We should also be clear concerning what type of object falls outside this class of generalized power series: First, essential singularities [effectively poles of infinite order, arising for example in functions such as $\exp(-1/x)$ considered in the neighbourhood of $x=0$] lie outside this classification, and secondly, certain variations on the notion of Puiseux series [specifically, series containing $(\ln x)^n$, $(\ln\ln x)^n$, $(\ln\ln\ln x)^n$,...] also lie outside this classification. We are not aware of any situations in which these exceptional cases become physically relevant.

\begin{definition}  \label{def_generic_a}
\emph{Generic finite-time cosmological milestones: } 
Suppose we have some unspecified generic cosmological milestone,  that  is defined in terms of the behaviour of the\index{scale factor} scale factor $a(t)$, and which occurs at some finite time $t_\generic$. We will assume that in the vicinity of the milestone the scale factor has a (possibly one-sided) generalized power series expansion of the form
\begin{equation}
a(t) = c_0 |t-t_\generic|^{\eta_0} + c_1  |t-t_\generic|^{\eta_1} + c_2  |t-t_\generic|^{\eta_2} 
+ c_3 |t-t_\generic|^{\eta_3} +\dots
\end{equation}
where the indicial exponents $\eta_i$ are generically real (but are often non-integer) and without loss of generality are ordered in such a way that they satisfy
\begin{equation}
\eta_0<\eta_1<\eta_2<\eta_3\dots
\end{equation}
Finally we can also without loss of generality set
\begin{equation}
c_0 > 0.
\end{equation}
There are no \emph{a priori} constraints on the signs of the other $c_i$, though by definition $c_i\neq0$.
\end{definition}  

 From a physical point of view, this definition is really generic and can be applied to any type of cosmological milestone.
 This generalized power series expansion is sufficient to encompass all the models we are aware of in the literature, and as a matter of fact, the indicial exponents $\eta_i$ will be used to classify the type of cosmological milestone we are dealing with.   
For many of the calculations in this chapter, the first term in the expansion is dominant, but even for the most subtle of the explicit calculations below  it will be sufficient to keep only the first three terms of the expansion:
\begin{equation}
a(t) = c_0 |t-t_\generic|^{\eta_0} + c_1  |t-t_\generic|^{\eta_1} + c_2  |t-t_\generic|^{\eta_2} \dots;
\qquad
\eta_0 < \eta_1<\eta_2; 
\qquad c_0 > 0.
\end{equation}
The lowest few of the indicial exponents are sufficient to determine the relationship between these cosmological milestones, the curvature singularities and even the energy \index{energy conditions}conditions of classical general relativity. 
\newline

Note that this expansion fails if the cosmological milestone is pushed into the infinite past or infinite future. We shall return to this point later when we discuss the total age of the universe.
\subsubsection{Big Bangs \index{singularity!big bang}and big crunches}\index{singularity!big crunch}\index{big crunch}
Big bangs and big crunches are some of the most basic cosmological milestones encountered in the literature. For these types of singularities, the\index{scale factor} scale factor $a(t) \rightarrow 0$ at some finite time as one moves to the past or future.

According to the Big Bang\index{singularity!big bang} theory, the universe originated in an infinitely dense and physically paradoxical singularity. Space has expanded with the passage of time, objects being moved farther away from each other.
In cosmology, the Big Bang \index{singularity!big bang}theory is the prevailing theory about the early development and shape of the universe. The central idea is that the observation that galaxies appear to be receding from each other can be combined with the theory of general relativity to extrapolate the conditions of the universe back in time. This leads to the conclusion that as one goes back in time, the universe becomes increasingly hot and dense. However, the big bang \index{singularity!big bang}represents the creation of the universe from a singular state, not explosion of matter into a pre-existing spacetime. It might be hoped that the perfect symmetry of the FRW universes was responsible for this singularity, but however, the singularity theorems predict (under relatively mild conditions) that any universe with $\rho >0$ and $p\geqslant 0$ must have begun at a singularity.

There are a number of consequences to this view. One consequence is that the universe now is very different than the universe in the past or in the future. The Big Bang\index{singularity!big bang} theory predicts that at some point, the matter in the universe was hot and dense enough to prevent light from flowing freely in space. That this period of the universe would be observable in the form of cosmic background radiation (CBR) was first predicted in the 1940s, and the discovery of such radiation in the 1960s swung most scientific opinion against the Big Bang\index{singularity!big bang} theory's chief rival, the steady state theory.
%

In cosmology, the Big Crunch \index{singularity!big crunch}\index{big crunch}is a hypothesis that states the universe will stop expanding and start to collapse upon itself; a counterpart to the Big Bang.
If the gravitational attraction of all the matter in the observable horizon \index{horizon}is high enough, then it could stop the expansion of the universe, and then reverse it. The universe would then contract, in about the same time as the expansion took. Eventually, all matter and energy would be compressed back into a gravitational singularity. It is meaningless to ask what would happen after this, as time would stop in this singularity as well.\\

We shall define the \index{scale factor}scale factor near a big bang \index{singularity!big bang}and a big crunch using the generalized power series expansion mentioned earlier.
\begin{definition}
 Let the time of the big bang\index{singularity!big bang} (if one occurs) be denoted by $t_\bang$ and the time of the big crunch (if one occurs) be denoted by $t_\crunch$. We shall say that the bang or crunch \index{singularity!big crunch}\index{big crunch}behaves with indicial exponents ($0<\eta_0<\eta_1\dots$) if the \index{scale factor}scale factor possesses a generalized power series in the vicinity of the singularity. That is, if
\begin{eqnarray}
a(t) =  c_0 (t-t_\bang)^{\eta_0} + c_1 (t-t_\bang)^{\eta_1} +\dots
\end{eqnarray}
or
\begin{eqnarray}
a(t) =  c_0 (t_\crunch-t)^{\eta_0} \; + c_1 (t_\crunch-t)^{\eta_1} + \dots
\end{eqnarray}
respectively.  Moreover, the series have been carefully constructed to make $a(t_\bang)=0$ and $a(t_\crunch)=0$.

\end{definition}

\subsubsection{Big rips}
A ``big rip" is a singularity for which $a(t)\rightarrow \infty$ at finite time. A big rip could occur in the future or in the past, however, the literature to date has solely considered future rips (as a past rip would be a most unusual and unexpected beginning to the history of the universe).  

The Big Rip is a cosmological hypothesis about the ultimate fate of the Universe.
The key to this hypothesis is the amount of dark energy in the universe. If the universe contains enough dark energy, it could end with all matter being pulled apart.
 First the galaxies would be separated from each other, then gravity would be too weak to hold individual galaxies together. 

We shall write the \index{scale factor}scale factor near a big rip using the generalized power series expansion.
\begin{definition}
 Let the time of the rip, if it occurs, be denoted $t_{\rip}$, then we define the indicial exponents of the rip (either future or past) to be
($\eta_0<\eta_1\dots$) if the scale factor possesses a generalized power series in the vicinity of the rip:
\begin{eqnarray}
a(t) = c_0  |t_{\rip}-t|^{\eta_0} + c_1  |t_{\rip}-t|^{\eta_1} + \dots,
\end{eqnarray}
with $\eta_0<0$ and $c_0>0$.
Moreover, the series has been carefully constructed to make $a(t_\rip)=\infty$.  
\end{definition}
Note the similarity to bangs \index{singularity!big bang}and \index{singularity!big crunch}\index{big crunch}crunches, with the only difference being in the \emph{sign} of the exponent $\eta_0$.

\subsubsection{Sudden singularities}\index{singularity!sudden singularity}
``Sudden singularities" are a recent type of cosmological milestones that have appeared in the literature. There could be past or future sudden singularities for which some time derivative of the \index{scale factor}scale factor diverges at finite time, while the scale factor itself remains finite. Again, future sudden singularities are more popular than past sudden singularities as those ones would be a most unusual and disturbing beginning to the history of the universe.

Recently it has been speculated that in an expanding FLRW universe a curvature singularity may occur at a finite time before a ``Big Crunch" for matter contents that
satisfy both weak and strong energy conditions\index{energy conditions}. This family of models has been further enlarged, and the same sort of behavior has also been found in inhomogeneous models~\cite{Barrow:2004hk, Barrow:2004xh}.\\

We shall define the\index{scale factor} scale factor near a\index{singularity!sudden singularity} sudden singularity using the generalized power series expansion.

\begin{definition}
 Let the time of the sudden \index{singularity!sudden singularity}singularity, if one occurs, be $t_{\sudden}$ (past or future). A suitable definition of the exponent of a sudden singularity is to take $\eta_0=0$ and $\eta_1>0$ to give
\begin{eqnarray}
a(t) =  c_0  + c_1 |t-t_\sudden|^{\eta_1} + \dots
\end{eqnarray}
with  $c_0>0$ and $\eta_1$ non-integer. Thus $a(t_\sudden)=c_0$ is finite and a sufficient number of differentiations yields
\begin{equation}
a^{(n)}(t\to t_\sudden) \sim c_0 \; \eta_1 (\eta_1-1)(\eta_1-2)\dots (\eta_1-n+1) \; |t-t_\sudden|^{\eta_1-n}\to\infty.
\end{equation}

\end{definition}

 The toy model considered by Barrow~\cite{Barrow:2004xh, Barrow:2004hk, Barrow:2004he} can be written 
\begin{equation}
a(t) = c_0 \left[ (t_\sudden-t)^\eta -1 \right] + \tilde c_0 (t-t_\bang)^{\tilde\eta}
\end{equation}
and falls into this classification when expanded around the time of the\index{singularity!sudden singularity} sudden singularity, $t_\sudden$,  while it falls into the classification of big bang\index{singularity!big bang} singularities considered above when expanded around the time of the big bang, $t_\bang$. 

\subsubsection{Extremality events}
\index{singularity!extremality events}
Let's consider some other common cosmological milestones (which are not singularities in any sense) that are referred to as ``extremality events". In particular, for these events, the \index{scale factor}scale factor $a(t)$ exhibits a local extremum at finite time, and whence, $\dot a \rightarrow 0$. In the vicinity of extremality events we can model the scale factor using ordinary Taylor series so that in terms of our generalized series we have $\eta_0=0$ and $\eta_i\in Z^+$. 
\begin{definition}
 A\index{singularity!extremality events!bounce} ``bounce'' \index{singularity!extremality events!bounce}is any local minimum of $a(t)$, the time of such an event being denoted by $t_\bounce$, so that  $a^{(1)}(t_\bounce)=0$. The ``order'' of the bounce is the first nonzero integer $n$ for which the $2n$'th time derivative is strictly positive:
\begin{equation}
a^{(2n)}(t_\bounce) > 0,
\end{equation}
so that 
\begin{equation}
a(t) = a(t_\bounce) + {1\over(2n)!} a^{(2n)}(t_\bounce)\; [t-t_\bounce]^{2n} + \dots.
\end{equation}
\end{definition}
Note that  $a^{(1)}(t_\bounce)=0$  at the bounce is in agreement with~\cite{Allen:2005xh, Bekenstein:1975ww, Brustein:2001xh, Gasperini:2004ss, Gordon:2002jw, Hochberg:1998vm, Hwang:2001zt, Medved:2003sk, Molina-Paris:1998tx, Parker:1990mk, Tippett:2004xj, Veneziano:2003sz} and that the $2n$'th time derivative strictly positive is also mentioned in~\cite{Molina-Paris:1998tx, Hochberg:1998vm}.

\begin{definition}
A\index{singularity!extremality events!turnaround} ``turnaround'' is any local maximum of $a(t)$, the time of such an event being denoted by $t_\turnaround$, 
so that  $a^{(1)}(t_\turnaround)=0$. The ``order'' of the turnaround is the first nonzero integer $n$ for which the $2n$'th time derivative is strictly negative:
\begin{equation}
a^{(2n)}(t_\turnaround) < 0,
\end{equation}
so that
\begin{equation}
a(t) = a(t_\turnaround) + {1\over(2n)!} a^{(2n)}(t_\turnaround)\; [t-t_\turnaround]^{2n} + \dots.
\end{equation}

\end{definition}

Note that turnarounds are mentioned in~\cite{Barrow:1988xi, Burnett:1993bn, Ford:1985qt, Miritzis:2005hg}.

\begin{definition}
An \index{singularity!extremality events!inflexion}``inflexion event'' is an extremality \index{singularity!extremality events}event that is neither a local maximum or a local minimum, the time of such an event being denoted by $t_\inflexion$. (Inflexion events can be thought of as an extreme case of ``loitering'', in the limit where the Hubble \index{Hubble!parameter} parameter momentarily vanishes at the\index{singularity!extremality events!inflexion} inflexion event.)
The order of the inflexion event is the first nonzero $n$ for which
\begin{equation}
a^{(2n+1)}(t_\inflexion) \neq 0,
\end{equation}
so that
\begin{equation}
a(t) = a(t_\inflexion) + {1\over(2n+1)!} a^{(2n+1)}(t_\inflexion)\; [t-t_\inflexion]^{2n+1} + \dots.
\end{equation}
\end{definition}

The loitering universe\index{loitering universe} scenario is an expanding Friedmann cosmology that undergoes a fairly recent phase of slow expansion.
It is during this semi-static phase that large-scale structure would be formed.
Loitering is also characterized by the fact that the Hubble parameter dips in value over a narrow redshift range referred to as the ``loitering epoch". During loitering, density perturbations would be expected to grow rapidly, and, since the expansion of the universe would slow down, its age near loitering would dramatically increase~\cite{Sahni:1991ks, Sahni:2004fb} . 

\begin{definition}
The ``emergent universe'' of~\cite{Ellis:2003qz, Ellis:2002we} can be thought of as an \index{singularity!extremality events}extremality event that has been pushed back into the infinite past.
\end{definition}
The ``emergent universe" \index{emergent universe}scenario consists of an inflationary universe that would emerge from a small static state that has within it the seeds for the development of the macroscopic universe.
The universe would have a finite initial size, with a finite amount of inflation occurring over an infinite time in the past, and with inflation would then coming to an end via reheating in the standard way. The scale-factor would be bounded away from zero in the past and there would be no horizon problem and no singularity, since the initial state would be Einstein static\index{Einstein!static universe}. Also, the initial static state could be chosen to have a radius above the Planck scale, so that these models could even avoid a quantum gravity regime~\cite{Ellis:2003qz, Ellis:2002we}.\\

These definitions have been chosen to match with and simplify the definitions in articles~\cite{Hochberg:1998vm} and ~\cite{Molina-Paris:1998tx}.
Note that for \index{singularity!extremality events!bounce}bounces these definitions imply that there will be some open interval such that
\begin{equation}
\forall t \in (t_\bounce-\Delta,t_\bounce)\union (t_\bounce,t_\bounce+\Delta); \qquad \ddot a(t) > 0,
\end{equation}
while for\index{singularity!extremality events!turnaround} turnarounds there will be some open interval such that
\begin{equation}
\forall t \in (t_\turnaround-\Delta,t_\turnaround)\union  (t_\turnaround,t_\turnaround+\Delta); \qquad \ddot a(t) < 0.
\end{equation}
Note that unless the \index{singularity!extremality events!bounce}bounce or turnaround is of order one we cannot guarantee that at the extremality event itself $\ddot a(t_\bounce)>0$ or $\ddot a(t_\turnaround)<0$.
For\index{singularity!extremality events!inflexion} inflexion events we can only assert the weaker condition of the existence of  some open interval such that
\begin{equation}
\forall t \in(t_\inflexion-\Delta,t_\inflexion)\union (t_\inflexion,t_\inflexion+\Delta); \qquad
\dot a(t) \hbox{ has fixed sign}.
\end{equation}

\subsubsection{Summary}
We have defined singular cosmological milestones (big bang,\index{singularity!big bang} big \index{singularity!big crunch}\index{big crunch}crunch, big rip, sudden \index{singularity!sudden singularity}singularity) and nonsingular cosmological milestones \index{singularity!extremality events}(extremality events) using a single framework based on generalized power series. From a kinematical point of view, major features will be deduced with the first two indicial exponents ($\eta_0$ and $\eta_1$).

The value of the first indicial exponent $\eta_0$ relates to the classification of the types of singular cosmological milestones as follows:
\begin{itemize}
\item $\eta_0>0$ for big bangs\index{singularity!big bang} or big crunches.
\item $\eta_0=0$ for sudden singularities.
\item $\eta_0<0$ for big rips.
\end{itemize}
 
For the nonsingular cosmological milestones (extremality\index{singularity!extremality events} events) , the parameterization is simpler: only one positive integer is needed, i.e. the order of the extremality event, to determine the qualitative behaviour of the Taylor series of the\index{scale factor} scale factor $a(t)$.

\subsection{Spacetime curvature}  \label{spacetime_curvature}

In this section, we will use the parameters define above to explore generic properties of the cosmological milestones. Do all cosmological milestones with scale\index{scale factor} factor singularities lead to curvature singularities?\\

\subsubsection{Cosmological parameters}

Before analyzing the spacetime curvature, we consider two  cosmological parameters:
\begin{itemize}
\item the Hubble \index{Hubble!parameter} parameter $H=\dfrac{\dot{a}}{a}$, characterizing the rate of expansion\footnote{See section \ref{cosmo_parameters}.}.
\item  the deceleration\index{deceleration parameter} parameter\footnote{See section \ref{cosmo_parameters}.}
 $q=-\dfrac{\ddot{a}}{a}H^{-2}=-\dfrac{a\ddot{a}}{\dot{a}^2}$. 
\end{itemize}

\begin{theorem} \label{theo_H_bangs}
 Consider a generalized power series expansion of the \index{scale factor}scale factor as defined in Definition \ref{def_generic_a}, for $\eta_0\neq 0$, that is for bangs\index{singularity!big bang}, \index{singularity!big crunch}\index{big crunch}crunches, and rips, the Hubble \index{Hubble!parameter} parameter exhibits a generic $1/(t-t_\generic)$ blow up, where $t_\generic$ is the the time of the event. Explicitly,
 \begin{equation}
\lim_{t\to t_\generic^+} H = \left\{ \begin{array}{ll}
+\infty & \eta_0>0;\\
-\infty &\eta_0<0.\\
\end{array}
\right.
\end{equation}
\end{theorem}

\begin{proof}
From the 
definition of a generic cosmological milestone
\begin{equation}
a(t) = c_0 |t-t_\generic|^{\eta_0} + c_1  |t-t_\generic|^{\eta_1} +\dots
\end{equation}
we have (assuming $t>t_\generic$ for simplicity, otherwise one need merely formally reverse the flow of time)\footnote{In fact for explicit calculations in the vicinity of any cosmological milestone it is always possible to choose the direction of time to force  $t>t_\generic$ and so dispense with the need to take the absolute value $|t-t_\generic|$, at least for one-sided calculations. This is not a physical restriction on the  cosmological milestone, just a mathematical convenience which we shall adopt henceforth without further explicit discussion.}
\begin{equation}
\dot a(t) = c_0 \eta_0 (t-t_\generic)^{\eta_0-1} + c_1  \eta_1 (t-t_\generic)^{\eta_1-1}  +\dots
\end{equation}
Keeping only the most dominant term, we have for $\eta_0\neq0$
\begin{equation}
H ={\dot a\over a} 
\sim 
{c_0 \eta_0 (t-t_\generic)^{\eta_0-1}\over c_0 (t-t_\generic)^{\eta_0}} 
=  
{\eta_0\over t-t_\generic};  \qquad (\eta_0\neq0).
\end{equation}

That is, for bangs\index{singularity!big bang}, crunches, and rips the Hubble \index{Hubble!parameter} parameter exhibits a generic $1/(t-t_\generic)$ blow up. For $\eta_0 > 0$, $\lim_{t\to t_\generic} \eta_0/(t-t_\generic) = +\infty$ whereas, for  $\eta_0 < 0$, $\lim_{t\to t_\generic} \eta_0/(t-t_\generic) = -\infty$. 
\end{proof}

Theorem \ref{theo_H_bangs} implies that for the particular cosmological milestones such as big bangs\index{singularity!big bang}, big crunches, and big rips, the Hubble \index{Hubble!parameter} parameter goes to infinity in the vicinity of the time of the event.

\begin{theorem} \label{theo_H_sudden}
 Consider a generalized power series expansion of the \index{scale factor}scale factor as defined in Definition \ref{def_generic_a}, for $\eta_0= 0$, that is either a \index{singularity!sudden singularity}sudden singularity or an \index{singularity!extremality events}extremality event, the Hubble  \index{Hubble!parameter}parameter does \emph{not necessarily} diverge at the cosmological milestone. As a matter of fact,
\begin{equation}
\lim_{t\to t_\generic} H = \left\{ \begin{array}{ll}
0 & \eta_0=0; \qquad \eta_1 > 1;\\
c_1/c_0 & \eta_0=0; \qquad \eta_1=1;\\
\mathrm{sign}(c_1)\infty  & \eta_0=0; \qquad \eta_1\in(0,1);\\

\end{array}
\right.
\end{equation}
where $t_\generic$ is the the time of the event.
\end{theorem}
 
 \begin{proof}
For $\eta_0=0$, we need to go to the next highest term in the numerator (a term which depends on $\eta_1$, which is guaranteed to be greater than zero by our definitions) to obtain
\begin{equation}
H \sim  {c_1 \eta_1 (t-t_\generic)^{\eta_1-1}\over c_0 } =  
\eta_1 \;{c_1\over c_0} \; (t-t_\generic)^{\eta_1-1};
\qquad (\eta_0=0; \eta_1>0).
\end{equation}

In particular, this guarantees that power law behaviour is completely generic near the cosmological milestone, and while  the value of the exponent is typically $-1$, there is an exceptional class of milestones (the \index{singularity!sudden singularity}sudden singularities and \index{singularity!extremality events}extremality events) for which the exponent will differ. It is \emph{not} automatic that the Hubble  \index{Hubble!parameter}parameter diverge at the cosmological milestone. In fact, 
\begin{equation}
\lim_{t\to t_\generic} H =  \eta_1 \;{c_1\over c_0} \; (t-t_\generic)^{\eta_1-1}  = 
 \left\{ \begin{array}{ll}
0  \; \; \; \; \; \; \qquad  \eta_0=0; \qquad \eta_1 > 1;\\                                                                                                                                     c_1/c_0 \qquad \eta_0=0; \qquad \eta_1=1;
 
\end{array}
\right.
\end{equation}
\end{proof}
Theorem \ref{theo_H_sudden} implies that for the particular cosmological milestones such as sudden singularities and extremality events, the Hubble \index{Hubble!parameter} parameter does not necessary diverge and in particular, the \index{Hubble!parameter} Hubble parameter has a finite limit iff $\eta_0=0$, $\eta_1\geq1$, corresponding to a particular subset of the\index{singularity!sudden singularity} sudden singularities.\\

The following theorem is a summary of the results encountered in Theorem \ref{theo_H_bangs} and Theorem \ref{theo_H_sudden}. 
\begin{theorem} \label{theo_H_general}
\emph{Summary:}\\
 Consider a generalized power series expansion of the\index{scale factor} scale factor as defined in Definition \ref{def_generic_a}, the  \index{Hubble!parameter}Hubble parameter does \emph{not necessarily} diverge at every different cosmological milestone. Indeed, 
\begin{equation}
\lim_{t\to t_\generic} H = \left\{ \begin{array}{ll}
+\infty & \eta_0>0;\\
0 & \eta_0=0; \qquad \eta_1 > 1;\\
c_1/c_0 & \eta_0=0; \qquad \eta_1=1;\\
\mathrm{sign}(c_1)\infty  & \eta_0=0; \qquad \eta_1\in(0,1);\\
-\infty &\eta_0<0,\\
\end{array}
\right.
\end{equation}
where $t_\generic$ is the the time of the event. 
\end{theorem}

\begin{proof}
Theorem \ref{theo_H_bangs} and Theorem \ref{theo_H_sudden} implies Theorem \ref{theo_H_general}.
\end{proof}

Now that we have analyzed the  \index{Hubble!parameter}Hubble parameter in detail, we will consider the so-called \index{deceleration parameter}deceleration \index{deceleration parameter}parameter $q=-\dfrac{\ddot{a}}{a}H^{-2}=-\dfrac{a\ddot{a}}{\dot{a}^2}$ and give some similar results.

\begin{theorem}  \label{theorem_q}
 Consider a generalized power series expansion of the\index{scale factor} scale factor as defined in Definition \ref{def_generic_a}, the deceleration \index{deceleration parameter}parameter $q$ can be \emph{either finite or infinite} as the cosmological milestone is approached:
 \begin{itemize}
\item For bangs\index{singularity!big bang}, \index{singularity!big crunch}\index{big crunch}crunches, and rips, ($\eta_0\neq 0$) the limit of $q$ is always finite.
\item For a certain subset of the \index{singularity!sudden singularity}sudden singularities ($\eta_0=0$) the limit $q$ is infinite.
\end{itemize}
  Indeed, 
 \begin{equation}
\lim_{t\to t_\generic} q = \left\{ \begin{array}{ll}
(1-\eta_0)/\eta_0 & \qquad\eta_0\neq0;\\
\mathrm{sign}(c_1[1-\eta_1])\infty & \qquad\eta_0=0; \qquad \eta_1\neq 1;\\
0 & \qquad\eta_0=0; \qquad \eta_1= 1; \qquad\eta_2 > 2;\\
-2c_2c_0/c_1^2& \qquad\eta_0=0; \qquad \eta_1= 1; \qquad\eta_2=2; \\
-\mathrm{sign}(c_2)\infty &\qquad \eta_0=0; \qquad \eta_1= 1; \qquad\eta_2\in(1,2), \\
\end{array}
\right.
\end{equation}
where $t_\generic$ is the the time of the event. 
\end{theorem}

\begin{proof}
 First, we will consider the cosmological acceleration $\ddot a$:
\begin{equation}
\ddot a(t) = c_0 \eta_0 (\eta_0-1) |t-t_\generic|^{\eta_0-2} 
+ c_1  \eta_1 (\eta_1-1) |t-t_\generic|^{\eta_1-2}  
+ c_2  \eta_2  (\eta_2-1) |t-t_\generic|^{\eta_2-2}  
+\dots
\end{equation}
Remember that for the \index{scale factor}scale factor $a(t)$, we have:
 \begin{equation}
a(t) = c_0 |t-t_\generic|^{\eta_0} + c_1  |t-t_\generic|^{\eta_1} + c_2  |t-t_\generic|^{\eta_2} 
+ c_3 |t-t_\generic|^{\eta_3} +\dots
\end{equation}
Then provided $\eta_0\neq 0$ and $\eta_0\neq 1$
\begin{equation}
{\ddot a\over a} \sim {\eta_0(\eta_0-1)\over (t-t_\generic)^2}.
\end{equation}
If $\eta_0=0$, then provided $\eta_1\neq 1$
\begin{equation}
{\ddot a\over a} \sim {\eta_1(\eta_1-1)c_1\over c_0} \;  (t-t_\generic)^{\eta_1-2}.
\end{equation}
If both $\eta_0=0$ and $\eta_1= 1$
\begin{equation}
{\ddot a\over a} \sim {\eta_2(\eta_2-1)c_2\over c_0} \;  (t-t_\generic)^{\eta_2-2}.
\end{equation}
Finally, if $\eta_0=1$, then (since $\eta_1>1$)
\begin{equation}
{\ddot a\over a} \sim {\eta_1(\eta_1-1)c_1\over c_0} \;  (t-t_\generic)^{\eta_1-3}.
\end{equation}
The behaviour of $\ddot a/a$ near the milestone will be \emph{some} power law, though the precise exponent of that power law will depend on the interplay between the various indicial exponents $\eta_i$. Note that there is at least one situation in which we have to calculate up  to the third exponent $\eta_2$.
We can now consider the so-called \index{deceleration parameter}deceleration parameter
\begin{equation}
q = - {\ddot a\over a} H^{-2} = - {a\; \ddot a\over \dot a^2}
\end{equation}
Following the same sort of analysis, for $\eta_0\neq0$ and $\eta_0\neq1$ we have the ``generic" result:
\begin{equation}
q \sim  {1-\eta_0\over \eta_0};  \qquad (\eta_0\neq 0,1).
\end{equation}
For the ``exceptional'' cases we easily see:\\
--- If $\eta_0=0$, then provided $\eta_1\neq 1$
\begin{equation}
q \sim -{(\eta_1-1)c_0\over \eta_1 c_1} \;  (t-t_\generic)^{-\eta_1}.
\end{equation}
--- If both $\eta_0=0$ and $\eta_1= 1$
\begin{equation}
q \sim -{c_2 c_0\over c_1^2} \eta_2(\eta_2-1) \;  (t-t_\generic)^{\eta_2-2}.
\end{equation}
--- Finally, if  $\eta_0=1$, then (since $\eta_1>1$)
\begin{equation}
q \sim - {\eta_1(\eta_1-1)c_1\over c_0} \;  (t-t_\generic)^{\eta_1-1}.
\end{equation}
Again we see the ubiquity of power law behaviour, with a ``generic'' case and a limited number of special cases. Using the generic case and the special cases, we can calculate the limit of the deceleration\index{deceleration parameter} parameter at the cosmological milestone.
For $\eta_0\neq0$ and $\eta_0\neq1$ we get the ``generic" result:
\begin{equation}
\lim_{t\to t_\generic} q =(1-\eta_0)/\eta_0   \qquad (\eta_0\neq 0,1).
\end{equation}
For the ``exceptional'' cases we get:\\
--- If $\eta_0=0$, then provided $\eta_1\neq 1$
\begin{equation}
\lim_{t\to t_\generic} q = \mathrm{sign}(c_1[1-\eta_1])\infty .
\end{equation}
--- If both $\eta_0=0$ and $\eta_1= 1$
\begin{equation}
\lim_{t\to t_\generic} q =  \left\{ \begin{array}{ll}
0 & \qquad\eta_2 > 2;\\
-2c_2c_0/c_1^2&  \qquad\eta_2=2; \\
-\mathrm{sign}(c_2)\infty &\qquad\eta_2\in(1,2). \\
\end{array}
\right.
\end{equation}
--- Finally, if  $\eta_0=1$, then (since $\eta_1>1$)
\begin{equation}
\lim_{t\to t_\generic} q =  0 .
\end{equation}
Note that in this case, the next dominant term $\eta_1$ goes to $0$ as $t \to t_\generic$. We can therefore write this limit as
\begin{equation}
\lim_{t\to t_\generic} q =  (1-\eta_0)/\eta_0 |_{\eta_0=1}=0 .
\end{equation}

Hence, we have for the\index{deceleration parameter} deceleration parameter:
 \begin{equation}
\lim_{t\to t_\generic} q = \left\{ \begin{array}{ll}
(1-\eta_0)/\eta_0 & \qquad\eta_0\neq0;\\
\mathrm{sign}(c_1[1-\eta_1])\infty & \qquad\eta_0=0; \qquad \eta_1\neq 1;\\
0 & \qquad\eta_0=0; \qquad \eta_1= 1; \qquad\eta_2 > 2;\\
-2c_2c_0/c_1^2& \qquad\eta_0=0; \qquad \eta_1= 1; \qquad\eta_2=2; \\
-\mathrm{sign}(c_2)\infty &\qquad \eta_0=0; \qquad \eta_1= 1; \qquad\eta_2\in(1,2). \\
\end{array}
\right.
\end{equation}

Therefore, the \index{deceleration parameter}deceleration parameter can be either finite or infinite, in particular, it has an infinite limit only for a certain subset of the \index{singularity!sudden singularity}sudden singularities $\eta_0=0$, and for bangs\index{singularity!big bang} and \index{singularity!big crunch}\index{big crunch}crunches the limit is always finite. This is largely because the definition of the deceleration parameter was carefully chosen to eliminate the leading $t$ behaviour whenever possible.
\end{proof}

\subsubsection{Polynomial curvature singularities} \index{singularity!polynomial curvature singularity}

\def\Rtt{{ R_{\hat t\hat t} }}
\def\Gtt{{ G_{\hat t\hat t} }}

Now to decide if the cosmological milestones we have defined are curvature singularities, we need to have a look at the\index{Riemann!tensor} Riemann tensor.
The curvature is measured by the Riemann tensor from which  can be constructed various scalar quantities, such as the Ricci\index{Ricci!scalar} scalar $R$, or higher order scalars $R^{ab}R_{ab}$, $R^{abcd}R_{abcd}$, etc...  If any of those scalars go to infinity when approaching some point $x$, there is a polynomial curvature singularity at this point $x$.

 Because of the symmetries of FRW geometry, the Weyl \index{Weyl tensor}tensor is automatically zero and so it suffices to consider the Ricci \index{Ricci!tensor}tensor (and implicitly the Ricci scalar). But because of spherical symmetry, and the perhaps less obvious translational symmetry, the only two non-zero orthonormal  independent components of the Ricci tensor are
\begin{equation}
R_{\hat t\hat t} \qquad\hbox{and} \qquad R_{\hat r\hat r}=R_{\hat\theta\hat\theta}=R_{\hat\phi\hat\phi}.
\end{equation}
Thus to test for all possible polynomial curvature singularities it suffices to test for singularities in, for instance, $R_{\hat t\hat t}$ and $R_{\hat r\hat r}$. Alternatively one could consider $\Rtt$ and the Ricci scalar\index{Ricci!scalar} $R$, or even $\Rtt$ and $\Gtt$.

From the metric \ref{metric_FRW}, we can calculate the $\Rtt$ and $\Gtt$ components in an orthonormal basis.  We find:
\begin{equation}
\Rtt = - 3 \;{\ddot a\over a};
\end{equation}
\begin{equation}
\Gtt = 3 \left( {\dot a^2\over a^2} + {k\over a^2}\right).
\end{equation}

The interest of using those particular two combinations of the orthonormal components of the Ricci \index{Ricci!tensor}tensor is that they satisfy interesting properties:
\begin{itemize}
\item they are linearly independent;
\item $\Rtt$ is independent of the curvature of space (no $k$ dependence);
\item $\Gtt$ is independent of $\ddot a$ (minimizing the number of derivatives involved);
\item testing these two objects for finiteness is sufficient to completely characterize \emph{all} polynomial curvature singularities in a FRW geometry. 
\end{itemize}

\begin{theorem}  \label{theo_Rtt}
\emph{The $\Rtt$ theorem:}

 Consider a generalized power series expansion of the \index{scale factor}scale factor as defined in Definition \ref{def_generic_a}, the $\Rtt$ component of the Ricci tensor in an orthonormal basis remains \emph{finite}, as the cosmological milestone is approached, only provided:
 \begin{itemize}
\item $\eta_0=0$, $\eta_1\geq 2$;
\item $\eta_0=0$, $\eta_1=1$, $\eta_2\geq2$;
\item $\eta_0=1$ and $\eta_1\geq 3$.
\end{itemize}
That is, except for these above rather limited cases, $\Rtt$ will blow up to \emph{infinity} as the cosmological milestone is approached.
\end{theorem}

\begin{proof}
The $\Rtt$ component of the Ricci \index{Ricci!tensor}tensor can be written as
\begin{equation}
\Rtt = - 3 \;{\ddot a\over a}.
\end{equation}
Using the generalized power series expansion we can rewrite the $\Rtt$ term only as a function of some $(t-t_\generic)$ factors in the vicinity of the event.\\
Then provided $\eta_0\neq 0$ and $\eta_0\neq 1$
\begin{equation}
\Rtt = - 3{\ddot a\over a} \sim-3 {\eta_0(\eta_0-1)\over (t-t_\generic)^2}.
\end{equation}
If $\eta_0=0$, then provided $\eta_1\neq 1$
\begin{equation}
\Rtt = - 3{\ddot a\over a} \sim -3{\eta_1(\eta_1-1)c_1\over c_0} \;  (t-t_\generic)^{\eta_1-2}.
\end{equation}
If both $\eta_0=0$ and $\eta_1= 1$
\begin{equation}
\Rtt = - 3{\ddot a\over a} \sim -3{\eta_2(\eta_2-1)c_2\over c_0} \;  (t-t_\generic)^{\eta_2-2}.
\end{equation}
Finally, if $\eta_0=1$, then (since $\eta_1>1$)
\begin{equation}
\Rtt = - 3{\ddot a\over a} \sim -3{\eta_1(\eta_1-1)c_1\over c_0} \;  (t-t_\generic)^{\eta_1-3}.
\end{equation}
The generic result is
\begin{equation}
\lim_{t\to t_\generic} \Rtt = \mathrm{sign}(\eta_0[1-\eta_0])\, \infty
\qquad \hbox{for} \qquad \eta_0\neq 0; \quad \eta_0\neq 1,
\end{equation}
For the ``exceptional'' cases we get:\\
--- If $\eta_0=0$, then provided $\eta_1\neq 1$
\begin{equation}
\lim_{t\to t_\generic} \Rtt = 
\left\{ \begin{array}{ll}
0 & \qquad\eta_0=0;  \qquad \eta_1 > 2;\\
-6c_1/c_0 & \qquad\eta_0=0;  \qquad \eta_1 = 2;\\
\mathrm{sign}(c_1)\infty & \qquad\eta_0=0; \qquad \eta_1\in(1,2);\\
-\mathrm{sign}(c_1)\infty & \qquad\eta_0=0; \qquad  \eta_1 \in(0,1);\\
\end{array}
\right.
\end{equation}
--- If both $\eta_0=0$ and $\eta_1= 1$
\begin{equation}
\lim_{t\to t_\generic} \Rtt = 
\left\{ \begin{array}{ll}
0 & \qquad\eta_0=0; \qquad \eta_1= 1; \qquad\eta_2 > 2;\\
-6c_2/c_0& \qquad\eta_0=0; \qquad \eta_1= 1; \qquad\eta_2=2; \\
-\mathrm{sign}(c_2)\infty &\qquad \eta_0=0; \qquad \eta_1= 1; \qquad\eta_2\in(1,2); \\
\end{array}
\right.
\end{equation}
--- Finally, if  $\eta_0=1$, then (since $\eta_1>1$)
\begin{equation}
\lim_{t\to t_\generic} \Rtt =  
\left\{ \begin{array}{ll}
0 & \qquad\eta_0=1; \qquad \eta_1>3;\\
-18c_1/c_0& \qquad\eta_0= 1; \qquad\eta_1=3; \\
-\mathrm{sign}(c_1)\infty &\qquad \eta_0= 1; \qquad\eta_1\in(0,3). \\
\end{array}
\right.
\end{equation}
To summarize, in the vicinity of the cosmological milestone, we can write 
\begin{equation}
\lim_{t\to t_\generic} \Rtt = \left\{ \begin{array}{ll}
-\infty & \qquad\eta_0>1; \\
0 & \qquad\eta_0=1; \qquad \eta_1>3;\\
-18c_1/c_0& \qquad\eta_0= 1; \qquad\eta_1=3; \\
-\mathrm{sign}(c_1)\infty &\qquad \eta_0= 1; \qquad\eta_1\in(0,3); \\
+\infty & \qquad\eta_0\in(0,1); \\
0 & \qquad\eta_0=0;  \qquad \eta_1 > 2;\\
-6c_1/c_0 & \qquad\eta_0=0;  \qquad \eta_1 = 2;\\
\mathrm{sign}(c_1)\infty & \qquad\eta_0=0; \qquad 
\eta_1\in(1,2);\\
0 & \qquad\eta_0=0; \qquad \eta_1= 1; \qquad\eta_2 > 2;\\
-6c_2/c_0& \qquad\eta_0=0; \qquad \eta_1= 1; \qquad\eta_2=2; \\
-\mathrm{sign}(c_2)\infty &\qquad \eta_0=0; \qquad \eta_1= 1; \qquad\eta_2\in(1,2); \\
-\mathrm{sign}(c_1)\infty & \qquad\eta_0=0; \qquad  \eta_1 \in(0,1);\\
-\infty & \qquad \eta_0<0.
\end{array}
\right.
\end{equation}
Therefore $\Rtt$ is finite provided
\begin{itemize}
\item $\eta_0=0$, $\eta_1\geq 2$;
\item $\eta_0=0$, $\eta_1=1$, $\eta_2\geq2$;
\item $\eta_0=1$ and $\eta_1\geq 3$.
\end{itemize}
\end{proof}

\begin{theorem}  \label{theo_Gtt}
\emph{The $\Gtt$ theorem:}

 Consider a generalized power series expansion of the\index{scale factor} scale factor as defined in Definition \ref{def_generic_a}, the $\Gtt$ component of the Ricci \index{Ricci!tensor}tensor in an orthonormal basis remains \emph{finite}, as the cosmological milestone is approached, only provided:
\begin{eqnarray}
\eta_0=0 & \eta_1\geq 1& \\
\eta_0=1 &  \eta_1\geq 3& \text{and} \qquad c_0^2+k=0 .
\end{eqnarray}
(Note that $c_0^2+k=0$ implies $k=-1$, and $c_0=1$.)

That is, except for these above rather limited cases, $\Gtt$ will blow up to \emph{infinity} as the cosmological milestone is approached.
\end{theorem}

\begin{proof}
The $\Gtt$ term can be written as a function of the\index{scale factor} scale factor, its first derivative and the curvature parameter $k$,
\begin{equation}
\Gtt = 3 \left( {\dot a^2\over a^2} + {k\over a^2}\right) =  3 \left( {H^2} + {k\over a^2}\right).
\end{equation}
To analyze $\Gtt$ recall that the condition that the Hubble \index{Hubble!parameter} parameter $H$ remain finite was $\eta_0=0$, $\eta_1 \geq 1$ from Theorem \ref{theo_H_general}. 
 In the vicinity of the milestone, we have
\begin{equation}
\Gtt \sim 3 \left[ {\eta_0^2\over(t-t_\generic)^2} + {k\over c_0^2(t-t_\generic)^{2\eta_0}} \right]
\qquad \eta_0\neq0,
\end{equation}
while
\begin{equation}
\Gtt \sim 3 \left[ {\eta_1^2c_1^2(t-t_\generic)^{2(\eta_1-1)} + k\over c_0^2} \right]
\qquad \eta_0=0.
\end{equation}
Whenever $\eta_0>1$, the term $k/ c_0^2(t-t_\generic)^{2\eta_0}$ with $k\neq 0$ is dominant and it tends to $\infty$. In the case of $\eta_0 >1$ and $k=0$, the term $\eta_0^2/(t-t_\generic)^2 $ is dominant and it tends to $\infty$.
Whenever $\eta_0<1$ and $\eta_0 \neq 0$, the term $\eta_0^2/(t-t_\generic)^2 $ is dominant and therefore, $\Gtt \to \infty$. For the special case where $\eta_0=1$, $\Gtt$ is a difference of squares and one term can be balanced against the other. Finally, when $\eta_0=0$ the first term needs to be developed further using the indicial exponent $\eta_1$.

Therefore, after further calculations, we find,
 \begin{equation}
\lim_{t\to t_\generic} \Gtt = \left\{ \begin{array}{ll}
\mathrm{sign}(k)\infty & \eta_0 > 1; \qquad k\neq 0;\\
+\infty & \eta_0 > 1; \qquad k= 0;\\
\mathrm{sign}(c_0^2+k)\infty & \eta_0 = 1; \qquad c_0^2+k\neq0;\\
0 &\eta_0=1; \qquad c_0^2+k=0; \qquad \eta_1>3;\\
18 c_1 &\eta_0=1; \qquad c_0^2+k=0; \qquad \eta_1=3; \\
\mathrm{sign}(c_1)\infty &\eta_0=1; \qquad c_0^2+k=0; \qquad \eta_1\in(1,3);\\
+\infty & \eta_0 \in(0,1);\\
3k/c_0^2 & \eta_0=0;  \qquad \eta_1 > 1;\\
3(c_1^2+k)/c_0^2 & \eta_0=0; \qquad \eta_1=1.\\
+\infty & \eta_0=0; \qquad \eta_1\in(0,1).\\
\end{array}
\right.
\end{equation}
Thus the necessary and sufficient conditions for $\Gtt$ to remain finite at the cosmological milestone are:
\begin{itemize}
\item $\eta_0=0$, $\eta_1\geq 1$;
\item $\eta_0=1$, $c_0^2+k=0$, and  $\eta_1\geq 3$.  \\
(Note that $c_0^2+k=0$ implies $k=-1$, and $c_0=1$.)
\end{itemize}
\end{proof}

\begin{theorem} \label{poly_curva_sing}
\emph{Polynomial curvature singularity theorem:}

 Consider a generalized power series expansion of the \index{scale factor}scale factor as defined in Definition \ref{def_generic_a}, every cosmological milestone is a polynomial curvature singularity at the event \emph{except} for the rather limited classes of cosmological milestones that satisfy the necessary and sufficient conditions:
 \begin{itemize}
\item $\eta_0=0 \qquad \eta_1\geq 2$ or;
\item     $ \eta_0=0  \qquad \eta_1=1  \qquad \eta_2\geq2$ or;
\item    $ \eta_0=1   \qquad \eta_1\geq 3\qquad \text{and} \qquad k=-1, \;  c_0=1 $.
\end{itemize}
That is, for a particular sub-class of\index{singularity!sudden singularity} sudden singularities $(\eta_0=0)$ and for a rather exceptional type of big\index{singularity!big bang} bang/crunch\index{singularity!big crunch}\index{big crunch} (as detailed above), there is \emph{no} polynomial curvature singularity. 
\end{theorem}

\begin{proof}
From Theorem \ref{theo_Rtt} and \ref{theo_Gtt}, we have determined the conditions for $\Rtt$ and $\Gtt$ (respectively) to remain finite at the time the cosmological milestone occurs. 

For $\Rtt$, those conditions are
\begin{itemize}
\item $\eta_0=0$, $\eta_1\geq 2$;
\item $\eta_0=0$, $\eta_1=1$, $\eta_2\geq2$;
\item $\eta_0=1$ and $\eta_1\geq 3$.
\end{itemize}

For $\Gtt$, those conditions are
\begin{itemize}
\item $\eta_0=0$, $\eta_1\geq 1$;
\item $\eta_0=1$, $c_0^2+k=0$, and  $\eta_1\geq 3$.  \\
(Note that $c_0^2+k=0$ implies $k=-1$, and $c_0=1$.)
\end{itemize}
Combining those two results, the necessary and sufficient conditions for both $\Rtt$ and $\Gtt$ to remain finite, so that a cosmological milestone is \emph{not} a polynomial curvature singularity, are:
\begin{itemize}
\item $\eta_0=0$, $\eta_1\geq 2$;
\item $\eta_0=0$, $\eta_1=1$, $\eta_2\geq2$;
\item $\eta_0=1$, $k=-1$, $c_0=1$, and  $\eta_1\geq 3$.
\end{itemize}
Note that the case $\eta_0=1$ that corresponds to an exceptional type of big \index{singularity!big bang}bang/crunch\index{singularity!big crunch}\index{big crunch} is asymptotic to a Milne universe. Indeed, the special case $a(t)= t $, $k=-1$ is called the Milne universe and is actually a disguised portion of Minkowski space. It corresponds to the interior of the future light cone based at some randomly specified point in Minkowski space, with a spatial foliation defined by the proper time hyperboloids based on that event. Since this spacetime is flat, it corresponds to a universe which on the largest scales is empty. This is not a popular cosmological model.

Any other conditions that are not mentioned above imply that $\Rtt$ and/or $\Gtt$ blow up at the time of the event; therefore, all other cases imply a polynomial curvature singularity.
\end{proof}

Even if there are a few classes of cosmological milestones that are not polynomial singularities, one can wonder what happens to them when looking at finite-order derivatives of the curvature tensor. Is it still possible to find an interesting type of singularity where the derivative curvature singularity does not blow up? 

\subsubsection{Derivative curvature singularities} \index{singularity!derivative curvature singularity}
\begin{definition}
A derivative curvature singularity is defined by some polynomial constructed from finite-order derivatives of the curvature tensor blowing up. 
\end{definition}
In our case, because of the symmetries of the FRW universe the only interesting derivatives will be time derivatives, and so the only objects we need to consider are
\begin{equation}
{\d^n \Rtt\over \d^n t} \qquad \hbox{and} \qquad {\d^n \Gtt\over \d^n t}.
\end{equation}

\begin{theorem} \label{derivative_curva_sing}
\emph{Derivative curvature singularity theorem}

The \emph{only} two situations in which a cosmological milestone is \emph{not} a derivative curvature singularity are if:
\begin{itemize}

\item $\eta_0=0$,\; $\eta_{i}\in Z^+$; corresponding to an \index{singularity!extremality events}extremality event (bounce\index{singularity!extremality events!bounce},\index{singularity!extremality events!turnaround} turnaround, or \index{singularity!extremality events!inflexion}inflexion event) rather than a \index{singularity!big bang}bang, \index{singularity!big crunch}\index{big crunch}crunch, rip, or \index{singularity!sudden singularity}sudden singularity;

\item $\eta_0=1$, $k=-1$, $c_0=1$, $\eta_{i}\in Z^+$, and  $\eta_1\geq 3$; corresponding to a FRW geometry that smoothly asymptotes near the cosmological milestone to the\index{Riemann} Riemann-flat  Milne universe. 
\end{itemize}
\end{theorem}

\begin{proof}
We can write $\d^n \Rtt / \d^n t$ and $ \d^n \Gtt/ \d^n t$ using the \index{scale factor}scale factor $a(t)$ and its derivatives $a^{(n)}(t)$,
\begin{equation}
{\d^n \Rtt\over \d^n t} = - 3 {a^{(n+2)}\over a} + \hbox{ (lower-order derivatives)};
\end{equation}
while
\begin{equation}
 {\d^n \Gtt\over \d^n t} = 3 {\dot a \; a^{(n+1)}\over a^2}  + \hbox{ (lower-order derivatives)}.
\end{equation}

Thus to avoid a $n$th-order derivative curvature singularity we must at the very least keep $a^{(n+2)}/a$ finite, and furthermore all related lower-order derivatives of the form $a^{(j)}/a$,  with $j\leq n+2$, must also be finite. 
To prevent any arbitrary-order derivative singularity from occurring, that is for both $\d^n \Rtt / \d^n t$ and $ \d^n \Gtt/ \d^n t$ to remain finite for all $n \in Z^+$, we must force all  $a^{(j)}/a$ to remain finite.  This condition holds in addition to the constraint coming from polynomial curvature singularities derived above.

The first constraint to hold, so that there is no polynomial curvature singularity, is $\eta_0=0$. In this case, the scale\index{scale factor} factor can be written as:
\begin{equation}
a(t)=c_0 +  c_1  |t-t_\generic|^{\eta_1} + c_2  |t-t_\generic|^{\eta_2} 
+ c_3 |t-t_\generic|^{\eta_3} +\dots
\end{equation}
Remember that the dominant term in the \index{scale factor}scale factor is $c_0$ which implies that $a(t)$ finite at the time of the event. Therefore, for $a^{(j)}/a$ to remain finite it suffices that $a^{(j)}$ be finite.

The second constraint to hold, so that there is no polynomial curvature singularity, is $\eta_0=1$. In this case, the \index{scale factor}scale factor can be written as:
\begin{equation}
a(t)=c_0|t-t_\generic| +  c_1  |t-t_\generic|^{\eta_1} + c_2  |t-t_\generic|^{\eta_2} 
+ c_3 |t-t_\generic|^{\eta_3} +\dots
\end{equation}
Remember that the dominant term in the \index{scale factor}scale factor is $c_0|t-t_\generic|$ which implies that $a(t)$ is not finite at the time of the event. However, the term $|t-t_\generic|$ from $a(t)$ can be ``absorbed" in the term $a^{(n+2)}$ as follows:
\begin{equation}
 {a^{(j)}\over a} = { \sum_{i} c_i \eta_i \dots \left( \eta_i -j +1 \right)  |t-t_\generic|^{\eta_i-j}   \over c_0|t-t_\generic| +  c_1  |t-t_\generic|^{\eta_1} + c_2  |t-t_\generic|^{\eta_2} + c_3 |t-t_\generic|^{\eta_3} +\dots} ,
\end{equation}
keeping only dominant terms, we have
\begin{equation}
 {a^{(j)}\over a} \sim  {  \sum_{i} c_i \eta_i \dots \left( \eta_i -j +1 \right)  |t-t_\generic|^{\eta_i-j}  \over c_0|t-t_\generic| } \sim {  \sum_{i} c_i \eta_i \dots \left( \eta_i -j +1 \right)  |t-t_\generic|^{\eta_i-j-1}  \over c_0} 
\end{equation}
 Therefore, for $a^{(j)}/a$ to remain finite it suffices that $a^{(j)}$ be finite.

These two constraints also imply that to force all  $a^{(j)}$ to remain finite, we must force all
the indicial exponents $\eta_i$ to be non-negative integers, thus making $a(t)$ a Taylor series.  
Indeed, we have
\begin{equation}
a^{(j)} = \sum_{i} c_i \eta_i \dots \left( \eta_i -j +1 \right)  |t-t_\generic|^{\eta_i-j}, 
\end{equation}
and $a^{(j)}$ is finite if and only if 
\begin{equation}
|t-t_\generic|^{\eta_i-j} \longrightarrow \text{finite},
\end{equation}
that is,
\begin{equation}
\forall \; i \qquad \eta_i- j \geqslant 0 \qquad \implies \qquad \eta_i \geqslant j .
\end{equation}
Thus, all $\eta_i$ must be positive or null. They must be integers as well for $a^{(j)}$ to remain finite: otherwise, there will be eventually a value of $j$ that will make the term $\eta_i-j<0$, to avoid that $\eta_i$ must be integers so the term $|t-t_\generic|^{\eta_i-j}$ disappear and do not tend to infinity. 
\end{proof}

Hence, we have demonstrated that almost all cosmological milestones are physical singularities, apart from a very limited sub-class corresponding to either \index{singularity!extremality events}extremality events (\index{singularity!extremality events!bounce}bounces, turnarounds, \index{singularity!extremality events!inflexion}inflexion events) or an asymptotically empty universe.

\section{Cosmological milestones and dynamics.}

In the previous section, we have considered only kinematics, from now on, we will use the Friedmann\index{Friedmann} equations and the Einstein \index{Einstein!field equations}equations to take our definitions to a dynamic level.
\subsection{Introduction}
To  now start to include dynamics we relate the geometry to the density and pressure using the Friedmann\index{Friedmann} equations and then ask what happens to the energy \index{energy conditions}conditions at the cosmological milestones. Even though we may have good reason to suspect that the energy conditions\index{energy conditions} are not truly fundamental~\cite{Barcelo:2002bv}: they make a very good first pass at the problem of quantifying just how ``strange'' physics gets at the cosmological milestone. \\

Using the Friedmann\index{Friedmann} equations, we can write the pressure $p(t)$ and the density $\rho(t)$ using the \index{scale factor}scale factor and its first two derivatives:
\begin{eqnarray}
\rho(t) & = & 3 \left(   \frac{\dot{a}^2}{a^2}+\frac{k}{a^2}  \right),   \\
p(t)      & = & -2\frac{\ddot{a}}{a}-\frac{\dot{a}^2}{a^2}-\frac{k}{a^2},   \\
\rho(t)+3p(t) & =& -6\frac{\ddot{a}}{a}.                                                   
\end{eqnarray}
Therefore, we have
\begin{equation}
\rho+p=2 \left(-\frac{\ddot{a}}{a}+  \frac{\dot{a}^2}{a^2}+\frac{k}{a^2} \right), \label{eq_rho_plus_p}
\end{equation}
and
\begin{equation}
\rho-p=2 \left(\frac{\ddot{a}}{a}+  2\frac{\dot{a}^2}{a^2}+2\frac{k}{a^2} \right).
\end{equation}

The standard energy\index{energy conditions} conditions are the \emph{null}, \emph{weak}, \emph{strong}, and \emph{dominant} energy conditions\footnote{See section \ref{section_energy_cond} for more details on the energy conditions.} which for a FRW spacetime specialise to~\cite{Hochberg:1998vm, Molina-Paris:1998tx, Barcelo:2002bv}:
\begin{itemize}

\item{}[NEC] $\rho+p \geq 0$. \\ \index{energy conditions!NEC}\index{NEC}
In view of equation (\ref{eq_rho_plus_p}) this reduces to
\begin{equation}
\ddot a \leq {\dot a^2+k \over a}; \qquad \hbox{that is} \qquad k \geq a \;\ddot a - \dot a^2.
\end{equation}

\item{}[WEC] This specializes to the NEC plus $\rho\geq0$.\\
This reduces to the NEC plus the condition\index{energy conditions!WEC}\index{WEC}
\begin{equation}
\dot a^2 + k \geq 0; \qquad \hbox{that is} \qquad k \geq - \dot a^2.
\end{equation}

\item{}[SEC] This specializes to the NEC plus $\rho+3p\geq0$.\\
This reduces to the NEC plus the \index{deceleration parameter}deceleration condition\index{energy conditions!SEC}\index{SEC}
\begin{equation}
\ddot a \leq 0.
\end{equation}

\item{}[DEC]\index{DEC} $\rho\pm p \geq 0$.\\
This reduces to the NEC plus the condition\index{energy conditions!DEC}\index{DEC}
\begin{equation}
\ddot a \geq -{2(\dot a^2+k) \over a}; \qquad \hbox{that is}
\qquad k \geq -{(a\; \ddot a +2\dot a^2)\over2}.
\end{equation}
\end{itemize}
The null energy condition NEC is the most interesting condition because it is the weakest of the standard energy conditions and it leads to the strongest theorems.

\subsection{NEC}\index{energy conditions!NEC}\index{NEC}
We have seen that the NEC is satisfied iff $k\geqslant a \;\ddot a - \dot a^2$.
\begin{theorem}  \label{NEC1}
The NEC\index{energy conditions!NEC}\index{NEC} is \emph{definitely satisfied} (meaning the inequality is \emph{strict}) at a generic cosmological milestone iff:
\begin{itemize}
\item \index{singularity!big bang}Big bangs/crunches:\index{singularity!big crunch}\index{big crunch}
\begin{itemize}
\item $\eta_0>1$, $k=+1$;
\item $\eta_0=1$: for $k=+1$ and $k=0$, and also for $k=-1$ with the proviso that $c_0> 1$;
\item $\eta_0\in(0,1)$: for any value of $k$; 
\end{itemize}
\item Sudden \index{singularity!sudden singularity}singularities/extremality\index{singularity!extremality events} events:
\begin{itemize}
\item $\eta_0=0$ subject to the additional constraints:
\begin{itemize}
\item $\eta_1>2$ and $k=+1$;
\item $\eta_1=2$ and $k>2c_0c_1$;
\item $\eta_1\in(1,2)$ and $c_1<0$;
\item $\eta_1=1$, $\eta_2>2$: for $k=+1$, $k=0$, and for $k=-1$ with the proviso $|c_1|>1$;
\item $\eta_1=1$, $\eta_2=2$: for $k>2c_0c_2-c_1^2$;
\item $\eta_1=1$, $\eta_2\in(1,2)$, and $c_2<0$;
\item $\eta_1\in(0,1)$, and $c_1>0$.
\end{itemize}
\end{itemize}
\end{itemize}
\end{theorem}

\begin{theorem} \label{NEC2}
The NEC\index{energy conditions!NEC}\index{NEC} is \emph{marginally satisfied} (meaning the non-strict inequality is actually an equality) at a generic cosmological milestone iff:
\begin{itemize}
\item \index{singularity!big bang} Big bangs/crunches:\index{singularity!big crunch}\index{big crunch}
\begin{itemize}
\item $\eta_0>1$, $k=0$;
\item $\eta_0=1$, $k=-1$ with the proviso that $c_0= 1$;
\end{itemize}
\item Sudden \index{singularity!sudden singularity}singularities/extremality \index{singularity!extremality events}events:
\begin{itemize}
\item $\eta_0=0$ subject to the additional constraints:
\begin{itemize}
\item $\eta_1>2$ and $k=0$;
\item $\eta_1=2$ and $k=2c_0c_1$ (which requires $k\neq0$);
\item $\eta_1=1$, $\eta_2>2$: for $k=-1$ and $c_1=\pm1$;
\item $\eta_1=1$, $\eta_2=2$, and $k=2c_0c_2-c_1^2$.
\end{itemize}
\end{itemize}
\end{itemize}
\end{theorem}

\begin{theorem} \label{NEC3}
 The NEC\index{energy conditions!NEC}\index{NEC} is \emph{definitely violated} (the inequality is \emph{strictly violated}) at a generic milestone iff:
\begin{itemize}
\item \index{singularity!big bang}Big bangs/crunches:\index{singularity!big crunch}\index{big crunch}
\begin{itemize}
\item $\eta_0>1$, $k=-1$;
\item $\eta_0=1$: for $k=-1$ with the proviso that $c_0< 1$;
\end{itemize}
\item Sudden\index{singularity!sudden singularity} singularities/extremality \index{singularity!extremality events}events:
\begin{itemize}
\item $\eta_0=0$ subject to the additional constraints:
\begin{itemize}
\item $\eta_1>2$ and $k=-1$;
\item $\eta_1=2$ and $k<2c_0c_1$;
\item $\eta_1\in(1,2)$ and $c_1<0$;
\item $\eta_1=1$, $\eta_2>2$:  for $k=-1$ with the proviso $|c_1|<1$;
\item $\eta_1=1$, $\eta_2=2$: for $k<2c_0c_2-c_1^2$;
\item $\eta_1=1$, $\eta_2\in(1,2)$, and $c_2>0$;
\item $\eta_1\in(0,1)$, and $c_1<0$.
\end{itemize}
\end{itemize}
\item Big rips:
\begin{itemize}
\item $\eta_0 < 0$.
\end{itemize}
\end{itemize}
In particular, ``big rips" $(\eta_0<0)$ will violate the NEC \index{energy conditions!NEC}\index{NEC}in the vicinity of the milestone and therefore, they will also violate every other energy conditions\index{energy conditions}.
 \end{theorem}

\begin{proof}
The NEC \index{energy conditions!NEC}\index{NEC}is satisfied if $k\geqslant a \;\ddot a - \dot a^2$. Remember that near the cosmological milestone we can write the \index{scale factor}scale factor as a generalized power series as defined in Defintion \ref{def_generic_a}:
\begin{equation}
a(t) = c_0 |t-t_\generic|^{\eta_0} + c_1  |t-t_\generic|^{\eta_1} + c_2  |t-t_\generic|^{\eta_2} 
+ c_3 |t-t_\generic|^{\eta_3} +\dots
\end{equation}
Therefore, near any generic cosmological milestone
\begin{equation}
a \;\ddot a - \dot a^2 \sim - \eta_0 \;c_0^2 \;t^{2(\eta_0-1)}; \qquad (\eta_0\neq0),
\end{equation}
while in the degenerate cases
\begin{equation}
a \;\ddot a - \dot a^2 \sim 
c_0 c_1 \; \eta_1(\eta_1-1) \; (t-t_\generic)^{(\eta_1-2)};  \qquad (\eta_0=0; \eta_1\neq1),
\end{equation}
and
\begin{equation}
a \;\ddot a - \dot a^2 \sim 
c_0 c_2 \; \eta_2(\eta_2-1) \; (t-t_\generic)^{(\eta_2-2)}  - c_1^2; \qquad (\eta_0=0; \eta_1=1).
\end{equation}
After further calculations, we find the limit of $( a \;\ddot a - \dot a^2)$ when the milestone occurs:
\begin{equation}
\lim_{t\to t_\generic} (a\;\ddot a-\dot a^2)  = 
\left\{ \begin{array}{ll}
0  & \qquad \eta_0>1;\\
-c_0^2 & \qquad\eta_0=1;\\
-\infty &\qquad \eta_0 \in(0,1);\\
 0 & \qquad \eta_0=0; \qquad \eta_1>2;\\
 2 c_0 c_1 & \qquad \eta_0=0; \qquad \eta_1=2;\\
+  \mathrm{sign}(c_1) \infty &\qquad \eta_0=0; \qquad \eta_1\in(1,2)\\
 -c_1^2 & \qquad \eta_0=0; \qquad \eta_1=1; \qquad \eta_2>2;\\
 2c_0c_2 -c_1^2 & \qquad \eta_0=0; \qquad \eta_1=1; \qquad \eta_2=2;\\
+  \mathrm{sign}(c_2) \infty   & \qquad \eta_0=0; \qquad \eta_1=1; \qquad \eta_2\in(1,2);\\
- \mathrm{sign}(c_1) \infty &\qquad \eta_0=0; \qquad \eta_1\in(0,1)\\
 +\infty &\qquad \eta_0 < 0.\\
\end{array}
\right.
\end{equation}
The NEC\index{energy conditions!NEC}\index{NEC} is \emph{definitely satisfied} iff:
\begin{equation}
k > a \;\ddot a - \dot a^2,
\end{equation}
that is, iff:
\begin{itemize}
\item $\eta_0>1$, $k=+1$;
\item $\eta_0=1$: for $k=+1$ and $k=0$, and also for $k=-1$ with the proviso that $c_0> 1$;
\item $\eta_0\in(0,1)$: for any value of $k$; 
\item $\eta_0=0$ subject to the additional constraints:
\begin{itemize}
\item $\eta_1>2$ and $k=+1$;
\item $\eta_1=2$ and $k>2c_0c_1$;
\item $\eta_1\in(1,2)$ and $c_1<0$;
\item $\eta_1=1$, $\eta_2>2$: for $k=+1$, $k=0$, and for $k=-1$ with the proviso $|c_1|>1$;
\item $\eta_1=1$, $\eta_2=2$: for $k>2c_0c_2-c_1^2$;
\item $\eta_1=1$, $\eta_2\in(1,2)$, and $c_2<0$;
\item $\eta_1\in(0,1)$, and $c_1>0$.
\end{itemize}
\end{itemize}
This proves Theorem \ref{NEC1}.\\
The NEC\index{energy conditions!NEC}\index{NEC} is \emph{marginally satisfied} iff:
\begin{equation}
k = a \;\ddot a - \dot a^2,
\end{equation}
 that is, iff
 \begin{itemize}
\item $\eta_0>1$, $k=0$;
\item $\eta_0=1$, $k=-1$ with the proviso that $c_0= 1$;
\item $\eta_0=0$ subject to the additional constraints:
\begin{itemize}
\item $\eta_1>2$ and $k=0$;
\item $\eta_1=2$ and $k=2c_0c_1$ (which requires $k\neq0$);
\item $\eta_1=1$, $\eta_2>2$: for $k=-1$ and $c_1=\pm1$;
\item $\eta_1=1$, $\eta_2=2$, and $k=2c_0c_2-c_1^2$.
\end{itemize}
\end{itemize}
This proves Theorem \ref{NEC2}.\\
The NEC\index{energy conditions!NEC}\index{NEC} is \emph{definitely violated} iff:
\begin{equation}
k < a \;\ddot a - \dot a^2,
\end{equation}
 that is, iff
\begin{itemize}
\item $\eta_0>1$, $k=-1$;
\item $\eta_0=1$: for $k=-1$ with the proviso that $c_0< 1$;
\item $\eta_0=0$ subject to the additional constraints:
\begin{itemize}
\item $\eta_1>2$ and $k=-1$;
\item $\eta_1=2$ and $k<2c_0c_1$;
\item $\eta_1\in(1,2)$ and $c_1<0$;
\item $\eta_1=1$, $\eta_2>2$:  for $k=-1$ with the proviso $|c_1|<1$;
\item $\eta_1=1$, $\eta_2=2$: for $k<2c_0c_2-c_1^2$;
\item $\eta_1=1$, $\eta_2\in(1,2)$, and $c_2>0$;
\item $\eta_1\in(0,1)$, and $c_1<0$.
\end{itemize}
\item $\eta_0 < 0$.
\end{itemize}
This proves Theorem \ref{NEC3}.
\end{proof}

Theorems \ref{NEC1}, \ref{NEC2}, and \ref{NEC3}, present a pattern that is in agreement with the standard folklore, and is a systematic expression of results that are otherwise scattered throughout the literature.  For instance, it is immediate that ``big rips'' \emph{always} violate the NEC\index{energy conditions!NEC}\index{NEC} (and hence all the other energy conditions\index{energy conditions}), and that for big bangs\index{singularity!big bang} and \index{singularity!big crunch}\index{big crunch}crunches the range $\eta_0\in(0,1)$ is preferred, since it is only in this range that the NEC\index{energy conditions!NEC}\index{NEC} holds \emph{independent} of the spatial curvature. 
In particular, if the NEC\index{energy conditions!NEC}\index{NEC} is to hold independent of the sign of space curvature $k$ all the way down to the singularity, then in the vicinity of the singularity the dominant term in the\index{scale factor} scale factor $a(t)$ is bounded from both above and below by
\begin{equation}
c_0 (t-t_\bang) < a_\mathrm{dominant}(t) \leq c_0.
\end{equation}
 Note that for sufficiently violent big bangs \index{singularity!big bang}($\eta_0>1$) and hyperbolic spatial curvature the NEC\index{energy conditions!NEC}\index{NEC} is violated: this indicates that ``phantom matter'' need not always lead to a ``big rip"~\cite{Caldwell:1999ew}; it might in contrast lead to a particularly violent bang or crunch.\index{singularity!big crunch}\index{big crunch}
It is for $\eta_0=0$ (which corresponds to either \index{singularity!sudden singularity}sudden singularities or extremality \index{singularity!extremality events}events) that the analysis becomes somewhat tedious. Certainly, many (though not all) of the sudden singularities violate the NEC.

\subsection{WEC}\index{energy conditions!WEC}\index{WEC}
We have seen that the WEC is satisfied iff 
\begin{itemize}
\item the NEC is satisfied, that is if $k\geqslant a \;\ddot a - \dot a^2$,
\item and, $\rho \geqslant 0$, that is if $k \geqslant -\dot{a}^2$.
\end{itemize}

\begin{theorem}  \label{WEC1}
The WEC\index{energy conditions!WEC}\index{WEC} is \emph{definitely satisfied} (meaning the inequality is \emph{strict}) at a generic cosmological milestone iff the NEC\index{energy conditions!NEC}\index{NEC} is \emph{definitely satisfied} (Theorem \ref{NEC1}).
\end{theorem}
\begin{theorem}  \label{WEC2}
The WEC\index{energy conditions!WEC}\index{WEC} is \emph{marginally satisfied}  (meaning the non-strict inequality is actually an equality) at a generic cosmological milestone iff the NEC\index{energy conditions!NEC}\index{NEC} is \emph{marginally satisfied} (Theorem \ref{NEC2}).
\end{theorem}
\begin{theorem}  \label{WEC3}
The WEC\index{energy conditions!WEC}\index{WEC} is \emph{definitely violated} (meaning the inequality is \emph{strict violated}) at a generic cosmological milestone for any cases that are not mentioned in Theorems \ref{NEC1} or \ref{NEC2}.
\end{theorem}

\begin{proof}
The WEC\index{energy conditions!WEC}\index{WEC} is satisfied if $k\geqslant a \;\ddot a - \dot a^2$ and $k \geqslant -\dot{a}^2$. 
The first condition $k\geqslant a \;\ddot a - \dot a^2$ has been dealt with in the proof of Theorems \ref{NEC1}, \ref{NEC2} and \ref{NEC3}. We need to look at $k \geqslant -\dot{a}^2$. 
Near the cosmological milestone we can write the\index{scale factor} scale factor as a generalized power series as defined in Definition \ref{def_generic_a}:
\begin{equation}
a(t) = c_0 |t-t_\generic|^{\eta_0} + c_1  |t-t_\generic|^{\eta_1} + c_2  |t-t_\generic|^{\eta_2} 
+ c_3 |t-t_\generic|^{\eta_3} +\dots
\end{equation}
Therefore, near any generic cosmological milestone
\begin{equation}
 - \dot a^2 \sim - \eta_0^2 c_0^2 (t-t_\generic)^{2(\eta_0-1)}; \qquad (\eta_0\neq0),
\end{equation}
while in the degenerate case
\begin{equation}
 - \dot a^2 \sim - \eta_1^2 c_1^2 (t-t_\generic)^{2(\eta_1-1)}; \qquad (\eta_0=0).
\end{equation}
Therefore
\begin{equation}
\lim_{t\to t_\generic} (-\dot a^2)  = 
\left\{ \begin{array}{ll}
0  & \qquad \eta_0>1;\\
-c_0^2 & \qquad\eta_0=1;\\
-\infty &\qquad \eta_0 \in(0,1);\\
 0 & \qquad \eta_0=0; \qquad \eta_1>1;\\
 -c_1^2 & \qquad \eta_0=0; \qquad \eta_1=1;\\
- \infty &\qquad \eta_0=0; \qquad \eta_1\in(0,1)\\
 -\infty &\qquad \eta_0 < 0.\\
\end{array}
\right.
\end{equation}
Since we want the condition $k>-\dot a^2$ to hold in addition to the NEC\index{energy conditions!NEC}\index{NEC}, it follows that the this condition
is equivalent to the NEC for bangs\index{singularity!big bang} \index{singularity!big crunch}\index{big crunch}crunches and rips, and that the only changes arise when dealing with $\eta_0=0$ (sudden\index{singularity!sudden singularity} singularities or \index{singularity!extremality events}extremality events), but overall the conditions on the NEC are dominant. Therefore, the WEC\index{energy conditions!WEC}\index{WEC} holds if and only if the NEC holds.\\
This proves Theorems \ref{WEC1}, \ref{WEC2} and \ref{WEC3}.
\end{proof}
Certainly, many (though not all) of the \index{singularity!sudden singularity}sudden singularities violate the WEC.\\
In particular, if the WEC \index{energy conditions!WEC}\index{WEC}is to hold independent of the sign of space curvature $k$ all the way down to the singularity, then in the vicinity of the singularity the dominant term in the \index{scale factor}scale factor $a(t)$ is bounded from both above and below by
\begin{equation}
c_0 (t-t_\bang) < a_\mathrm{dominant}(t) \leq c_0.
\end{equation}

\subsection{SEC}
\index{energy conditions!SEC}\index{SEC}
We have seen that the SEC is satisfied iff 
\begin{itemize}
\item the NEC is satisfied, that is if $k\geqslant a \;\ddot a - \dot a^2$,
\item and, $\rho +3p \geqslant 0$, that is if $\ddot{a}\leqslant 0$.
\end{itemize}

\begin{theorem}  \label{SEC1}
The SEC\index{energy conditions!SEC}\index{SEC} is \emph{definitely satisfied} (meaning the inequality is \emph{strict}) at a generic cosmological milestone iff:
\begin{itemize}
\item \index{singularity!big bang} Big bangs/crunches:\index{singularity!big crunch}\index{big crunch}
\begin{itemize}
\item  $\eta_0=1$: for $k=+1$ and $k=0$;
\item $\eta_0\in(0,1)$: for any value of $k$; 
\end{itemize}
\item Sudden\index{singularity!sudden singularity} singularities/ \index{singularity!extremality events}extremality events:
\begin{itemize}
\item $\eta_0=0$ subject to the additional constraints:
\begin{itemize}
\item $\eta_1>2$, $k=+1$ and $c_1<0$;
\item $\eta_1=2$, $k>2c_0c_1$ and $c_1<0$;
\item $\eta_1\in(1,2)$ and $c_1<0$;
\item $\eta_1=1$, $\eta_2>2$ and $c_2<0$: for $k=+1$, $k=0$, and for $k=-1$ with the proviso $|c_1|>1$;
\item $\eta_1=1$, $\eta_2=2$ and $c_2<0$: for $k>2c_0c_2-c_1^2$;
\item $\eta_1=1$, $\eta_2\in(1,2)$, and $c_2<0$;
\item $\eta_1\in(0,1)$, and $c_1>0$.
\end{itemize}
\end{itemize}
\end{itemize}
\end{theorem}

\begin{theorem}  \label{SEC2}
The SEC\index{energy conditions!SEC}\index{SEC} is \emph{marginally satisfied} (meaning the non-strict inequality is actually an equality) at a generic cosmological milestone iff:
\begin{itemize}
\item \index{singularity!big bang}Big bangs/crunches:\index{singularity!big crunch}\index{big crunch}
\begin{itemize}
\item  $\eta_0=1$, $c_0=1$, $c_1<0$, and $k=-1$;
\end{itemize}
\item Sudden\index{singularity!sudden singularity} singularities/ extremality \index{singularity!extremality events}events:
\begin{itemize}
\item $\eta_0=0$ subject to the additional constraints:
\begin{itemize}
\item $\eta_1>2$, $k=0$ and $c_1<0$;
\item $\eta_1=2$, $k=2c_0c_1$ and $c_1<0$;\\
(which requires that $c_0=-1/(2c_1)$)
\item $\eta_1=1$, $\eta_2>2$ and $c_2<0$, $k=-1$, $c_1=\pm 1$ and $c_2<0$;
\item $\eta_1=1$, $\eta_2=2$ and $c_2<0$, and $k=2c_0c_2-c_1^2$;\\
(which requires that $k=-1$ and therefore $2 c_0 c_2 -c_1^2=-1$)
\end{itemize}
\end{itemize}
\end{itemize}
\end{theorem}

\begin{theorem}  \label{SEC3}
The SEC\index{energy conditions!SEC}\index{SEC} is \emph{definitely violated} (meaning the inequality is \emph{strict violated}) at a generic cosmological milestone for any cases that are not mentioned in Theorems \ref{SEC1} or \ref{SEC2}.
\end{theorem}

\begin{proof}
The SEC\index{energy conditions!SEC}\index{SEC} is satisfied if $k\geqslant a \;\ddot a - \dot a^2$ and $\ddot{a}\leqslant 0$. 
The first condition $k\geqslant a \;\ddot a - \dot a^2$ has been dealt with in the proof of Theorems \ref{NEC1}, \ref{NEC2} and \ref{NEC3}. We need to look at $\ddot{a}\leqslant 0$. 
Near the cosmological milestone we can write the \index{scale factor}scale factor as a generalized power series as defined in Defintion \ref{def_generic_a}:
\begin{equation}
a(t) = c_0 |t-t_\generic|^{\eta_0} + c_1  |t-t_\generic|^{\eta_1} + c_2  |t-t_\generic|^{\eta_2} 
+ c_3 |t-t_\generic|^{\eta_3} +\dots
\end{equation}
Therefore, near any generic cosmological milestone
\begin{equation}
\ddot a  \sim \eta_0(\eta_0-1) \;c_0 \;t^{(\eta_0-2)}; \qquad (\eta_0\neq0, \;\;\; \eta_0\neq 1),
\end{equation}
while in the degenerate cases
\begin{equation}
\ddot a  \sim 
\eta_1(\eta_1-1) \; c_1 \;  (t-t_\generic)^{(\eta_1-2)};  \quad 
\left\{ \begin{array}{ll}
\eta_0=0; \qquad \eta_1\neq1; \\
\eta_0=1;
 \end{array}\right.
\end{equation}
\begin{equation}
\ddot a  \sim 
\eta_2(\eta_2-1) \; c_2 \;  (t-t_\generic)^{(\eta_2-2)};  \qquad (\eta_0=0; \;\;\; \eta_1=1; ),
\end{equation}
Therefore
\begin{equation}
\lim_{t\to t_\generic^+} [\mathrm{sign}(\ddot a)] = 
\left\{ \begin{array}{ll}
+1  & \qquad \eta_0>1\;\\
\mathrm{sign}(c_1) & \qquad \eta_0=1;\\
-1 & \qquad \eta_0\in(0,1):\\
\mathrm{sign}(\eta_1[\eta_1-1] c_1 ) & \qquad\eta_0=0;\qquad  \eta_1\neq 1;\\
\mathrm{sign}(c_2)  &\qquad \eta_0 =0; \qquad \eta_1=1;\\
 +1 &\qquad \eta_0 < 0.\\
\end{array}
\right.
\end{equation}
Since we want the condition $\ddot a \leq 0$ to hold in addition to the NEC\index{energy conditions!NEC}\index{NEC}, it follows that the SEC\index{energy conditions!SEC}\index{SEC} is definitely satisfied when those conditions below hold:
\begin{itemize}
\item  $\eta_0=1$: for $k=+1$ and $k=0$;
\item $\eta_0\in(0,1)$: for any value of $k$; 
\item $\eta_0=0$ subject to the additional constraints:
\begin{itemize}
\item $\eta_1>2$, $k=+1$ and $c_1<0$;
\item $\eta_1=2$, $k>2c_0c_1$ and $c_1<0$;
\item $\eta_1\in(1,2)$ and $c_1<0$;
\item $\eta_1=1$, $\eta_2>2$ and $c_2<0$: for $k=+1$, $k=0$, and for $k=-1$ with the proviso $|c_1|>1$;
\item $\eta_1=1$, $\eta_2=2$ and $c_2<0$: for $k>2c_0c_2-c_1^2$;
\item $\eta_1=1$, $\eta_2\in(1,2)$, and $c_2<0$;
\item $\eta_1\in(0,1)$, and $c_1>0$.
\end{itemize}
\end{itemize}
The SEC \index{energy conditions!SEC}\index{SEC}is marginally satisfied when:
\begin{itemize}
\item  $\eta_0=1$, $c_0=1$, $c_1<0$, and $k=-1$;
\item $\eta_0=0$ subject to the additional constraints:
\begin{itemize}
\item $\eta_1>2$, $k=0$ and $c_1<0$;
\item $\eta_1=2$, $k=2c_0c_1$ and $c_1<0$;\\
(which requires that $c_0=-1/(2c_1)$)
\item $\eta_1=1$, $\eta_2>2$ and $c_2<0$, $k=-1$, $c_1=\pm 1$ and $c_2<0$;
\item $\eta_1=1$, $\eta_2=2$ and $c_2<0$, and $k=2c_0c_2-c_1^2$;\\
(which requires that $k=-1$ and therefore $2 c_0 c_2 -c_1^2=-1$)
\end{itemize}
\end{itemize}
These conditions prove Theorems \ref{SEC1}, \ref{SEC2} and \ref{SEC3}
\end{proof}

 The SEC\index{energy conditions!SEC}\index{SEC} is definitely violated for all ``rips'', and is also definitely violated for all ``violent'' 
bangs\index{singularity!big bang} and \index{singularity!big crunch}\index{big crunch}crunches with $\eta_0>1$. The SEC\index{energy conditions!SEC}\index{SEC} is definitely satisfied for bangs and crunches in the range $\eta_0\in(0,1)$. 
In particular, if the SEC is to hold independent of the sign of space curvature $k$ all the way down to the singularity, then in the vicinity of the singularity the dominant term in the\index{scale factor} scale factor $a(t)$ is bounded from both above and below by
\begin{equation}
c_0 (t-t_\bang) < a_\mathrm{dominant}(t) \leq c_0.
\end{equation}

For \index{singularity!extremality events!turnaround}turnarounds or \index{singularity!extremality events!bounce}bounces  $\eta_1=2n$, $n\in Z^+$, and the \emph{sign} of $c_1$ governs possible SEC\index{energy conditions!SEC}\index{SEC} violations: Turnarounds satisfy the SEC\index{energy conditions!SEC}\index{SEC} while bounces violate the SEC\footnote{See also~\cite{Hochberg:1998vm, Molina-Paris:1998tx}.}. For\index{singularity!extremality events!inflexion} inflexion events,  where  $\eta_1=2n+1$, $n\in Z^+$, the SEC\index{energy conditions!SEC}\index{SEC} is violated either just before or just after the inflexion event\footnote{See also~\cite{Hochberg:1998vm, Molina-Paris:1998tx}.}. \\

\subsection{DEC} \index{energy conditions!DEC}\index{DEC}
We have seen that the DEC\index{DEC} is satisfied iff 
\begin{itemize}
\item the NEC is satisfied, that is if $k\geqslant a \;\ddot a - \dot a^2$,
\item and, $\rho -p \geqslant 0$, that is if $ k \geq -\dfrac{(a\; \ddot a +2\dot a^2)}{2}$.
\end{itemize}

\begin{theorem}  \label{DEC1}
The DEC\index{energy conditions!DEC}\index{DEC}is \emph{definitely satisfied} (meaning the inequality is \emph{strict}) at a generic cosmological milestone iff:
\begin{itemize}
\item \index{singularity!big bang} Big bangs/crunches:\index{singularity!big crunch}\index{big crunch}
\begin{itemize}
\item $\eta_0>1$, $k= +1$.
\item $\eta_0=1$,  $k=0, +1$, and $k=-1$ if $c_0> 1$.
\item $\eta_0\in(1/3,1)$, for all $k$.
\item $\eta_0=1/3$,
\begin{itemize}
\item $\eta_1>5/3$, $k=+1$.
\item  $\eta_1=5/3$, $k>-{14\over9}c_0c_1$.
\item  $\eta_1\in(1/3,5/3)$, $c_1>0$, for all $k$.
\end{itemize}
\end{itemize}
\item Sudden \index{singularity!sudden singularity}singularities/extremality \index{singularity!extremality events}events:
\begin{itemize}
\item $\eta_0=0$,
\begin{itemize}
 \item $\eta_1>2$, $k= +1$.
\item  $\eta_1=2$, with $k> c_0\;\mathrm{max}\{2c_1,-c_1\}$.
\item  $\eta_1=1$, $\eta_2>2$, $k=0, +1$, and $k=-1$ if $|c_1|>1$.
\item  $\eta_1=1$, $\eta_2=2$, $k> c_0\;\mathrm{max}\{2c_1,-c_1\} - c_1^2$.
\end{itemize}
\end{itemize}
\end{itemize}
\end{theorem}

\begin{theorem} \label{DEC2}
The DEC\index{energy conditions!DEC}\index{DEC} is \emph{marginally satisfied} (meaning the non-strict inequality is actually an equality) at a generic cosmological milestone iff:
\begin{itemize}
\item  \index{singularity!big bang}Big bangs/crunches:\index{singularity!big crunch}\index{big crunch}
\begin{itemize}
\item $\eta_0>1$, $k=0$.
\item $\eta_0=1$, $k=-1$ if $c_0= 1$.
\item $\eta_0=1/3$, 
\begin{itemize}
\item $\eta_1>5/3$, $k=0$.
\item $\eta_1=5/3$, $k=-{14\over9}c_0c_1$.
\end{itemize}
\end{itemize}
%
\item Sudden\index{singularity!sudden singularity} singularities/extremality \index{singularity!extremality events}events:
\begin{itemize}
\item $\eta_0=0$, 
\begin{itemize}
\item $\eta_1>2$, $k=0$.
\item $\eta_1=2$, with $k= c_0\;\mathrm{max}\{2c_1,-c_1\}$.
\item  $\eta_1=1$, $\eta_2>2$, $k=-1$ if $|c_1|=1$.
\item  $\eta_1=1$, $\eta_2=2$, $k= c_0\;\mathrm{max}\{2c_1,-c_1\} - c_1^2$.
\end{itemize}
\end{itemize}
\end{itemize}
\end{theorem}
\begin{theorem}  \label{DEC3}
The DEC\index{energy conditions!DEC}\index{DEC} is \emph{definitely violated} (meaning the inequality is \emph{strict violated}) at a generic cosmological milestone for any cases that are not mentioned in Theorems \ref{DEC1} or \ref{DEC2}.
\end{theorem}

\begin{proof}
The DEC\index{energy conditions!DEC}\index{DEC} is satisfied if $k\geqslant a \;\ddot a - \dot a^2$ and $k \geq -(a\; \ddot a +2\dot a^2)/2$. 
The first condition $k\geqslant a \;\ddot a - \dot a^2$ has been dealt with in the proof of Theorems \ref{NEC1}, \ref{NEC2} and \ref{NEC3}. We need to look at $k \geq -(a\; \ddot a +2\dot a^2)/2$. 
Near the cosmological milestone we can write the\index{scale factor} scale factor as a generalized power series as defined in Defintion \ref{def_generic_a}:
\begin{equation}
a(t) = c_0 |t-t_\generic|^{\eta_0} + c_1  |t-t_\generic|^{\eta_1} + c_2  |t-t_\generic|^{\eta_2} 
+ c_3 |t-t_\generic|^{\eta_3} +\dots
\end{equation}
Therefore, near any generic cosmological milestone

\begin{equation}
-{(a \ddot a  + 2 \dot a^2)\over 2} \sim 
-{\eta_0(3\eta_0-1) \;c_0^2\over 2}
 \;(t-t_\generic)^{2(\eta_0-1)}; \qquad (\eta_0\neq0, \;\;\; \eta_0\neq 1/3),
\end{equation}
while in the various degenerate cases
\begin{equation}
-{(a \ddot a  + 2 \dot a^2)\over 2} \sim 
-{\eta_1(\eta_1-1) \;c_0 \; c_1\over 2}
 \;(t-t_\generic)^{(\eta_1-2)};
 \qquad 
(\eta_0=0; \;\;\; \eta_1\neq1),
\end{equation}
\begin{equation}
-{(a \ddot a  + 2 \dot a^2)\over 2} \sim 
-{\eta_2(\eta_2-1) \;c_0 \; c_2\over 2}
 \;(t-t_\generic)^{(\eta_2-2)}-c_1^2;
\qquad (\eta_0=0; \;\;\; \eta_1=1),
\end{equation}
\begin{equation}
-{(a \ddot a  + 2 \dot a^2)\over 2} \sim 
-{(3\eta_1+2)(3\eta_1-1) \;c_0 \; c_1\over 18}
 \;(t-t_\generic)^{(\eta_1-5/3)};
\qquad (\eta_0=1/3).
\end{equation}
Therefore
\begin{equation}
\lim_{t\to t_\generic} 
\left[-{(a \ddot a  + 2 \dot a^2)\over 2} \right]
= 
\left\{ \begin{array}{ll}
0  & \qquad \eta_0>1\;\\
- c_0^2 & \qquad\eta_0=1;\\
-\infty &\qquad \eta_0 \in(1/3,1);\\
%
0 & \qquad \eta_0=1/3; \qquad \eta_1>5/3;\\
-{14\over9} c_0 c_1 & \qquad \eta_0=1/3; \qquad \eta_1=5/3;\\
-\mathrm{sign}(c_1)\infty & \qquad \eta_0=1/3; \qquad \eta_1<5/3;\\
+\infty &\qquad \eta_0 \in(0,1/3);\\
0 & \qquad \eta_0=0; \qquad \eta_1>2;\\
-c_0c_1 & \qquad \eta_0=0; \qquad \eta_1=2;\\
-\mathrm{sign}(c_1)\infty   & \qquad \eta_0=0; \qquad \eta_1\in(1,2)\\
-c_1^2 & \qquad \eta_0=0; \qquad \eta_1=1; \qquad \eta_2 >2;\\
-c_0c_2 - c_1^2& \qquad \eta_0=0; \qquad \eta_1=1; \qquad \eta_2=2;\\
-\mathrm{sign}(c_2)\infty & \qquad \eta_0=0; \qquad \eta_1=1; \qquad \eta_2\in(1,2);\\
+\mathrm{sign}(c_1)\infty & \qquad \eta_0=0; \qquad \eta_1\in(0,1);\\
-\infty &\qquad \eta_0 <0.\\
\end{array}
\right.
\end{equation}
Remember that to satisfy the DEC\index{energy conditions!DEC}\index{DEC} one needs to satisfy the NEC\index{energy conditions!NEC}\index{NEC} in addition to the constraint coming from the above. Let us define
\begin{equation}
K = \mathrm{max} \left\{   
\lim_{t\to t_\generic} 
\left[ a \ddot a - \dot a^2 \right],
\lim_{t\to t_\generic} 
\left[-{(a \ddot a  + 2 \dot a^2)\over 2} \right]  
\right\}.
\end{equation}
Then satisfying the DEC\index{energy conditions!DEC}\index{DEC} at the cosmological milestone is equivalent to the constraint
\begin{equation}
k \geq K.
\end{equation}
--- For \index{singularity!big bang}bangs and \index{singularity!big crunch}\index{big crunch}crunches we calculate:
\begin{equation}
K = 
\left\{ \begin{array}{ll}
0  & \qquad \eta_0>1;\\
- c_0^2 & \qquad\eta_0=1;\\
-\infty &\qquad \eta_0 \in(1/3,1);\\
%
0 & \qquad \eta_0=1/3; \qquad \eta_1>5/3;\\
-{14\over9} c_0 c_1 & \qquad \eta_0=1/3; \qquad \eta_1=5/3;\\
-\mathrm{sign}(c_1)\infty & \qquad \eta_0=1/3; \qquad \eta_1<5/3;\\
+\infty &\qquad \eta_0 \in(0,1/3).\\
\end{array}
\right.
\end{equation}

--- For rips we have:
\begin{equation}
K = +\infty \qquad \eta_0 <0.
\end{equation}
Note that rips, because they violate the NEC\index{energy conditions!NEC}\index{NEC}, always violate the DEC\index{energy conditions!DEC}\index{DEC}, so the particular constraint derived above is not the controlling feature. 
\\
--- Finally, for sudden\index{singularity!sudden singularity} singularities and \index{singularity!extremality events}extremality events:
\begin{equation}
K = 
\left\{ \begin{array}{ll}
0 & \qquad \eta_0=0; \qquad \eta_1>2;\\
c_0\;\mathrm{max}\{2c_1,-c_1\} & \qquad \eta_0=0; \qquad \eta_1=2;\\
+\infty   & \qquad \eta_0=0; \qquad \eta_1\in(1,2)\\
-c_1^2 & \qquad \eta_0=0; \qquad \eta_1=1; \qquad \eta_2 >2;\\
c_0\;\mathrm{max}\{2c_1,-c_1\} - c_1^2& \qquad \eta_0=0; \qquad \eta_1=1; \qquad \eta_2=2;\\
+\infty & \qquad \eta_0=0; \qquad \eta_1=1; \qquad \eta_2\in(1,2);\\
+\infty & \qquad \eta_0=0; \qquad \eta_1\in(0,1).\\
%
%
\end{array}
\right.
\end{equation}
As we have now seen (several times), the $\eta_0=0$ case is ``special'' and requires extra care and delicacy in the analysis --- this is the underlying reason why\index{singularity!sudden singularity} ``sudden singularities'' are so ``fragile'', and so dependent on the specific details of the particular model. Indeed there are several classes of cosmological milestone for which the DEC\index{energy conditions!DEC}\index{DEC} is \emph{satisfied}. 
 For instance, a complete catalogue of the bangs \index{singularity!big bang}and \index{singularity!big crunch}\index{big crunch}crunches for which the DEC\index{DEC} is satisfied is:
\begin{itemize}
\item $\eta_0>1$, $k=0, +1$.
\item $\eta_0=1$,  $k=0, +1$, and $k=-1$ if $c_0\geq 1$.
\item $\eta_0\in(1/3,1)$, for all $k$.
\item $\eta_0=1/3$, $\eta_1>5/3$, $k=0, +1$.
\item  $\eta_0=1/3$, $\eta_1=5/3$, $k\geq-{14\over9}c_0c_1$.
\item  $\eta_0=1/3$, $\eta_1\in(1/3,5/3)$, $c_1>0$, for all $k$.
\end{itemize}

Similarly, a complete catalogue of the \index{singularity!sudden singularity}sudden singularities for which the DEC\index{energy conditions!DEC}\index{DEC} is satisfied is:
\begin{itemize}
\item $\eta_0=0$, $\eta_1>2$, $k=0, +1$.
\item $\eta_0=0$, $\eta_1=2$, with $k\geq c_0\;\mathrm{max}\{2c_1,-c_1\}$.\\
(In particular this requires $k=+1$ though that is not sufficient.)
\item  $\eta_0=0$, $\eta_1=1$, $\eta_2>2$, $k=0, +1$, and $k=-1$ if $|c_1|\geq1$.
\item  $\eta_0=0$, $\eta_1=1$, $\eta_2=2$, $k\geq c_0\;\mathrm{max}\{2c_1,-c_1\} - c_1^2$.
\end{itemize}
These catalogues prove Theorem \ref{DEC1}, \ref{DEC2}, and \ref{DEC3}.
\end{proof}
If one wishes to use ``normal'' matter (that is, matter satisfying all of the energy conditions\index{energy conditions}) to drive a bang\index{singularity!big bang} or a \index{singularity!big crunch}\index{big crunch}crunch independent of the sign of space curvature $k$, then one is forced in a model independent manner into the range $\eta_0\in(1/3,1)$. 
In particular, if the DEC\index{energy conditions!DEC}\index{DEC} is to hold independent of the sign of space curvature $k$ all the way down to the singularity, then in the vicinity of the singularity the dominant term in the \index{scale factor}scale factor $a(t)$ is bounded from both above and below by
\begin{equation}
c_0 (t-t_\bang) < a_\mathrm{dominant}(t) \leq c_0 (t-t_\bang)^{1/3}.
\end{equation}

The fact that we find special cases of\index{singularity!sudden singularity} sudden singularities that do not violate the DEC\index{energy conditions!DEC}\index{DEC} \emph{appears} at first glance, to contradict the analysis of Lake~\cite{Lake:2004fu}, who claimed that \emph{all} sudden singularities violate the DEC\index{DEC}.  The resolution of this apparent contradiction lies in the question ``\emph{just how sudden is the sudden singularity?}'' Lake takes as his definition $\ddot a(t\to t_\generic) = -\infty$, corresponding in our analysis to $\eta_0=0$ with either $\eta_1\in(0,1)\cup(1,2)$ or $\eta_1=1$ with $\eta_2<2$. In this situation our results certainly agree with those of Lake: if $\ddot a(t\to t_\generic) = -\infty$ then the DEC\index{energy conditions!DEC}\index{DEC} is certainly violated. However, our  ``counterexamples'' where the DEC\index{DEC} is satisfied all satisfy $\ddot a(t\to t_\generic) = \hbox{finite}$, and for these counterexamples it is only some higher derivative  $a^{(n)}(t\to t_\generic)$ for $n\geq3$ that diverges. Sudden singularities of this type are sufficiently ``gentle'' that at least for some of them even the DEC\index{energy conditions!DEC}\index{DEC} can be satisfied all the way to the singularity: a conclusion completely in agreement with Barrow and Tsagas~\cite{Barrow:2004he}, though now we have a complete characterization of those situations for which the DEC\index{energy conditions!DEC}\index{DEC} can be satisfied.

\section{Total age of the universe}\index{universe!total age of the universe}
In this section, we express the density as a function of the scale factor $\rho(a(t))$ and analyze the Friedmann equation (\ref{Fried_eq1}) rewritten with $\rho(a(t))$. This equation can then be formally solved by integration to find $t(a)$, time as a function of the scale parameter.\\

We adopt a \textit{barotropic equation of state} by assuming the statement that the pressure is a function of density, $p(\rho)$. If we express the density as a function of the scale factor, we can write $\rho=\rho(a(t))$. We can then write,
\begin{equation}
\frac{d \rho}{\d a}=\frac{d \rho / dt}{d a /d t}.
\end{equation}
Now rewriting the conservation equation, in units where $8 \pi G_{N}=1$ and $c=1$, we obtain,
\begin{equation}
\frac{d\rho}{da}+3\left[ \frac{\rho+p(\rho) }{a}\right]=0,
\end{equation}
so that 
\begin{equation}
-\frac{d\rho}{3[\rho+p(\rho)]}=\frac{da}{a}.
\end{equation}
Integrating, we get
\begin{equation}
-\int \frac{d\rho}{3[\rho+p(\rho)]}=\int \frac{da}{a}.
\end{equation}

\begin{definition} \label{def_relativistic_enthalpy}
Define,
\begin{equation}
h(\rho) = \int_{\rho_{p=0}}^\rho {d\bar\rho \over [\bar\rho+p(\bar\rho)]} 
\end{equation}
where $\rho_{p=0}$ is the density at zero pressure which is defined by
\begin{equation}
p(\rho_{p=0}) = 0.
\end{equation}
The function $h(\rho)$ is called the \emph{relativistic specific enthalpy}.
\end{definition}
 Using Definition \ref{def_relativistic_enthalpy}, then in terms of the present day density $\rho_0$ and the present day scale factor $a_0$
\begin{equation}
-\frac{1}{3}h(\rho) - h(\rho_0) = \ln(a/a_0).
\end{equation}
We can now write
\begin{equation}
a=a_0 \exp \left[-\frac{1}{3}(h(\rho)-h(\rho_0))\right],
\end{equation}
and deduce the existence of a function $f$ such as
\begin{equation}
a=f(\rho). \label{eq_f}
\end{equation}
Note that when the density increases, assuming the NEC holds so does the enthalpy and therefore, the scale factor decreases.
Now, apply the inverse function theorem to the function $f$ in equation (\ref{eq_f}). After all 
\begin{equation}
\frac{d h}{d \rho} =\frac{1}{\rho+p(\rho)} \neq 0
\end{equation}
except for very peculiar isolated cases. Therefore, the conditions for employing the inverse function theorem apply, and thus we can write
\begin{equation}
\rho=f^{-1}(a)
\end{equation}
which we often just abbreviate by writing $\rho(a)$.  \\
The Friedmann equation can now be written as:
\begin{equation}
\rho(a(t))=3\left(\frac{\dot{a}^{2}(t)}{a^{2}(t)}+\frac{k}{a^{2}(t)}\right) \label{Fried_a}
\end{equation}

Equation (\ref{Fried_a}) can be rearranged into equation (\ref{Friedmann_a_rearranged}) and formally solved by integration to find $t(a)$, time as a function of scale parameter $a$. First note
\begin{equation}
\frac{1}{3}\rho(a(t)) \; a^{2}(t)-\dot{a}^{2}(t)=k.  \label{Friedmann_a_rearranged}
\end{equation}
\begin{theorem}
Let the time now be $t_{0}$ and the size of the universe now be $a_{0}$, the standard solution of equation (\ref{Friedmann_a_rearranged}) in the current epoch is:
\begin{equation}
 t(a)=\int ^{a}_{a_{0}} \frac{d \bar{a}}{\sqrt{\frac{1}{3} \rho(\bar{a}) \bar{a}^{2}-k}}+t_{0}. \label{time_int}
\end{equation}

\end{theorem}
\begin{proof}
There are three solutions to equation (\ref{Friedmann_a_rearranged}):
\begin{enumerate}

\item 
\begin{equation}
\dot{a}(t)=0 \Rightarrow  \rho(a(t)) \; a^{2}(t)/3=k,
\end{equation}
 this means that the size of the universe does not change over time. This case is the Einstein static universe.
A more complicated analysis would show that for ``normal'' matter the Einstein static universe is always unstable. It might be possible to find some form of  ``unusual'' matter to make the Einstein static universe stable. 

\item 
\begin{equation}
 t(a)=-\int ^{a} \frac{d \bar{a}}{\sqrt{\frac{1}{3} \rho(\bar{a}) \bar{a}^{2}-k}}+\mathrm{constant},
 \end{equation}
in this case 
\begin{equation}
\frac{1}{\sqrt{\frac{1}{3} \rho(\bar{a}) \bar{a}^{2}-k}} > 0  \Rightarrow \int ^{a} \frac{d \bar{a}}{\sqrt{\frac{1}{3} \rho(\bar{a}) \bar{a}^{2}-k}} > 0. 
\end{equation}

Time decreases as the integral increases. That is time increases as the integral decreases (as the scale factor $a$ decreases). So time increases as the universe gets smaller. This represents a contracting universe. Our universe is not currently contracting, but it is conceivable that the cosmological expansion could achieve a maximum, then turn around, and start to contract. If this happens then we would want to use this solution during the contracting phase.

\item 
\begin{equation}
t(a)=\int ^{a} \frac{d \bar{a}}{\sqrt{\frac{1}{3} \rho(\bar{a}) \bar{a}^{2}-k}}+\mathrm{constant},
\end{equation} 
 this solution is the standard solution appropriate to the current cosmological epoch.
\end{enumerate}
Therefore, if we define $t_{0}$ as the time now and $a_{0}$ as the size of the universe now, the solution of equation (\ref{Friedmann_a_rearranged}) is:
\begin{equation}
 t(a)=\int ^{a}_{a_{0}} \frac{d \bar{a}}{\sqrt{\frac{1}{3} \rho(\bar{a}) \bar{a}^{2}-k}}+t_{0}
\end{equation}
\end{proof}

\subsubsection{Domains of reality} 
We have shown that the current standard solution to equation (\ref{Friedmann_a_rearranged}) contains an integral of a square root $ \sqrt{ \rho(a) a^2/3- k }$. However, this integral exists and is \emph{real} only if the term inside the square root is positive.\\

There are at least 5 situations wherein the square root $ \sqrt{ \rho(a) a^2/3- k }$ is real:
\begin{enumerate}
\item The square root is real only on $[0,a_\mathrm{max}]$. (This corresponds to a big bang, and then a turnaround and a possible recollapse, or a future sudden singularity.)
\item The square root is real only on $[a_\star,\infty]$. (This corresponds to a bounce or a past sudden singularity then followed by infinite expansion.)
\item The square root is real on the entire positve axis $[0,\infty]$. (This is the most common assumption in cosmology, it corresponds to a big bang, then followed by infinite expansion).
\item The square root is real only on a finite interval bounded away from zero, $[a_\star,a_\mathrm{max}]$. (This corresponds to a bounce or past sudden singularity subsequently followed by a turnaround or future sudden singularity; a possible oscillating universe...)
\item
Disjoint unions of various of the above. (Not physically interesting.)
\end{enumerate}

\subsection{Age of the universe $\tau$}\index{universe!age of the universe}
In this section, we will concentrate on the lower bound of the intervals of the different cases mentioned above, where the square root $ \sqrt{ \rho(a) a^2/3- k }$ is real.
But first, we will present a generic definition and a generic theorem on the age of the universe. \begin{definition}
 The \emph{age of the universe} $\tau$ is the time elapsed since $a=a_{\generic}$.
\end{definition}
Here $a_{\generic}$ can represent a big bang (in this case $a_{\generic}=0$) or a minimum value (for a bounce or past sudden singularity in this case $a_{\generic}=a_{\star}$).

\begin{theorem} \label{theo_age}
The age of the universe $\tau$ is given by,
\begin{equation}
 \tau=\int ^{a_{0}}_{a_{\generic}} \frac{d \bar{a}}{\sqrt{\frac{1}{3} \rho(\bar{a}) \bar{a}^{2}-k}}.   \label{age_universe_generic}
\end{equation}
\end{theorem}
\begin{proof}
Define $t_\generic = t(a=a_{\generic})$ then
\begin{equation}
 t_\generic=\int ^{a_{\generic}}_{a_{0}} \frac{d \bar{a}}{\sqrt{\frac{1}{3} \rho(\bar{a}) \bar{a}^{2}-k}}+t_{0},
\end{equation}
where $t_0$ is the time now. The age of the universe is therefore given by
\begin{equation}
 \tau=t_{0}-t_\generic=\int ^{a_{0}}_{a_{\generic}} \frac{d \bar{a}}{\sqrt{\frac{1}{3} \rho(\bar{a}) \bar{a}^{2}-k}}.   
 \end{equation}
\end{proof}

\begin{corollary}
Assuming the universe starts with a big bang, the \emph{age of the universe} $\tau$ is the time elapsed since $a=0$, and $\tau$ is given by,
\begin{equation}
 \tau=\int ^{a_{0}}_{0} \frac{d \bar{a}}{\sqrt{\frac{1}{3} \rho(\bar{a}) \bar{a}^{2}-k}}.  
\end{equation}
\end{corollary}

\begin{corollary}
Assuming the universe starts with a bounce or a past sudden singularity, the \emph{age of the universe} $\tau$ is the time elapsed since a minimum value $a=a_{\star}$, and $\tau$ is given by,
\begin{equation}
 \tau=\int ^{a_{0}}_{a_{\star}} \frac{d \bar{a}}{\sqrt{\frac{1}{3} \rho(\bar{a}) \bar{a}^{2}-k}}.  
\end{equation}
\end{corollary}

\begin{theorem} \label{theo_real_finite}
The age of the universe $\tau$ is \emph{real} and \emph{finite}, that is, $\tau < \infty$ and $ \tau \in\mathbb{R}$, if only if the integral 
\begin{equation}
\int ^{a_{0}}_{a_{\generic}} \frac{d \bar{a}}{\sqrt{\frac{1}{3} \rho(\bar{a}) \bar{a}^{2}-k}}   
\end{equation}
is both real and convergent. \\
That is, we demand both:
\begin{enumerate}
\item for reality:
\begin{equation}
 \rho(a)>\frac{3 k }{a^{2}};
\end{equation}
\item for convergence: \\
there exist constants $\alpha<1$, $c>0$ and $a_\epsilon>0$ such that: $\forall a \in(a_{\generic},a_\epsilon)$
\begin{equation}
\rho(a)  \geq {3k\over  a^2} + c^2 a^{2\alpha-2};
\end{equation}
and $\forall a \in(a_\epsilon,a_0)$ 
\begin{equation}
\rho(a) \geq \rho(a_\epsilon). 
\end{equation}
\end{enumerate}
\end{theorem}
\index{universe!age of the universe}
\begin{proof}
\begin{enumerate}
\item First, $\tau \in\mathbb{R}$ if 
\begin{equation}
\int ^{a_{0}}_{a_{\generic}} \frac{d \bar{a}}{\sqrt{\frac{1}{3} \rho(\bar{a}) \bar{a}^{2}-k}}   \; \in \mathbb{R},
\end{equation}
that is, if
\begin{equation}
 \rho(a) a^{2}/3-k >0,
 \end{equation} 
 which leads to the reality condition:
\begin{equation}
 \rho(a)>\frac{3 k }{a^{2}}.
\end{equation}
 
\item Second, $\tau < \infty$ if there exist constants $\alpha<1$, $c>0$ and $a_\epsilon>0$ such that: $\forall a \in(a_{\generic},a_\epsilon)$
\begin{equation}
\sqrt{  \rho(a) a^2/3- k } \geq c a^\alpha \geq 0,
\end{equation}
that is 
\begin{equation}
\rho(a)  \geq {3k\over  a^2} + c^2 a^{2\alpha-2};
\end{equation}
and, if furthermore $\forall a \in(a_\epsilon,a_0)$ 
\begin{equation}
\sqrt{  \rho(a) a^2/3 - k } \geq \sqrt{\rho(a_\epsilon) a_\epsilon^2/3 - k },
\end{equation}
that is
\begin{equation}
\rho(a) \geq \rho(a_\epsilon).
\end{equation}
\end{enumerate}
\end{proof}

\subsection{Remaining lifetime of the universe $T$}\index{universe!remaining lifetime of the universe}
\begin{definition}
The \emph{remaining lifetime of the universe} $T$ is the remaining time from now, where $a=a_0$, until $a_{\infty}=a_{\endu}$.
\end{definition}
Here $a_{\endu}$ can represent a big rip or eternal expansion at a finite non zero rate (in this case $a_{\endu}=\infty$), or a maximum value (for a turnaround or future sudden singularity or eternal expansion asymptotic to zero velocity in this case $a_{\endu}=a_\mathrm{max}$).

\begin{theorem}  \label{theo2}
The remaining lifetime of the universe $T$ is given by,
\begin{equation}
 T=\int ^{a_{\endu}}_{a_{0}} \frac{d \bar{a}}{\sqrt{\frac{1}{3} \rho(\bar{a}) \bar{a}^{2}-k}}
\end{equation}
\end{theorem}
\begin{proof}
Define $ t(a_{\infty})$ the time when $a_{\infty}=a_{\endu}$, then
\begin{equation}
 t(a_{\infty})=\int ^{a_{\endu}}_{a_{0}} \frac{d \bar{a}}{\sqrt{\frac{1}{3} \rho(\bar{a}) \bar{a}^{2}-k}}+t_{0}.
\end{equation}
Therefore, the remaining lifetime of the universe $T$ is given by
\begin{equation}
 T=t(a_{\infty})-t_{0}=\int ^{a_{\endu}}_{a_{0}} \frac{d \bar{a}}{\sqrt{\frac{1}{3} \rho(\bar{a}) \bar{a}^{2}-k}}.
\end{equation}
\end{proof}
\index{remaining lifetime of the universe}
\begin{corollary}
Assuming the universe ends with a big rip, the \emph{remaining lifetime of the universe} $T$ is the remaining time from now $a=a_0$ until $a_{\endu}=\infty$, and $T$ is given by,
\begin{equation}
 T=\int _{a_{0}}^{\infty} \frac{d \bar{a}}{\sqrt{\frac{1}{3} \rho(\bar{a}) \bar{a}^{2}-k}}.  
\end{equation}
\end{corollary}
\index{universe!remaining lifetime of the universe}
\begin{corollary}
Assuming the universe continues to expand at a finite nonzero rate for all eternity then, the \emph{remaining lifetime of the universe} $T$ \emph{diverges}. Here $T$ is the remaining time from now $a=a_0$ until $a_{\endu}=\infty$, and $T$ is given by,
\begin{equation}
 T=\int _{a_{0}}^{\infty} \frac{d \bar{a}}{\sqrt{\frac{1}{3} \rho(\bar{a}) \bar{a}^{2}-k}}. 
 \end{equation}
\end{corollary}

\begin{corollary}
Assuming the universe ends by reaching a maximum value (possible turnaround or future sudden singularity), the \emph{remaining lifetime of the universe} $T$ is the remaining time from now $a=a_0$ until $a_{\endu}=a_{\mathrm{max}}$, and $T$ is given by,
\begin{equation}
 T=\int _{a_{0}}^{a_{\mathrm{max}}} \frac{d \bar{a}}{\sqrt{\frac{1}{3} \rho(\bar{a}) \bar{a}^{2}-k}}.  
\end{equation}
\end{corollary}


\begin{theorem} \label{theo_real_finite_T}
The remaining lifetime of the universe $T$ is \emph{real} and \emph{finite}, that is, $T < \infty$ and $ T \in\mathbb{R}$, if and only if the integral 
\begin{equation}
\int _{a_{0}}^{a_{\endu}} \frac{d \bar{a}}{\sqrt{\frac{1}{3} \rho(\bar{a}) \bar{a}^{2}-k}}   
\end{equation}
is both real and convergent.\\
That is, we demand both
\begin{enumerate}
\item for reality:
\begin{equation}
 \rho(a)>\frac{3 k }{a^{2}};
\end{equation}
\item for convergence:\\
\begin{enumerate}
\item \emph{case $a_{\endu}=\infty$}:
 If there exist constants $\beta>1$, $c>0$ and $a_\epsilon>0$ such that: $\forall a \in(a_\epsilon,a_{\endu})$
\begin{equation}
\rho(a)  \geq {3k\over  a^2} + c^2 a^{2\beta-2}.
\end{equation}

\index{universe!remaining lifetime of the universe}

\item \emph{case $a_{\endu}=a_{\mathrm{max}}$}:
If there exists constants $\alpha < 1$, $K>0$ and $a_\epsilon>0$ such that: $\forall a \in(a_\epsilon,a_{max})$
\begin{equation}
\rho(a)  \geq {3k\over  a^2} + \dfrac{3(a_{max}-a)^{2 \alpha}}{K^{2}a^{2}} > {3k\over  a_{max}^2} + \dfrac{3(a_{max}-a)^{2 \alpha}}{K^{2}a_{max}^{2}}.
\end{equation}
\end{enumerate}
And finally, $\forall a \in(a_0,a_\epsilon)$ 
\begin{equation}
\rho(a) \geq \rho(a_\epsilon). 
\end{equation}

\end{enumerate}
\end{theorem}

\begin{proof}
\begin{enumerate}
\item First, $T \in\mathbb{R}$ if 
\begin{equation}
\int _{a_{0}}^{a_{\endu}} \frac{d \bar{a}}{\sqrt{\frac{1}{3} \rho(\bar{a}) \bar{a}^{2}-k}}   \; \in \mathbb{R},
\end{equation}
that is, if
\begin{equation}
 \rho(a) a^{2}/3-k >0,
 \end{equation} 
 which leads to the reality condition:
\begin{equation}
 \rho(a)>\frac{3 k }{a^{2}}.
\end{equation}
 
\item Second, 
\begin{enumerate}
\item \emph{case $a_{\endu}=\infty$}: 
$T < \infty$ if there exist constants $\beta>1$, $c>0$ and $a_\epsilon>0$ such that: $\forall a \in(a_\epsilon,a_{\endu})$
\begin{equation}
\sqrt{  \rho(a) a^2/3- k } \geq c a^\beta \geq 0,
\end{equation}
that is 
\begin{equation}
\rho(a)  \geq {3k\over  a^2} + c^2 a^{2\beta-2}.
\end{equation}

\item \emph{case $a_{\endu}=a_{\mathrm{max}}$}: 
$T < \infty$ if there exist constants  $\alpha < 1$, $K>0$ and $a_\epsilon>0$ such that: $\forall a \in(a_\epsilon,a_{max})$
\begin{equation}
\vert f(a) \vert \leq \dfrac{K}{(a_{max}-a)^{\alpha}},
\end{equation}
that is,
\begin{equation}
\rho(a)  \geq {3k\over  a^2} + \dfrac{3(a_{max}-a)^{2 \alpha}}{K^{2}a^{2}} > {3k\over  a_{max}^2} + \dfrac{3(a_{max}-a)^{2 \alpha}}{K^{2}a_{max}^{2}}. 
\end{equation}

\end{enumerate}
And, if furthermore $\forall a \in(a_0, a_\epsilon)$ 
\begin{equation}
\sqrt{  \rho(a) a^2/3 - k } \geq \sqrt{\rho(a_\epsilon) a_\epsilon^2/3 - k },
\end{equation}
that is
\begin{equation}
\rho(a) \geq \rho(a_\epsilon).
\end{equation}
\end{enumerate}
\end{proof}

\subsection{Total age of the universe $T_{\mathrm{total}}$}\index{universe!total age of the universe}
\begin{definition}
The \emph{total age of the universe} $T_{\mathrm{total}}$ is the time elapsed since $a=a_{\generic}$ until $a_{\infty}=a_{\endu}$.
\end{definition}
\begin{lemma}
The total age of the universe $T_{\mathrm{total}}$ is the sum of the age of the universe $\tau$ and the remaining lifetime of the universe $T$, that is, 
\begin{equation}
T_{\mathrm{total}} =\tau +T.
\end{equation}
\end{lemma}
\begin{lemma}
The total age of the universe $T_{\mathrm{total}}$ is \emph{real} and \emph{finite} if and only if:
\begin{enumerate}
\item the present age of the universe is real and finite $\tau \in\mathbb{R}$ and $\tau<\infty$,
\item and the remaining time of the universe is real and finite $T \in\mathbb{R}$ and $T<\infty$.
\end{enumerate}
  \end{lemma}
\index{universe!total age of the universe}
\begin{theorem} \label{theo3}
The total age of the universe $T_{\mathrm{total}}$ is given by,
\begin{equation}
 T_{\mathrm{total}}  =\int ^{a_{\endu}}_{a_{\generic}} \frac{d \bar{a}}{\sqrt{\frac{1}{3} \rho(\bar{a}) \bar{a}^{2}-k}}.  
\end{equation}
\end{theorem}

\begin{proof}
Theorem \ref{theo3} is deduced from Theorem \ref{theo_age} and Theorem \ref{theo2}.
\end{proof}

\begin{theorem} \label{theo_total}
The total age of the universe $T_{\mathrm{total}}$ is \emph{real} and \emph{finite}, that is, $T_{\mathrm{total}} < \infty$ and $ T_{\mathrm{total}} \in\mathbb{R}$, if and only if the integral 
\begin{equation}
\int ^{a_{\endu}}_{a_{\generic}} \frac{d \bar{a}}{\sqrt{\frac{1}{3} \rho(\bar{a}) \bar{a}^{2}-k}}   
\end{equation}
is both real and convergent.\\
That is, we demand both
\begin{enumerate}
\item for reality:
\begin{equation}
 \rho(a)>\frac{3 k }{a^{2}};
\end{equation}
\item for convergence:\\
\begin{enumerate}
\item if there exist constants $\alpha<1$, $c>0$ and $a_\epsilon>0$ such that: $\forall a \in(a_{\generic},a_\epsilon)$
\begin{equation}
\rho(a)  \geq {3k\over  a^2} + c^2 a^{2\alpha-2};
\end{equation}
\item 
\begin{enumerate}
\item \emph{case $a_{\endu}=\infty$}:
 If there exist constants $\beta>1$, $c>0$ and $a_\epsilon>0$ such that: $\forall a \in(a_\epsilon,a_{\endu})$
\begin{equation}
\rho(a)  \geq {3k\over  a^2} + c^2 a^{2\beta-2}.
\end{equation}
\item \emph{case $a_{\endu}=a_{\mathrm{max}}$}:
If there exists constants $\alpha < 1$, $K>0$ and $a_\epsilon>0$ such that: $\forall a \in(a_\epsilon,a_{max})$
\begin{equation}
\rho(a)  \geq {3k\over  a^2} + \dfrac{3(a_{max}-a)^{2 \alpha}}{K^{2}a^{2}} > {3k\over  a_{max}^2} + \dfrac{3(a_{max}-a)^{2 \alpha}}{K^{2}a_{max}^{2}}.
\end{equation}
\end{enumerate}
\end{enumerate}
\end{enumerate}

\end{theorem}
\index{universe!total age of the universe}

\begin{proof}
Theorem \ref{theo_total} is deduced from Theorem \ref{theo_real_finite} and Theorem \ref{theo_real_finite_T}.
\end{proof}

\index{universe!total age of the universe}

Now, applying the generic Theorem \ref{theo_total} to the five cases where the square root $ \sqrt{ \rho(a) a^2/3- k }$ is real, we obtain the following corollaries:

\begin{corollary} \label{coro1}
Assume that the square root is real only on $[0,a_\mathrm{max}]$ and $a_\mathrm{max}> a_{0}$. 

If there exist constants $\alpha<1$, $c>0$ and $a_\epsilon>0$ such that: $\forall a \in(0,a_\epsilon)$
\begin{equation}
\rho(a)  \geq {3k\over  a^2} + c^2 a^{2\alpha-2} > {3k\over  a^2};
\end{equation}
and if there exists constants $\alpha < 1$, $K>0$ and $a_\epsilon>0$ such that: $\forall a \in(a_\epsilon,a_{max})$
\begin{equation}
\rho(a)  \geq {3k\over  a^2} + \dfrac{3(a_{max}-a)^{2 \alpha}}{K^{2}a^{2}} > {3k\over  a_{max}^2} + \dfrac{3(a_{max}-a)^{2 \alpha}}{K^{2}a_{max}^{2}};
\end{equation}
then
\begin{equation}
T_{\mathrm{total}} < \infty \qquad \hbox{and} \qquad T_{\mathrm{total}} \in\mathbb{R}.
\end{equation} 
\end{corollary}
Corollary \ref{coro1} deals with a universe that starts with a big bang ($a_{\generic}=0$), and tends to a maximum size ($a_\endu=a_{\mathrm{max}}$) in a finite time. This universe could possibly have a turnaround and contracting phase until a big crunch $a=0$ or endure a bounce $a=a_{\star}$.

\begin{corollary}  \label{coro2}
Assume that the square root is real only on $[a_\star,\infty]$.

Here $a=0$ is not possible, hence there is a bounce or a past sudden singularity. 

If there exist constants $\alpha<1$, $c>0$ and $a_\epsilon>0$ such that: $\forall a \in(a_\star,a_\epsilon)$
\begin{equation}
\rho(a)  \geq {3k\over  a^2} + c^2 a^{2\alpha-2} > {3k\over  a^2};
\end{equation}
and if there exists constants $\beta>1$, $c>0$ and $a_\epsilon>0$ such that: $\forall a \in(a_\epsilon,\infty)$
\begin{equation}
\rho(a)  \geq {3k\over  a^2} + c^2 a^{2\beta-2} > {3k\over a^2}; 
\end{equation}
then
\begin{equation}
T_{\mathrm{total}} < \infty \qquad \hbox{and} \qquad T_{\mathrm{total}} \in\mathbb{R}.
\end{equation}
\end{corollary}
Corollary \ref{coro2} deals with a universe that starts with a bounce or a past sudden singularity (minimum value $a_{\generic}=a_\star$), and ends with a big rip (infinite value $a_{\endu}=\infty$) in a finite time.
\index{universe!total age of the universe}
\begin{corollary} \label{coro3}
Assume that the square root is real on the entire positive axis $[0,\infty]$. 

If there exist constants $\alpha<1$, $c>0$ and $a_\epsilon>0$ such that: $\forall a \in(0,a_\epsilon)$
\begin{equation}
\rho(a)  \geq {3k\over  a^2} + c^2 a^{2\alpha-2} > {3k\over  a^2};
\end{equation}
and if there exists constants $\beta>1$, $c>0$ and $a_\epsilon>0$ such that: $\forall a \in(a_\epsilon,\infty)$
\begin{equation}
\rho(a)  \geq {3k\over  a^2} + c^2 a^{2\beta-2} > {3k\over a^2};
\end{equation}
then
\begin{equation}
T_{\mathrm{total}} < \infty \qquad \hbox{and} \qquad T_{\mathrm{total}} \in\mathbb{R}.
\end{equation}
\end{corollary}

Corollary \ref{coro3} deals with a universe that starts with a big bang ($a_{\generic}=0$) and ends with a big rip ($a_{\endu}=\infty$) in a finite time. 

\begin{corollary} \label{coro4}
Assume that the square root is real only on a finite interval bounded away from zero, $[a_\star,a_\mathrm{max}]$ and $ a_\star<a_{0}<a_\mathrm{max}$.


If there exist constants $\alpha<1$, $c>0$ and $a_\epsilon>0$ such that: $\forall a \in(a_\star,a_\epsilon)$
\begin{equation}
\rho(a)  \geq {3k\over  a^2} + c^2 a^{2\alpha-2} > {3k\over  a^2};
\end{equation}
and if there exists constants $\alpha < 1$, $K>0$ and $a_\epsilon>0$ such that: $\forall a \in(a_\epsilon,a_{max})$
 \begin{equation}
\rho(a)  \geq {3k\over  a^2} + \dfrac{3(a_{max}-a)^{2 \alpha}}{K^{2}a^{2}} > {3k\over  a_{max}^2} + \dfrac{3(a_{max}-a)^{2 \alpha}}{K^{2}a_{max}^{2}} ;
\end{equation}
then
\begin{equation}
T_{\mathrm{total}} < \infty \qquad \hbox{and} \qquad T_{\mathrm{total}} \in\mathbb{R}.
\end{equation}
\end{corollary}
Corollary \ref{coro4} deals with a universe that starts with a bounce or a past sudden singularity (minimum value $a_{\generic}=a_{\star}$) and ends with a turnaround or a future sudden singularity (maximum value $a_{\endu}=a_{\mathrm{max}}$). This universe could possibly have a turnaround and contracting phase until a big crunch $a=0$ or endure a bounce $a=a_{\star}$.
\index{universe!total age of the universe}
\begin{corollary}
Disjoint unions of various of the above.

This corollary corresponds to some of the cases described in the above corollaries. 
\end{corollary}
This case is not physically interesting since it corresponds to several unrelated universes that are totally independent of each other.

\subsection{Summary}
In this section, we have presented definitions and theorems regarding the conditions that should be placed on the the density as a function of the scale factor in order for the integral 
\begin{equation}
T_{\mathrm{total}}=\int ^{a_{\endu}}_{a_{\generic}} \frac{d \bar{a}}{\sqrt{\frac{1}{3} \rho(\bar{a}) \bar{a}^{2}-k}}   
\end{equation}
to converge. We obtain model independent results that place constraints on $\rho(a)$ under a minimum of technical assumptions.
Whether or not the total age of the universe is a real and finite number depends on whether these conditions on the density $\rho(a)$ hold or not.

\section{Results and discussion}

In this chapter we have explored three issues concerning singularities in a FRW universe using only classical general relativity\index{universe!FRW}\index{universe!FLRW} and an absolute minimum of technical input assumptions:

\begin{itemize}
\item First we have developed an extensive catalogue of the various ``cosmological milestones'' found in the literature in terms of a generalized power series expansion of the FRW\index{scale factor} scale factor. \\
If in the vicinity of any cosmological milestone, the input scale factor $a(t)$ is a generalized power series, then all physical observables ($H$, $q$, the\index{Riemann!tensor} Riemann tensor, \emph{etc}.) will likewise be generalized power series, with related indicial exponents that can be calculated from the indicial exponents of the scale factor. Whether or not the particular physical observable then diverges at the cosmological milestone is ``simply'' a matter of calculating its dominant indicial exponent in terms of those occuring in the scale factor.  \\
 Indeed, for some unspecified generic cosmological milestone,  that  is defined in terms of the behaviour of the scale factor $a(t)$, and which occurs at some finite time $t_\generic$, we can write in the vicinity of the milestone:
 \begin{equation}
a(t) = c_0 |t-t_\generic|^{\eta_0} + c_1  |t-t_\generic|^{\eta_1} + c_2  |t-t_\generic|^{\eta_2} 
+ c_3 |t-t_\generic|^{\eta_3} +\dots
\end{equation}
The $\eta_i$ are generically real and ordered such that $\eta_0<\eta_1<\eta_2<\eta_3\dots$ Also $c_0 > 0$ though there are no \emph{a priori} constraints on the signs of the other $c_i$, except by definition $c_i\neq0$.

 This generalized power series expansion is sufficiently general to accommodate all commonly occurring models considered in the literature. Specifically,
 \begin{itemize}
\item $\eta_0>0$ holds for the class of cosmological milestones such as big \index{singularity!big bang}bangs or big crunches;\index{singularity!big crunch}\index{big crunch}
\item $\eta_0<0$ holds for the class of cosmological milestones such as big rips;
\item $\eta_0=0$ holds for the class of cosmological milestones such as \index{singularity!sudden singularity}sudden singularities or \index{singularity!extremality events}extremality events (\index{singularity!extremality events!bounce}bounces, \index{singularity!extremality events!turnaround}turnarounds, \index{singularity!extremality events!inflexion}inflexions);\\
 Note that for extremality events the\index{scale factor} scale factor can be modeled using ordinary Taylor series with $\eta_i \; \in Z^+$.
\end{itemize}

\item Second, with the notion of a generalized power series in hand, it is possible at a purely kinematic level to address the question of when a ``cosmological milestone'' corresponds to a curvature singularity, and what type of singularity is implied.\\
In section \ref{spacetime_curvature}, we have shown that the  \index{Hubble!parameter}Hubble parameter $H(t)$ \emph{blows up} in the vicinity of cosmological milestones such as big \index{singularity!big bang}bangs/crunches \index{singularity!big crunch}\index{big crunch}and big rips (Theorem \ref{theo_H_bangs}), however it is \emph{finite} for a specific class of \index{singularity!sudden singularity}sudden singularities/extremality \index{singularity!extremality events}events (Theorem \ref{theo_H_sudden}), specifically,
\begin{equation}
\lim_{t\to t_\generic} H = \left\{ \begin{array}{ll}
+\infty & \eta_0>0;\\
0 & \eta_0=0; \qquad \eta_1 > 1;\\
c_1/c_0 & \eta_0=0; \qquad \eta_1=1;\\
\mathrm{sign}(c_1)\infty  & \eta_0=0; \qquad \eta_1\in(0,1);\\
-\infty &\eta_0<0.
\end{array}
\right.
\end{equation}
where $t_\generic$ is the the time of the event.\\
In the same section, we have also demonstrated that the \index{deceleration parameter}deceleration parameter $q(t)$ is \emph{always finite} for bangs\index{singularity!big bang}, crunches\index{singularity!big crunch}\index{big crunch} and rips ($\eta_0\neq 0$) but it is \emph{infinite} only for a certain subset of the\index{singularity!sudden singularity} sudden singularities
 \begin{equation}
\lim_{t\to t_\generic} q = \left\{ \begin{array}{ll}
(1-\eta_0)/\eta_0 & \qquad\eta_0\neq0;\\
\mathrm{sign}(c_1[1-\eta_1])\infty & \qquad\eta_0=0; \qquad \eta_1\neq 1;\\
0 & \qquad\eta_0=0; \qquad \eta_1= 1; \qquad\eta_2 > 2;\\
-2c_2c_0/c_1^2& \qquad\eta_0=0; \qquad \eta_1= 1; \qquad\eta_2=2; \\
-\mathrm{sign}(c_2)\infty &\qquad \eta_0=0; \qquad \eta_1= 1; \qquad\eta_2\in(1,2), \\
\end{array}
\right.
\end{equation}
where $t_\generic$ is the the time of the event. \\
We have considered the\index{Riemann!tensor} Riemann tensor and analysed its behaviour, in this way it is possible to classify all ``cosmlogical milestones'' as to whether they are polynomial curvature singularities or not: \emph{there are only a very few cases of cosmological milestones that are not polynomial curvature singularities.} In Theorem \ref{poly_curva_sing}, we mention that those cases correspond to sudden\index{singularity!sudden singularity} singularities/extremality\index{singularity!extremality events} events ($\eta_0=0$) and asymptotically Milne big \index{singularity!big bang}bangs/crunches\index{singularity!big crunch}\index{big crunch}. In detail, the only cases where cosmological milestones are not polynomial curvature singularities are:
 \begin{itemize}
\item $\eta_0=0 \qquad \eta_1\geq 2$ or;
\item     $ \eta_0=0  \qquad \eta_1=1  \qquad \eta_2\geq2$ or;
\item    $ \eta_0=1   \qquad \eta_1\geq 3\qquad \text{and} \qquad k=-1, \;  c_0=1 $.
\end{itemize}
However, when looking at time derivatives of the curvature tensor, most of these limited cases disappear:
the \emph{only} two situations in which a cosmological milestone is \emph{not} a derivative curvature singularity are if:
\begin{itemize}

\item $\eta_0=0$,\; $\eta_{i}\in Z^+$; corresponding to an\index{singularity!extremality events} extremality event;
\item $\eta_0=1$, $k=-1$, $c_0=1$, $\eta_{i}\in Z^+$, and  $\eta_1\geq 3$; corresponding to a FRW geometry that smoothly asymptotes near the cosmological milestone to the Riemann-flat  Milne universe. 
\end{itemize}

\item Third, this definition of cosmological milestones in terms of generalized power series enables us to perform a complete model-independent check on the validity or otherwise of the classical energy conditions\index{energy conditions}. 

In particular we provide a complete catalogue of those bangs/crunches\index{singularity!big crunch}\index{big crunch}, \index{singularity!sudden singularity}sudden singularities and \index{singularity!extremality events}extremality events for which the NEC\index{energy conditions!NEC}\index{NEC}, the WEC\index{energy conditions!WEC}\index{WEC}, the SEC\index{energy conditions!SEC}\index{SEC} and the DEC\index{energy conditions!DEC}\index{DEC} are \emph{satisfied}.

Depending on one's attitude towards the energy \index{energy conditions}conditions~\cite{Barcelo:2002bv}, one could use this catalogue as a guide towards deciding on potentially interesting scenarios to investigate. In particular,
\begin{itemize}
\item The NEC, the WEC and the SEC hold independent of the sign of space curvature $k$ all the way down to the singularity, for big\index{singularity!big bang} bangs/crunches\index{singularity!big crunch}\index{big crunch} for which $0<\eta_0<1$. That is, in the vicinity of the singularity the dominant term in the \index{scale factor}scale factor $a(t)$ is bounded from both above and below by
\begin{equation}
c_0 (t-t_\bang)  < a_\mathrm{dominant}(t) \leq c_0.
\end{equation}

\item  ``Big rips" $(\eta_0<0)$ will violate the NEC in the vicinity of the milestone and therefore, they will also violate every other energy conditions.

\item The SEC will be violated for all ``violent'' bangs and\index{singularity!big crunch}\index{big crunch} crunches with $\eta_0>1$, though the NEC and the WEC will still be satisfied in this case.

\item The DEC\index{DEC} holds independent of the sign of space curvature $k$ all the way down to the singularity, for big \index{singularity!big bang}bangs/crunches for which $1/3<\eta_0<1$. That is, in the vicinity of the singularity the dominant term in the \index{scale factor}scale factor $a(t)$ is bounded from both above and below by
\begin{equation}
c_0 (t-t_\bang)< a_\mathrm{dominant}(t) \leq c_0 (t-t_\bang)^{1/3} .
\end{equation}

\item The DEC\index{DEC} is also satisfied for a limited range of \index{singularity!sudden singularity}sudden singularities $(\eta_0=0)$.
Sudden singularities of this type are sufficiently ``gentle'' that at least for some of them even the DEC\index{DEC} can be satisfied all the way to the singularity: a conclusion completely in agreement with Barrow and Tsagas~\cite{Barrow:2004he}.

\end{itemize}

\end{itemize}

Writing the scale factor as a generalized power series expansion has allowed us to derive important and very general results on the various cosmological milestones both on a kinematical and dynamical levels. These theorems have been developed with an absolute minimum of technical imput. Simply by calculating the dominant indicial exponent of the scale factor we have been able to determine whether or not some particular physical observable (expressed as power series) diverges at the cosmological milestone.

\chapter{Gravastars}

 This chapter introduces the reader to the idea of gravastars, how they are configured, as well as some of their properties. We present results deduced from these properties but we do so from an agnostic point of view as to the existence or non-existence of gravastars.

The study and understanding of the features of gravastars are important to comprehend what observational data regarding astrophysical black holes\index{black hole} are telling us.

\section{Definitions}

\subsection{GRAvitational VAcuum STARS}
The concept of black\index{black hole} hole\footnote{See section \ref{black_holes_event_horizon} for more details.} is generally accepted in the general relativity community, but one sometimes encounter some scepticism concerning the reality of the mathematical solution and wariness regarding the interpretation of observational data~\cite{Abramowicz:2002vt}. The Schwarzschild \index{Schwarzschild!black hole} solution presents a central singularity at $r=0$ and an event \index{horizon}horizon at the Schwarzschild\index{Schwarzschild!radius} radius\footnote{See section \ref{Schwarzschild_solution} for more details.} $R_{ Schwarzschild }=2M$. 

The GRAvitational VAcuum STAR (\textit{gravastar})  model is a proposal by Emil Mottola and Pawel Mazur~\cite{Mazur:2001fv, Mazur:2004ku, Mazur:2004fk} to replace black\index{black hole} holes. Instead of having a star contracting and collapsing until matter arrives at a singularity at the centre, the gravastar model suggests that a gravitationally collapsing star would force spacetime itself to undergo a phase transition that would prevent further collapse. Thus the star would be transformed into a spherical ``quantum vacuum" surrounded by a form of super-dense matter. 

It has been speculated that violent creation of a gravastar might be an alternate explanation for gamma ray bursts. Gravastars could also be a solution for the black \index{black hole}hole information paradox\footnote{The black hole information paradox results from the combination of quantum mechanics and general relativity.
In 1975, Stephen Hawking showed that black\index{black hole} holes should slowly radiate away energy, which poses a problem. From the \textit{no hair theorem} one would expect the Hawking radiation to be completely independent of the material entering the black hole\index{black hole}. However, if the material entering the black hole\index{black hole} were a pure quantum state, the transformation of that state into the mixed state of Hawking radiation would destroy information about the original quantum state. This violates the rules of standard quantum mechanics and presents a seeming physical paradox.}. 
The Gravastar is theorized to have very low amounts of entropy, on the contrary, a black hole\index{black hole} apparently has a billion times more entropy than the star it formed from. A problem with the theory regarding the creation of a gravastar is whether or not a star would be capable of shedding enough entropy upon implosion. Externally, a gravastar appears similar to a black hole\index{black hole}: it is visible only by the high-energy emissions it creates while consuming matter. Astronomers observe the sky for X-rays emitted by infalling matter to detect black holes\index{black hole}, and a gravastar would produce a similar signature.



\subsection{Common models}   \label{models}
In the Mazur--Mottola model, the concept of Bose--Einstein \index{Einstein}condensation\footnote{In Bose--Einstein condensate, all matter (protons, neutrons, electrons, etc...) goes into what is called a quantum state creating a ``super-atom". } in gravitational systems is extended to compact object with an interior de Sitter \index{de Sitter}space (with an equation of state $\rho = -p > 0$) , and an outer region of the gravastar consisting of a (relatively thin) finite-thickness shell of stiff matter ($p = \rho$). The exterior of the gravastar is surrounded by a Schwarzschild\index{Schwarzschild!exterior solution} vacuum geometry ($p = \rho = 0$).

In addition to these three layers, the model requires two infinitesimally-thin shells with surface densities $\sigma_{\pm}$, and surface tensions $\vartheta_{\pm}$. These compensate the discontinuities in the pressure profile and stabilize this 5-layer construction, introducing delta-function anisotropic pressures ~\cite{Mazur:2001fv, Mazur:2004ku, Mazur:2004fk}.

\subsection{Problem}
Is it possible to replace the thin shell (from the Mazur--Mottola model) completely with a continuous layer of finite thickness? For physical reasons, we find it useful to minimize the use of thin shells as they really are a mathematical abstraction. 

Is the pressure\index{anisotropy} anisotropy (implicit in the Mazur--Mottola infinitesimally thin shell)  a \textit{necessity} for any gravastar-like objects? Is it really possible to build a gravastar-like objects using \textit{only perfect fluid}\index{perfect fluid} with a continuous layer of finite thickness? Could a \index{horizon}horizon or naked\index{naked singularity} \index{singularity!naked singularity}singularity form and why?

Assuming pressure is continuous and differentiable, we will first analyze the resulting static geometry with the resulting isotropy\index{isotropy} TOV equation (with isotropic pressure) and then the resulting anisotropy\index{anisotropy} TOV equation (anisotropic pressure).

\section{Properties}  \label{arbitrary_gravastar_def}

\subsection{A static spherically symmetric geometry}
To describe the geometry of the gravastar, we use spherical coordinates $(t,r,\theta, \phi)$ and we assume that the geometry is \textit{static, and spherically symmetric}. 
\begin{itemize}
\item A \textit{static} solution means that there exists a hypersurface orthogonal\index{Killing!vector field} Killing vector that is timelike near spatial infinity. This implies that the spacetime metric $ds^2$ can be chosen to be invariant under a time reversal about any origin of time, e.g. all cross terms $dx^t dx^i \; \forall \; i \neq t$ vanish in the chosen coordinate system.

\item A \textit{spherically symmetric} solution means that there exists a privileged point, e.g. the origin, such that the system is invariant under spatial rotations about the origin.
\end{itemize}

The most general static spherically symmetric line element in four dimensions can be written in the canonical form~\cite{Wald:1984rg}:
\begin{equation}
ds^2=-f(r) dt^2+h(r) dr^2+r^2 d\Omega,
\end{equation}
where $d\Omega^2=d\theta^2+\sin^2 \theta d\phi^2$. 

For any spherically symmetric stress-energy tensor, we have:
\begin{equation}
T_{ab}=\rho u_a u_b +p_r n_a n_b+p_t  \left(  g_{ab}+u_a u_b  -n_an_b \right), \label{eq_Tab}
\end{equation}
where $p_r$ is the radial pressure, $p_t$ the tangential pressure, $u^a$ the fluid 4-velocity pointing in the same direction as the static \index{Killing!vector field}Killing vector field $\xi^a$ (in order to be compatible with the static symmetry spacetime) and $n_a$ is a unit radius vector. This property of $u^a$ leads to a relation between the 4-velocity and the function $f$ appearing in the general metric:
\begin{equation}
u^a=-f^{\frac{1}{2}}(dt)^a
\end{equation}
Now, introducing the Einstein's equations\index{Einstein!field equations} developed in section \ref{Einsteins_equations}, we obtain relations between elements of the stress energy tensor and the functions $f$ and $h$ from the metric.
Specifically, the equation in the $tt$ components, using ``geometrized units"\footnote{Geometrized units mean that $G$ and $c$ are set equal to one, see section  \ref{Einsteins_equations} for more details. } and with orthonormal components, involve:
\begin{equation}
G_{\hat{t}\hat{t}}=8\pi T_{\hat{t}\hat{t}} =8 \pi \rho  \label{eq_G_tt}
\end{equation}
Equation (\ref{eq_G_tt}) can be rewritten involving the function $h$:
\begin{equation}
\frac{1}{r^2}\frac{d}{r} \left(   r (1-h^{-1}) \right)=8 \pi \rho.
\end{equation}
Hence, the solution for $h$ can be calculated: 
\begin{equation}
h(r)=\left(    1-\frac{2m(r)}{r}  \right)^{-1},
\end{equation}
where \footnote{Note that in order to avoid a ``conical singularity" in the metric at $r=0$, (in other words if we require the geometry to be regular at $r=0$), then the integration constant $m_0$ in the equation $m(r)=4 \pi \int^r _0 \rho(\tilde{r})\tilde{r}^2d\tilde{r} +m_0$ must be set equal to $0$.}
\begin{equation}
m(r)=4 \pi \int^r _0 \rho(\tilde{r})\tilde{r}^2d\tilde{r}. \label{eq_m} 
\end{equation}
Therefore, the $tt$-field equation yields equation (\ref{eq_m}).
The term $m(r)$ represents the total mass-energy inside the radius $r$ integrated from the centre $r=0$ until arbitrary $r$. Note that if $r=0$ cannot be defined (e.g., in the case of wormhole throats), the integration would be from an arbitrary constant $r_0$ to $r$. Also note that this equation of the mass is precisely
the equation you would expect for a spherically symmetric mass in an Euclidean space.

The $rr$-field equation from the relation (\ref{einstein_field_eq}) of Einstein \index{Einstein!field equations}field equations, in geometrized units and with orthonormal components, gives:
\begin{equation} 
G_{\hat{r}\hat{r}}=8 \pi T_{\hat{r}\hat{r}}=8 \pi p_r.  \label{eq_Grr}
\end{equation}
 Equation (\ref{eq_Grr}) involves the $g_{tt}=-f(r)$ term of the metric, but it is more physically convenient to express $f(r)$ as a function of the quantity $g(r)$ defined below.
 
 \begin{definition}
Let $g(r)$ be related to the $g_{tt}$ component of the metric as follows:
\begin{eqnarray}
g(r)&=&\frac{1}{2}\frac{d}{dr}ln(-g_{tt})\\
      &=& a^r g_{rr}
\end{eqnarray}
where $a^r$ is the radial acceleration of an observer ``at rest", which means that the 4 velocity vector is parallel to the\index{Killing!vector field} Killing field \footnote{In a Killing vector field, $(r,\theta, \phi)$ is constant.}.
\end{definition}
Now, after rearranging this definition, we can write the $g_{tt}$ component of the metric in terms of $g(r)$ :
\begin{equation}
g_{tt}=-\exp \left( 2  \int_r ^\infty g(\tilde{r})d\tilde{r}  \right).
\end{equation}
Now, after further calculations, equation (\ref{eq_Grr}), that is the $G_{rr}$ equation implies that
\begin{equation}
g(r)=\frac{m(r)+4 \pi p_r(r)r^3}{r^2 \left(    1-\frac{2m(r)}{r}\right)}.   \label{eq_g}
\end{equation}
The physical meaning of $g(r)$ is that this function represents the locally measured gravitational acceleration, which is pointing inwards for positive $g(r)$.
Note that equation (\ref{eq_m}) also means that 
\begin{equation}
\frac{dm}{dr}=4 \pi \rho r^2.         \label{eq_dm}
\end{equation}

\begin{definition} \label{def2}
We define the average density $\bar{\rho}=\dfrac{m(r)}{\frac{4 \pi}{3}r^3}$, so that the gravitational acceleration $g(r)$ can be written:
\begin{equation}
g(r)=\frac{4 \pi r}{3}\frac{\bar{\rho} +3p_r(r)}{1-\frac{2m(r)}{r}}.
\end{equation}
\end{definition}

Finally, adopting $(t,r,\theta, \phi)$ coordinates we write any static spherically symmetric geometry, and in particular for gravastars, in the form\footnote{This spacetime geometry is also referred to as an ``interior solution".} \label{interior_sol}
\begin{equation}
ds^2=-\exp  \left( 2  \int_r ^\infty g(\tilde{r})d\tilde{r}  \right) dt^2 +\frac{dr^2}{\left(    1-\frac{2m(r)}{r}  \right)} +r^2 d\Omega^2.
\end{equation}
Of course, the metric\index{metric!Schwarzchild} of the exterior of the gravastar is obviously the Schwarzschild \index{Schwarzschild!exterior solution}exterior solution (\ref{Schwarzschild_metric})\footnote{Remember that the unique vacuum solution to the Einstein's equations\index{Einstein!field equations} is the Schwarzschild exterior solution by Birkhoff stated in Theorem \ref{Birkhoff_th}.}. 

\subsection{A TOV equation}
The remaining field equation $G_{\theta \theta}$ from the Einstein\index{Einstein!field equations} field equations (\ref{Einsteins_equations}), gives an expression for the radial pressure $p_r$, however, the rather messy algebra can be circumvented by the use of the Bianchi identity\index{Bianchi identity} (\ref{Bianchi}) to replace it with the covariant conservation equation of the stress-energy tensor:
\begin{equation}
\nabla _b T^{ab}=0.
\end{equation}
The stress-energy tensor is given in equation (\ref{eq_Tab}), after differentiating, this yields to
\begin{equation}
\frac{d p_r}{d r} = -(\rho+p_r) g + \frac{2\left( p_t - p_r \right)}{r}.  \label{TOV}
\end{equation}
Equation (\ref{TOV}) is known as the anisotropic Tolman--Oppenheimer--Volkoff (TOV) equation. In the case of isotropic pressure ($p=p_r=p_t$) this leads to the more familiar isotropic TOV equation:
  \begin{equation}
\frac{d p}{d r} = -(\rho+p) g,  
\end{equation}
which can be rewritten, using Definition \ref{def2} and the expression of the average density $\bar{\rho}$, as
\begin{equation}
\frac{d p}{d r} =  -\frac{(\rho + p)  (m+4\pi p r^3)}{r^2 \left[ 1-2m(r)/r \right] } 
= -\frac{4\pi\,r}{3} \frac{(\rho+p) (\bar{\rho}+3p)}{1-2m(r)/r}.  \label{TOV_iso}
\end{equation}

\begin{definition}
Let's define the dimensionless \index{anisotropy}anisotropy parameter $\Delta$ for anisotropic pressures by
\begin{equation}
\Delta =\frac{p_t-p_r}{\rho}.
\end{equation}
\end{definition}
Therefore, in terms of this dimensionless parameter, the anisotropic TOV equation becomes:
\begin{equation}
\frac{d p_r}{d r} = -\frac{4\pi r}{3} \frac{(\rho+p_r) (\bar{\rho}+3p_r)}{1-2m(r)/r}+\frac{2 \rho \Delta}{r}. \label{eq_full_TOV}
\end{equation}
The TOV equation shows that pressure contributes as a source of gravity, and one can notice that it follows that  the pressure is higher than in Newtonian stars of the same density profile. It also means that for a given $\rho(r)$, the central pressure $p_c$ required for equilibrium is always higher in general relativity than in Newtonian theory. Therefore, it is harder to maintain equilibrium in general relativity.

\subsubsection{How to solve the TOV equation?}
For full determination of the variables $p_r$, $p_t$, $\rho$, $m$, one needs the TOV equation (\ref{eq_full_TOV}) but also 
\begin{itemize}
\item equation (\ref{eq_dm}) which gives  the derivative of the mass $dm/dr$ as a function of the density $\rho$,
\item an equation of state for the radial pressure of the stellar material 
\begin{equation}
\rho = \rho(p_r)
\end{equation}
\item an equation of state for the tangential pressure of the stellar material
\begin{equation}
\rho = \rho(p_t)
\end{equation}
\item boundary conditions: 
\begin{itemize}
\item  spacetime must be everywhere locally Lorentzian and in particular at the centre of the star, it is therefore necessary that the mass vanishes at the centre\footnote{This condition (locally Lorentz) is actually responsible for leading exactly to equation (\ref{eq_m}), that is $m(r)=\int_0 ^r 4 \pi \tilde{r}^2 \rho d\tilde{r}$.}:
\begin{equation}
m=0 \qquad \text{at} \qquad r=0
\end{equation}
  \item at the star's surface the interior spacetime geometry (\ref{interior_sol}) must join smoothly to the exterior Schwarzschild\index{Schwarzschild!exterior solution} geometry (\ref{Schwarzschild_metric}), therefore
  \begin{equation}
m=M  \qquad \text{at} \qquad  r=R_{surface},
\end{equation}
and by definition,  the radial pressure $p_r$ is zero at the surface of the star. 
\end{itemize}
\end{itemize}

The method to produce a model is straightforward providing that:
\begin{itemize}
\item the equations of state mentioned before are specified,
\item central density $\rho_c$ or central pressure $p_c$\footnote{Spherical symmetry requires that $p_r=p_t=p_c$ at the centre.} for the star are given.
\end{itemize}
One can integrate the coupled hydrostatic equilibrium equation (\ref{eq_full_TOV}) and the mass equation  (\ref{eq_dm}) outward from the centre, beginning with the initial conditions $m = 0$ and $p_r=p_c$ at the center. The integration terminates when the pressure falls to zero\footnote{Note that for gravastars, theoretically, the pressure starts off negative, passes through a first zero, increases to a maximum pressure before falling down to zero at the surface.} at the surface of the star. The value of the mass at this surface radius represents the total mass-energy $M$ that appear in the \index{Schwarzschild!exterior solution}Schwarzschild exterior solution (\ref{Schwarzschild_metric}).

\subsubsection{Qualitative differences introduced by $\Delta$.} 
Let's have a qualitative look at the difference that \index{anisotropy}anisotropy introduces in the TOV equation with dimensionless anisotropy\index{anisotropy} parameter $\Delta$. Remember that if we assume the WEC\index{energy conditions!WEC}\index{WEC}\footnote{See section \ref{NEC}.} holds, then $\rho \geqslant 0$ everywhere in the star.
\begin{enumerate}
\item \underline{case $\Delta > 0$:} (e.g. $p_r< p_t$)

 Typically, this condition holds for rigid solids, in telluric planets like the earth for example. If the radial pressure is negative, and equivalently pushing inwards, the tangential pressure, created by the rigidity of the surface, can be responsible for preventing the object from collapsing. In other words, the tangential pressure is doing more to stop gravitational collapse than the radial pressure.
 
\item  \underline{case $\Delta < 0$, and $0<p_t<p_r$:} 

This particular case occurs in fluids with weak surface tension and where gravity plays an important role to hold the fluid together. More importantly, the radial pressure must be greater at the centre than it would be in a completely isotropic object as the pressure profile is steeper.

\item \underline{special case: $p_t<0$:}

 Objects satisfying this condition are fluids with strong surface tension, such as raindrops for example. The radial pressure must also be greater at the centre than it would be in a completely isotropic object as the pressure profile is steeper. However, in this case, the radial pressure does not necessarily go to zero smoothly at the surface.
\end{enumerate}

\subsection{Specific key features}
In this section, we present the class of spacetime geometries that we are particularly interested in for the gravastar model.\\

\begin{figure}[htb]
\begin{center}
\input{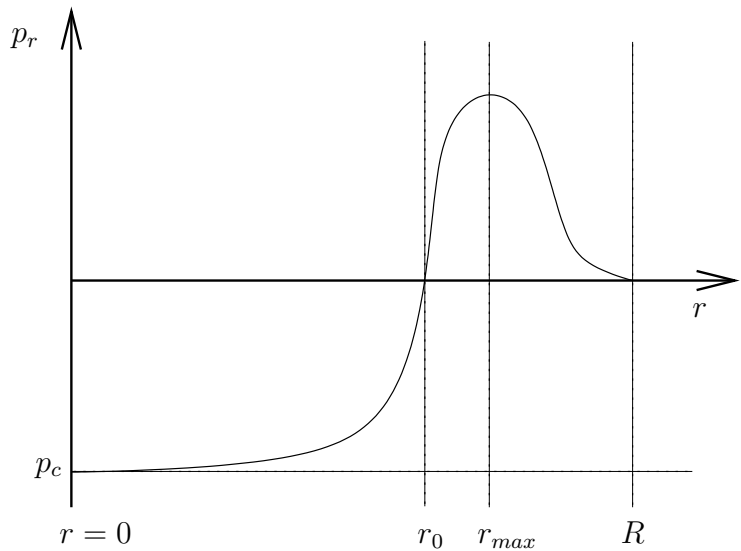}
\end{center}
\caption{\label{F:pressure}
Qualitative sketch of radial pressure as a function of $r$ for a gravastar .}
\label{fig1}
\end{figure}

In the same spirit of Mazur and Mottola ~\cite{Mazur:2001fv, Mazur:2004ku, Mazur:2004fk}, and Laughlin \textit{et al} ~\cite{Chapline:2000en,Chapline:2002}, we assume that the weak energy condition \index{energy conditions}holds through out the configuration (i.e. the density is positive), but we permit the pressure to become negative in the gravastar interior. The density does not need to be continuous\footnote{Typically, the density is not continuous at the surface of the gravastar.} but to avoid infinitesimally thin shells (purpose of this model), one must demand that the radial pressure $p_r$ is continuous. The radial pressure is drawn qualitatively in Figure \ref{fig1}.\\

To ``smooth out''  the infinitesimally thin shells of the Mazur and Mottola gravastar model, we consider static spherically symmetric geometries such that:
\begin{itemize}
\item Inside the gravastar, $ r < R$, the density is everywhere positive and finite.

\item The  central pressure is negative, $p_c<0$, and in fact $p_c = - \rho_c$.
\\
(We do \emph{not} demand $\rho=-p_r=-p_t$ except at the centre \footnote{Remember that because of the spherical symmetry $p_r=p_t$ at the centre.} .)

Positive energy density but negative pressure, and origin free of any mass singularity, are the characteristic features of \textit{de Sitter} \index{de Sitter}space.
The idea of replacing the Schwarzschild singularity with de Sitter\index{de Sitter} vacuum goes back at least to 1965. 
Gliner~\cite{Gliner:1966} interpreted $p = -\rho$ as corresponding to a ``vacuum" and suggested that it could be a final state in a gravitational collapse.


 Indeed for the the specific metric
\begin{equation}
ds^2=-c^2\gamma(r)dt^2+\gamma^{-1}(r) dr^2 +r^2d\Omega^2,
\end{equation}
and the choice of cosmological-constant-like matter leads to
\begin{equation}
R_{ab}-\frac{1}{2}g_{ab}R=\left\{ \begin{array}{cc}  \frac{3}{r_s^2}g_{ab}& r<r_s \;  ; \\0 & r>r_s \;  ; \end{array}\right.
\end{equation}
where $r_s=\dfrac{2GM}{c^2}$, with $M$ the total mass of the object, and with
\begin{equation}
\gamma(r)=\left\{ \begin{array}{cc}  1-\left( \dfrac{r}{r_s} \right)^2 & r<r_s  \; ; \\ 1- \dfrac{r_s}{r} &  r > r_s \;  ; \end{array}\right.
\end{equation}
Note that the interior de Sitter \index{de Sitter}region may be also interpreted as a cosmological spacetime, with the horizon\index{horizon} of the expanding universe replaced by a quantum phase interface. Also note the unpleasantness at $r=r_s$: Laughlin \emph{et al} call this a ``phase transition"; Mazur--Mottola seek to smooth out the unpleasant region.

Hence, the characteristics of the gravastar \textit{at the centre}, are those of a de-Sitter \index{de Sitter}spacetime geometry.

\item The spacetime is assumed to \emph{not} possess an event \index{horizon}horizon. \\
This implies that $\forall r$ we have $2m(r)<r$.

\end{itemize}
These three features, positive density, negative central pressure, and the absence of \index{horizon}horizons, are the three most important features characterizing a gravastar.

 Other important features are:
\begin{itemize}
\item To keep the centre of the spacetime regular, we enforce both $p_r'(0)=0$ 
and $p_c = p_r(0)=p_t(0)$.

\item There should be a pressure maximum in the general vicinity of the\index{Schwarzschild!radius} Schwarzschild radius\footnote{See section \ref{Schwarzschild_solution}}, $r_\mathrm{max}\approx~R_\mathrm{Schwarzschild}$, satisfying  $p_r(r_\mathrm{max})>0$, and $p_r'(r_\mathrm{max})=0$. \\
(This permits the physics in the region $r\gg r_\mathrm{max}$ to be more or less standard.)

\item There should be exactly two radii where the radial pressure vanishes:
\begin{itemize}
\item  The first pressure zero $p_r(r_0)=0$, where $p_r'(r_0) > 0$, and 
\item
the second pressure zero $p_r(R)=0$, where $p_r'(R)\leq 0$. \\
The point $R$,  (which by construction must satisfy $R>r_\mathrm{max}>r_0$), is called the surface of the gravastar.
\end{itemize}

\item The pressure profile $p_r(r)$ should be continuous.\\
 (In contrast, it is sometimes useful to  allow $p_t(r)$ to be discontinuous.)
 
\item The strong energy condition \index{energy conditions!SEC}\index{SEC}[SEC;  $\rho+p_r+2p_t\geq 0$] is definitely violated, at least near the centre of the gravastar.

At the centre of the star, $p_c=p_r=p_t=-\rho$ with $\rho>0$, therefore
\begin{equation}
\rho+p_r+2p_t=\rho -\rho-2\rho= -2\rho<0 .
\end{equation}

\item We choose to enforce the null energy condition\index{energy conditions!NEC}\index{NEC} [NEC; $\rho+p_i\geq 0$] throughout the gravastar. In view of our first comment that density is everywhere positive, this implies that we are enforcing the  weak energy condition\index{energy conditions!WEC}\index{WEC} [WEC; $\rho+p_i\geq 0$ and $\rho\geq 0$]. 

\item We impose no restriction regarding the dominant energy condition\index{energy conditions!DEC}\index{DEC} [DEC; $\rho\geq0$ and $|p_i|\leq \rho$], and in fact we shall see that the DEC\index{energy conditions!DEC}\index{DEC} must fail in parts of the gravastar that are sufficiently ``close'' to forming a horizon\index{horizon}.
\end{itemize}

\subsubsection{Why does the radial pressure goes through two zeros?}
It is clear on Figure \ref{fig1} that our model describes a radial pressure that starts negative at the centre, then increases till $p_r=0$, becomes positive, and starts decreasing after a maximum $p_r=p_\mathrm{max}$ at $r=r_\mathrm{max}$ eventually reaching a second zero at the surface.

By definition, when the pressure is zero it corresponds to the surface of the star so we could have  smoothly joined the vacuum interior to an exterior \index{Schwarzschild!exterior solution}Schwarzschild solution (\ref{Schwarzschild_metric}). 
 The reason for not doing so is a purely pragmatic one based on the fate of infalling positive-pressure matter. If we start without a positive-pressure region of type $(r_0,R)$, then any infalling positive-pressure  matter that accumulates above $r_0$ will automatically generate a positive-pressure region of type $(r_0,R)$. The only way to avoid a positive-pressure region of type $(r_0,R)$ is if the negative-pressure matter in the $(0,r_0)$ region immediately catalyzes any infalling positive-pressure matter into negative-pressure matter. In this entire chapter, we will keep the negative-pressure matter deep in the core, discretely hidden behind a layer of positive-pressure matter.
 
\subsubsection{How does our model relate to others?}
We have highlighted specific features that a \index{perfect fluid!sphere}perfect fluid sphere gravastar, with continuous pressure, should have. But how does our gravastar model relate to other features in previous models?
\begin{itemize}
\item In the Mazur--Mottola model , the infinitesimally thin shell interface is the limiting case where $r_\mathrm{max}\to r_0$, while $\rho=-p_r=-p_t$ is strictly enforced for $r<r_0$. Also, an additional thin shell is placed at the surface so that the radial pressure is positive on the left limit of the surface radius $p_r(R^-)>0$. 

\item A simplified variant of the Mazur--Mottola 3-region model is also considered in ~\cite{Visser:2003ge} but the surface radius tends to $r_\mathrm{max}$, e.g. $R \to r_\mathrm{max}$. So in this model  $r_0=r_\mathrm{max}=R$ and there is a single thin shell at $r_\mathrm{max}$ with de Sitter \index{de Sitter}geometry inside and Schwarzschild geometry outside.

\item In the Laughlin \emph{et al}  model, $r_0\to R_\mathrm{Schwarzschild}$ from below, while $R\to R_\mathrm{Schwarzschild}$ from above. And in the region  $r<R_\mathrm{Schwarzschild}$, the equality  $\rho=-p_r=-p_t$ remains all the way through. There is implicitly a singular infinitesimally thin shell located exactly at $R_\mathrm{Schwarzschild}$ with infinite surface tension.
                 
 \item The Gliner ~\cite{Gliner:2002yh} and Dymnikova~\cite{ Dymnikova:1992ux, Dymnikova:2001fb,Dymnikova:2003vt,  Dymnikova:2004qg, Dymnikova:1999cz} proposals all satisfy the constraint  $\rho=-p_r$ everywhere throughout the configuration. We assume $\rho=-p_r$ only at the centre, this equation does not need to hold everywhere.
\end{itemize}

\section{The problem of the negative-pressure \textit{perfect fluid} sphere gravastar.}  \label{no_isotropy}\index{perfect fluid!sphere}
 In this section, we will only consider perfect fluids, and therefore the pressure is now isotropic $p=p_r=p_t$, which means that  the dimensionless anisotropy \index{anisotropy}parameter is null throughout this whole section $\Delta=0$. The TOV equation is described by equation (\ref{TOV_iso}) in this section.
 
\subsection{Four inconsistencies}  \label{inconsistencies}
 In this section, we will have a look at four cases where some inconsistency within the isotropic TOV equation arise for perfect fluid sphere\index{perfect fluid!sphere} gravastars.
 
 \subsubsection{First inconsistency}
 \begin{theorem}
  The isotropic TOV equation {\bf cannot} hold at the point $r_0$ where the radial pressure is zero $p(r_0)=0$.
 \end{theorem}
 
 \begin{proof}
By assumption, the LHS of the TOV isotropic equation, $dp/dr$, is positive. As shown on Figure (\ref{fig1}), the pressure is increasing in the region $r<r_0$.  However, we can calculate the exact value of the pressure gradient at $r=r_0$ where $p=0$:
\begin{equation}
\left. \frac{dp}{dr} \right|_{r=r_0} =- \frac{4\pi r_0}{3} \frac{\rho\,\bar{\rho}}{1-2m/r_0}. \label{TOV_r0}
\end{equation}
The density is positive (as we assume the WEC\index{energy conditions!WEC}\index{WEC}) and therefore the average density $\bar{\rho}$ is positive as well. Furthermore $r_0< R_{\mathrm{Schwarzschild}}$, which implies that $2m/r_0<1$.
As a consequence, it follows that the RHS of equation (\ref{TOV_r0}) is negative.

Therefore, there is an inconsistency, the LHS cannot be positive while the RHS is negative. Thus, the isotropic TOV \textit{cannot} hold at the point $r_0$.
 \end{proof}
 
 \subsubsection{Second inconsistency}
  \begin{theorem}
  The isotropic TOV equation {\bf cannot} hold at the point $r_\mathrm{max}$ where the radial pressure is positive and reaches a maximum $p(r_\mathrm{max})>0$.
 \end{theorem}
 
  \begin{proof}
 Now let's consider the point of maximum radial pressure $r_\mathrm{max}$. We can calculate the value of the TOV equation at this point:
 \begin{equation}
\left. \frac{\d p}{\d r}\right|_{r=r_\mathrm{max}}  =
 - \frac{4\pi r_\mathrm{max}}{3} \;
   \frac{(\rho+p)(\bar{\rho}+3p)}{1-2m/r_\mathrm{max}}.
\end{equation}
At $r_\mathrm{max}$ the pressure is a maximum and positive for the class of models we consider, therefore $p(r_\mathrm{max})>0$ and its derivative is zero $\left. \d p/ \d r\right|_{r=r_\mathrm{max}}  =0$. But the density is also positive everywhere $\rho>0$ by assumption, therefore the LHS is zero while the RHS is negative.

This is another inconsistency, the isotropic TOV \textit{cannot} hold at the point $r_\mathrm{max}$. 
  \end{proof}
  
  \subsubsection{Third inconsistency}
   \begin{theorem}
  The isotropic TOV equation {\bf cannot} hold in the region $r_0<r<r_\mathrm{max}$ where the radial pressure is positive and increasing.
 \end{theorem}
 
   \begin{proof}
  Now if we look at the region between the first and second inconsistency, $r_0<r<r_\mathrm{max}$, we can notice that the object under investigation has an increasing positive radial pressure, therefore the LHS of the TOV equation is positive in this entire region. However, because we assume $\rho>0$ which also implies that $\bar{\rho}>0$, and that there is no horizon\index{horizon} for at least $r<r_\mathrm{max}$, the RHS of the TOV equation is negative for this entire region.
\begin{equation}
 - \frac{4\pi r}{3} \;
   \frac{(\rho+p)(\bar{\rho}+3p)}{1-2m/r}<0
\end{equation}  
  
  This is the third inconsistency and it follows that isotropic pressure in that interval $r_0\leqslant r \leqslant r_\mathrm{max}$ is not able to satisfy the TOV equation, and therefore, we conclude that a static spacetime geometry can only be obtained with the introduction of tangential pressures, that is, anisotropic pressures.
    \end{proof}
    
  \subsubsection{Fourth inconsistency}

 \begin{definition} \label{def_rg}
 Let's define $r_g$ the location where the quantity $\bar{\rho} +3p$ in the isotropic TOV equation (\ref{TOV_iso}) changes sign.
 \end{definition}
  
  \begin{theorem}
  The isotropic TOV equation {\bf cannot} hold in the region $r_g<r<r_0$ where the radial pressure is negative and increasing and with $\left (\bar{\rho} +3p \right)_{r_g}$.
 \end{theorem}
  
   \begin{proof}
  The same arguments hold for a larger interval below $r_0$ and into the negative pressure region.
  
  Assuming the WEC\index{energy conditions!WEC}\index{WEC} through the entire configuration implies that NEC\index{energy conditions!NEC}\index{NEC}\footnote{The WEC implies the NEC, see sections \ref{NEC} and \ref{WEC}} holds, from which follows that 
  \begin{equation}
\rho+p \geqslant 0  \qquad  \forall \; r:  \; r < r_0.
\end{equation}
But remember that the LHS of the TOV equation is positive, therefore
\begin{equation}
 - \frac{4\pi r}{3} \;
   \frac{(\rho+p)(\bar{\rho}+3p)}{1-2m/r}<0  \qquad  \Longrightarrow  \qquad \bar{\rho} +3p<0  \qquad \forall \; r < r_0
\end{equation}  
Indeed, at the centre of the gravastar, we have
\begin{equation}
\left (\bar{\rho} +3p \right)_c = \rho_c+3 p_c=-2\rho_c <0.
\end{equation}
However, exactly at the point $r_0$, where the pressure is zero, we have
\begin{equation}
\left (\bar{\rho} +3p \right)_{r_0} =\bar{\rho}_{r_0}>0. 
\end{equation}
Therefore, it means that the term $\bar{\rho} +3p$ changes sign somewhere in the interval $0<r<r_0$.

 In the region $r_g<r<r_0$, we have both:
 \begin{equation}
 \bar{\rho} +3p >0 \; \;  \text{and}  \; \; \frac{\d p}{\d r} \geqslant 0
 \end{equation}      
 This is another inconsistency.  
 Therefore, the pressure isotropy\index{isotropy} fails for the entire region $r_g<r<r_0$ and quite possibly fails for an even larger region.
     \end{proof}

 \subsubsection{Qualitative definitions}
 The following definitions relate the qualitative physics to the different regions of the gravastar described in the previous sections.
 \begin{definition}
The region $0<r<r_g$, where the physics is qualitatively similar to that  of de Sitter\index{de Sitter} space, will be referred as the {\bf``core"}. In the core, the local acceleration of equation (\ref{eq_g}) due to gravity is outward.
  \end{definition}

 \begin{definition}
 The region $r_g<r<r_{max}$, where the physics is still definitely ``unusual", will be referred as the {\bf ``crust"}. In the crust, the local acceleration due to gravity is inward, but the pressure still rises as one moves outward.
  \end{definition}
 
 \begin{definition}
 The region $r_{max}<r<R$, where the physics is still definitely ``normal", will be referred as the {\bf ``atmosphere"}. In the atmosphere, the local acceleration due to gravity is inward, and the pressure decreases as one moves outward.
  \end{definition}

\begin{figure}[htb]
\begin{center}
\input{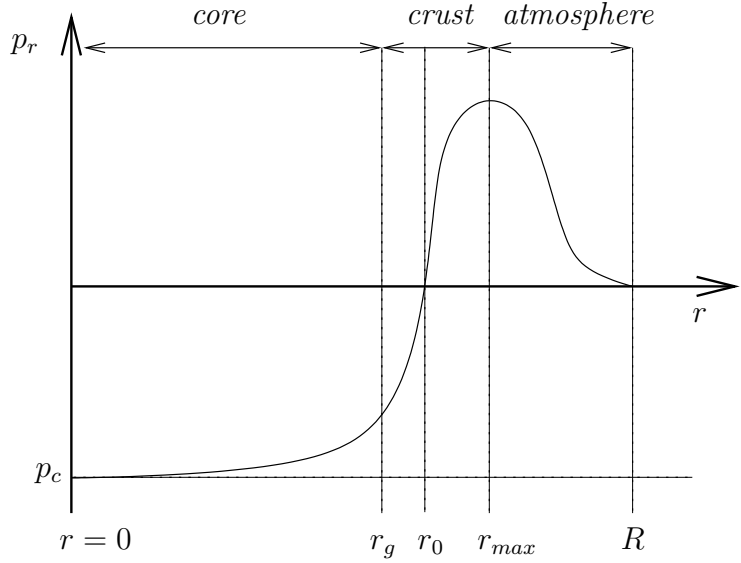}
\end{center}
\caption[Qualitative sketch of  gravastar: ``core'', ``crust'', and  ``atmosphere''.]
{\label{fig2}
Qualitative sketch of  gravastar labelling the ``core'', ``crust'', and ``atmosphere''.}
\end{figure}

Figure (\ref{fig2}) presents a qualitative sketch of the previous defintions.

  With these definitions, we see that pressure is guaranteed to be anisotropic throughout the ``crust''. Note that even if we were to dispense with the entire positive-pressure region by chopping the gravastar off at $r_0$, there is still an anisotropic crust in the region $(r_g,r_0]$.\\

In conclusion, we have shown that there are inconsistencies within the isotropic TOV equation from which follows that a static spherically symmetric object with positive density, negative central pressure and vanishing pressure at the surface cannot be supported by isotropic pressures alone.

There are \textit{no perfect fluid} gravastars.

\subsection[Case study: Schwarzschild interior and Buchdahl-Bondi bound]{A particular gravastar case study: the Schwarzschild\index{Schwarzschild!interior solution} interior solution and the Buchdahl-Bondi bound}   
\label{case_study1}
The Schwarzschild interior solution illustrates what can go wrong when trying to build a perfect fluid gravastar. 

  The Schwarzschild solution in standard coordinates $(t,r,\theta,\phi)$ is given by equation (\ref{Schwarzschild_metric}). But the general Schwarzschild interior solution for values of $r$ smaller than the Schwarzschild radius $R_{\mathrm{Schwarzschild}}$ is given by:

\begin{equation}
ds^2=-  \left(  A +B \sqrt{1-\frac{2m_*(r)}{r}} \right)^2 dt^2 + \left(  \frac{1}{1-\frac{2m_*(r)}{r}} \right) dr^2+r^2 d\Omega^2   \label{general_interior}
\end{equation}
where $d\Omega^2=d\theta^2+\sin^2 \theta d\phi^2$ and $m_*(r) = M (r/R)^3$.
This general solution describes static spherically symmetric stars with constant density independent of $r$ from the centre to the surface of the star.

Using boundary conditions, we can determine $A$ and $B$ as functions of the surface radius $R$ and the total mass $M$ of the star. 
The density can be calculated with the $G_{tt}$ term of the Einstein tensor\index{Einstein!tensor} and  $g_{tt}$ part of the metric (\ref{general_interior}) due to the field equations (\ref{einstein_field_eq}):
\begin{eqnarray}
8 \pi \rho_* g_{tt}   & = & G_{tt}\\
\Longrightarrow \qquad \rho_*  & = & \frac{3M}{4 \pi R^3}.
\end{eqnarray}
 The value of the density is constant and depends only on $M$ and $R$ that are fixed for a specific star.

The radial pressure can also be calculated exactly with the field equations in the $rr$ components as a function of $A$ and $B$ from the general \index{metric}metric. However, the expression of $p^*$ is quite complicated and messy, but the important feature is, that the surface radius $R$ can be deduced as a function of $A$ and $B$ from the boundary condition:
\begin{equation}
\text{at}\; R_{\mathrm{surface}} \qquad p^*(R)=0 \qquad \Longrightarrow \qquad f_1(A,B,R,M)=0.
\end{equation}
Furthermore, the total mass of the star is:
\begin{equation}
M=4 \pi \int _{r=0}^{r=R(A,B,R,M)}  \rho^* r^2 dr \qquad \Longrightarrow  \qquad f_2 (A,B,R,M)=0.
\end{equation}
Therefore, solving the following system (\ref{system}) of two equations for $A$ and $B$
\begin{equation}
\left\{\begin{array}{c}     
f_1 (A,B,R,M)=0     \\ 
f_2 (A,B,R,M)=0     
     \end{array}\right.
     \label{system}
\end{equation}
 leads to $A(R,M)$ and $B(R,M)$ which determines the metric of the\index{Schwarzschild!interior solution} Schwarzschild interior solution entirely as a function of the total mass $M$ and the surface radius $R$.
Indeed, it can be found that:
\begin{equation}
\left\{\begin{array}{ccc}     
 A &=&   \sqrt{1-\dfrac{2M}{R}}  \\ 
B  &= &    -\dfrac{1}{3}    .
     \end{array}\right.
\end{equation}

In this case the metric (\ref{general_interior}) becomes:
\begin{equation}
ds^2=-  \left(  \sqrt{1-\frac{2M}{R}}-\frac{1}{3} \sqrt{1-\frac{2m_*(r)}{r}} \right)^2 dt^2 + \left(  \frac{1}{1-\frac{2m_*(r)}{r}} \right) dr^2+r^2 d\Omega^2      \label{interior}
\end{equation}
where $m_*(r) = M (r/R)^3$.

 In 1916, Schwarzschild \index{Schwarzschild}integrated exactly by hand the isotropic equation (\ref{TOV_iso}) yeilding
  \begin{equation}
p_*(r) = \rho_*\;
 \frac{\sqrt{1-2m_*(r)/r}-\sqrt{1-2M/R}}{3\sqrt{1-2M/R}-\sqrt{1-2m_*(r)/r}}   \label{p*_profile}
\end{equation} 
where $M$ and $R$ are the total mass and surface radius. Note that the denominator of the pressure $p_*$ is exactly $3 g_{tt}$ from the $tt$-component of the metric (\ref{interior}). 

\subsubsection{Central pressure $p_c^*$ and the Buchdahl-Bondi bound}
The central pressure required for equilibrium of a uniform density star is
\begin{equation}
p_c^* = \rho_* \frac{1-\sqrt{1-2M/R}}{3\sqrt{1-2M/R}-1}   \label{eq_pc}
\end{equation}

For $R\gg M$, equation (\ref{eq_pc}) reduces to the Newtonian value (\ref{eq_pc_newton})\footnote{Note that the isotropic TOV equation in a Newtonian star is
\begin{equation}
\frac{d p}{d r} =  -g(r)\rho(r) =   -\frac{m(r)}{r^2 } \rho(r) 
= \frac{4}{3}\pi r \bar{\rho}.
\end{equation}}  \footnote{Value given in~\cite{Wald:1984rg}} :
\begin{equation}
  p_c^* =\left(   \frac{\pi}{6}  \right)^{\frac{1}{3}}M^{\frac{2}{3}} \rho_*^{\frac{4}{3}}           \label{eq_pc_newton}
\end{equation}

The sign of the central pressure $p_c^*$ is determined by specific conditions only on the denominator as the numerator is always positive:
\begin{equation}
1-\sqrt{1-2M/R}>0   \qquad \Longleftrightarrow \qquad 0> - \frac{2M}{R} \qquad \text{(always true)}.
\end{equation}

The central pressure $p_c^*$ in the gravastar becomes infinite when
\begin{equation}
3\sqrt{1-2M/R}=1       \qquad  \Longrightarrow \qquad \frac{2M}{R}=\frac{8}{9},
\end{equation}
and therefore, $p_c^*>0$ when 
\begin{equation}
3\sqrt{1-2M/R}>1       \qquad \Longleftrightarrow \qquad \frac{2M}{R}<\frac{8}{9}.
\end{equation}
This result on the compactness, $\chi=2M/R$, is known as the Buchdahl-Bondi bound: the compactness of a static spherically symmetric fluid $\chi$ is bounded above by $8/9$. This is quite significant as it shows that a star cannot get arbitrarily close to forming a black\index{black hole} hole $(\chi =1)$.

However, in our particular gravastar model, we deal with negative central pressures $p_c^*<0$, and this implies that we have the relation:
\begin{equation}
 p_c^*<0      \qquad \Longleftrightarrow \qquad \frac{8}{9}<\frac{2M}{R}<1.   \label{bound}
\end{equation}

Of course, while this is a perfectly sensible solution in the \textit{mathematical sense} one would generally rule it out physically  because of the negative central pressure. If we leave our prejudice against negative pressure aside, adopting the gravastar philosophy, we will find some interesting results coming from the choice of (\ref{bound}).
  
\subsubsection{First order pole}  
  
 \begin{theorem} \label{theo_pole}
 For a Schwarzschild \index{Schwarzschild!interior solution}interior solution, with constant positive density $\rho_*>0$, and a negative central pressure $p_c^*<0$, the pressure profile given by equation (\ref{p*_profile}) will have a first order pole at
 \begin{equation}
r_\mathrm{pole}=3 R \sqrt{1-\frac{8/9}{2M/R}}    \qquad \text{and furthermore}  \qquad r_\mathrm{pole}<R.
\end{equation}
 \end{theorem} 
  Note that Theorem \ref{theo_pole} assumes constant positive density $\rho_*>0$ and a central negative pressure $p_c^*<0$, but it does not assume that at the centre $p_c^*=-\rho_*$. As a matter of fact, the only way to obtain exactly $p_c^*=-\rho_*$ at the centre is if $R=2M$ that is if there is a Schwarzschild radius. Indeed, $p_c^*=-\rho_*$ if and only if:
  \begin{equation}
\frac{p_c^*}{\rho_*}= \frac{1-\sqrt{1-2M/R}}{3\sqrt{1-2M/R}-1}=-1,
\end{equation}
that is, if,
\begin{equation}
2\sqrt{1-2M/R}=0,
\end{equation}
finally, if
\begin{equation}
R=2M.
\end{equation}

  \begin{proof}
  From the pressure profile given in equation (\ref{p*_profile}), there is a pole when its denominator cancels out or, in this case, when the $tt$-component of the metric (\ref{interior}) is zero. 
  \begin{equation}
 g_{tt}=0 \qquad \Longrightarrow \qquad  3\sqrt{1-2\frac{M}{R}}-\sqrt{1-2m_*(r)/r}=0,
\end{equation}
where $m_*(r)=M (r/R)^3$. We can extract the radius $r$ from this relation and obtain a value for $r$ at the pole.
\begin{eqnarray}
\sqrt{1-2\frac{Mr^2}{R^3}}& = & 3\sqrt{1-2\frac{M}{R}}\\
 \Longrightarrow \; \; \;          1-2\frac{Mr^2}{R^3} & = &  9 \left(  1-2\frac{M}{R} \right)   \\
  \Longrightarrow \; \; \;  \; \;  \;     -2\frac{Mr^2}{R^3} & = &  \left(  8-9\times 2\frac{M}{R} \right)   \\
   \Longrightarrow \; \; \;  \; \; \;  \; \;  \;    \; \;  \;    \; \;  \;     r^2 &=& 9 R^2  \left(  1-\frac{8/9}{2M/R} \right)   \\
   \Longrightarrow \; \; \;    \; \;  \;    \; \;  \;    \; \;      r_\mathrm{pole} &=&3R\sqrt{1-\frac{8/9}{2M/R}}
\end{eqnarray}
Now we can see what the conditions to have the upper bound $R$ on $r_\mathrm{pole}$ are:
\begin{eqnarray}
3R\sqrt{1-\frac{8/9}{2M/R}} &< & R \\
 \Longrightarrow \; \; \;  9R^2 \left( 1-\frac{8/9}{2M/R} \right)  &< & R^2\\
\Longrightarrow \; \; \;    \; \;  \;    \; \;  \;   \left( 8/9-\frac{8/9}{2M/R} \right)  &< & 0\\
\Longrightarrow \; \; \; \; \; \;    \; \;  \;    \; \;  \; \; \; \;    \; \;  \;    \; \;  \;  \dfrac{2M}{R}  & < & 1  \; \; \; \;  \text{true for no event horizon}\index{horizon}
\end{eqnarray}
Therefore, for a Schwarzschild \index{Schwarzschild!interior solution}interior solution, with negative central pressure and constant positive density, that has no event horizon, there exists a pole such as 
\begin{equation}
 r_\mathrm{pole}  <R,
\end{equation}
where $R$ is the surface.
  \end{proof}
  
\begin{figure}[htb]
\begin{center}
\input{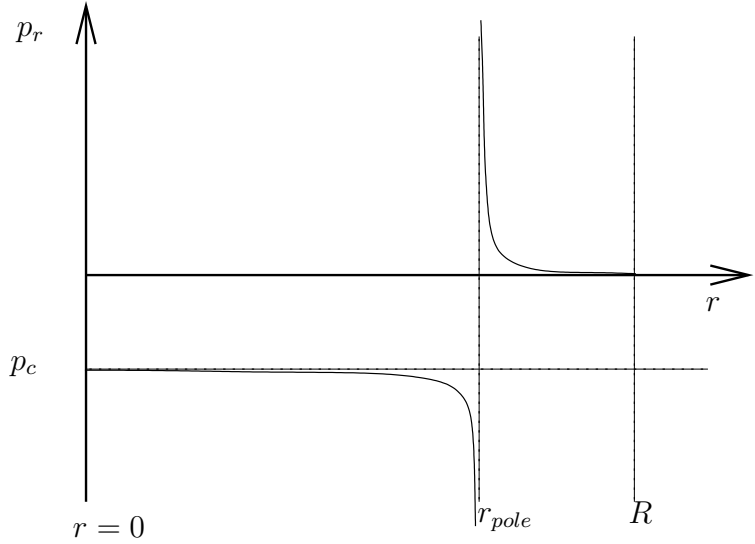}
\end{center}
\caption[Qualitative sketch of pressure pole in the interior 
Schwarzschild solution.]{\label{fig3}
Qualitative sketch of pressure pole in the interior 
Schwarzschild solution for $2M/R > 8/9$.}
\end{figure}

  Note that as $2M/R$ goes from 8/9 to 1, the position of this pole moves from the centre of the star to the surface of the star.
The situation is qualitatively sketched in figure \ref{fig3}.

\begin{theorem} 
 The NEC\index{energy conditions!NEC}\index{NEC} must be violated sufficiently close to the pole at $r_\mathrm{pole}=3 R \sqrt{1-\frac{8/9}{2M/R}} $.
\end{theorem}

\begin{proof}
The compactness is bounded $8/9<2M/R<1$ which implies that the denominator term in equation (\ref{p*_profile}) is negative, and therefore, that the pressure $p^*$ tends to $-\infty$ as $r$ approaches $r_\mathrm{pole}$.

The NEC\index{energy conditions!NEC}\index{NEC} holds, in the vicinity of the pole, if and only if:
\begin{equation}
\rho_* +p^* \geqslant 0.
\end{equation}
However, near the pole,
\begin{equation}
\lim_{r \longrightarrow r_\mathrm{pole}} \left( \rho_* +p^* \right) =-\infty.
\end{equation}
 Therefore, the NEC\index{energy conditions!NEC}\index{NEC} must be violated sufficiently close to the pole.
 \end{proof}

\subsubsection{Curvature singularity}\index{singularity!curvature singularity}
We are going to see now that \index{Schwarzschild!interior solution}Schwarzschild interior solution gravastars are clearly unphysical, even if one is willing to accept negative pressures and even violations of the NEC. It is unphysical because the pressure pole implies a curvature singularity.
With the metric given by equation (\ref{interior}), we can calculate the \index{Riemann!tensor}Riemann tensor in an orthonormal basis $R_{\hat{a} \hat{b}\hat{c}\hat{d}}$. 
In particular, for isotropic pressures, we know that,
\begin{eqnarray}
R_{\hat t \hat  r \hat t \hat r } &= &-\dfrac{R_{trtr}}{g_{tt} g_{rr}} = \dfrac{4 \pi}{3} \left( \rho+3p \right) \\
R_{\hat t \hat  \theta \hat t \hat \theta } & = &-\dfrac{R_{t \theta t \theta}}{g_{tt} g_{\theta \theta}} = \dfrac{4 \pi}{3} \left( \rho+3p \right) \\ 
R_{\hat t \hat  \phi \hat t \hat \phi }& = & -\dfrac{R_{t\phi t \phi }}{g_{tt} g_{\phi \phi}} = \dfrac{4 \pi}{3} \left( \rho+3p \right).
\end{eqnarray}

Now, we know that those specific terms of the\index{Riemann} Riemann tensor will have a problem at $r=r_{\mathrm{pole}}$.
As shown previously for the pressure pole, the term $g_{tt}$ is zero at $r=r_{\mathrm{pole}}$ and 
\begin{equation}
g_{tt}= O \left(  \left[  r-r_{\mathrm{pole}}  \right]^2  \right),
\end{equation}
whereas $g_{rr}$ is finite at $r=r_{\mathrm{pole}}$ and
\begin{equation}
g_{rr}(r_{\mathrm{pole}}) =\dfrac{1}{9 \left(  1-2M/R \right)} > 0.
\end{equation}
Concerning the $trtr$-term of the Riemann tensor, we have
\begin{eqnarray}
\lim_{r \rightarrow r_{\mathrm{pole}} } R_{trtr}  & = &0, \\
\lim_{r \rightarrow r_{\mathrm{pole}} } \dfrac{d R_{trtr}}{dr}  & = & \frac{3}{81R^3} \left( \frac{2M}{R}\right)^2\frac{\left( 1-\frac{8/9}{2M/R}\right)^{1/2}}{\left( 1-2M/R\right)}
\end{eqnarray}
Hence, we have for the $trtr$-term of the\index{Riemann!tensor} Riemann tensor at the pole:
\begin{equation}
\lim_{r \rightarrow r_{\mathrm{pole}} } R_{trtr} =  O \left(  \left[  r-r_{\mathrm{pole}}  \right]  \right).
\end{equation}

Therefore, the $\hat t \hat  r \hat t \hat r $-term of the Riemann tensor becomes infinite at $r=r_{\mathrm{pole}}$ , as $g_{tt}$ becomes the dominant term:
\begin{equation}
\lim_{r \rightarrow r_{\mathrm{pole}}}  R_{\hat t \hat  r \hat t \hat r } = \infty . \label{R1212}
\end{equation}
On a similar way, we have $g_{\theta \theta}$ and $g_{\phi \phi}$ finite at the pole, and
\begin{eqnarray}
g_{\theta \theta}(r_{\mathrm{pole}}) & = &9 R^2  \left(  1-\frac{8/9}{2M/R} \right)  > 0, \\
g_{\phi \phi}(r_{\mathrm{pole}}) & = & 9 \sin^2 \theta R^2 \left(  1- \frac{8/9}{2M/R}\right) > 0.
\end{eqnarray}

Concerning the $t \theta t \theta$  and $t \phi t \phi$ terms of the \index{Riemann!tensor}Riemann tensor, we have,
\begin{eqnarray}
\lim_{r \rightarrow r_{\mathrm{pole}} } R_{t \theta t \theta} &=& 0 \; , \\
\lim_{r \rightarrow r_{\mathrm{pole}} }  \dfrac{dR_{t \theta t \theta}}{dr} &=& \frac{3}{R} \left( \frac{2M}{R}\right)^2 \left( 1-\frac{8/9}{2M/R}\right)^{1/2}  \; , \\
\lim_{r \rightarrow r_{\mathrm{pole}} } R_{t \phi t \phi} &= &0 \; , \\
\lim_{r \rightarrow r_{\mathrm{pole}} } \dfrac{dR_{t \phi t \phi}}{dr} &=& \frac{3}{R} \left( \frac{2M}{R}\right)^2 \left( 1-\frac{8/9}{2M/R}\right)^{1/2} \sin^2 \left( \theta \right).
\end{eqnarray}

Hence, we have for the  $t \theta t \theta$  and $t \phi t \phi$ terms of the Riemann tensor at the pole:
\begin{eqnarray}
\lim_{r \rightarrow r_{\mathrm{pole}} } R_{t \theta t \theta}  =  O \left(  \left[  r-r_{\mathrm{pole}}  \right]  \right), \\
\lim_{r \rightarrow r_{\mathrm{pole}} } R_{t \phi t \phi}  =  O \left(  \left[  r-r_{\mathrm{pole}}  \right]  \right).
\end{eqnarray}
Therefore, the $\hat t \hat  \theta \hat t \hat \theta $-term and the $\hat t \hat  \phi \hat t \hat \phi$-term of the \index{Riemann!tensor}Riemann tensor become infinite at $r=r_{\mathrm{pole}}$, as $g_{tt}$ becomes the dominant term:
\begin{eqnarray}
\lim_{r \rightarrow r_{\mathrm{pole}}}  R_{ \hat t \hat  \theta \hat t \hat \theta}& =& \infty ,  \label{R1313} \\
\lim_{r \rightarrow r_{\mathrm{pole}}}  R_{\hat t \hat  \phi \hat t \hat \phi } & = &\infty . \label{R1414}
\end{eqnarray} 
However, the rest of the non-zero orthonormal components of the Riemann tensor are finite,
\begin{eqnarray}
R_{\hat r \hat  \theta \hat r \hat \theta } & =& \dfrac{2M}{R^3}\\
R_{\hat r \hat  \phi \hat r \hat \phi }  &=& \dfrac{2M}{R^3}\\
R_{\hat \theta \hat  \phi \hat \theta \hat \phi }& = & \dfrac{2M}{R^3}.
\end{eqnarray}
To conclude, we have shown that 3 orthonormal components of the \index{Riemann!tensor}Riemann tensor ($R_{\hat t\hat r\hat t\hat r}$ (\ref{R1212}), $R_{\hat t\hat\theta\hat t\hat\theta}$ (\ref{R1313}), $R_{\hat \theta \hat  \phi \hat \theta \hat \phi }$ (\ref{R1414})) are infinite at $r=r_{\mathrm{pole}}$, and therefore, the pressure pole implies a naked\index{naked singularity} \index{singularity!naked singularity}singularity.

The reason why we mention this specific example is because we shall soon see that this ``pressure pole'' behaviour is generic. Continuous solutions with isotropic pressure are not possible\footnote{Consider for example~\cite{Bilic:2005sn}, where a \index{perfect fluid}perfect fluid Chaplygin gas $\rho\propto 1/p$ is considered. The surface of their configuration occurs at $\rho=0$ where $p=-\infty$, at least as one approaches the surface from below. So the surface of their configuration is a\index{singularity!naked singularity} naked\index{naked singularity} singularity, in agreement with the observations above.}.

 \subsection[Case study: the fate of a negative pressure perfect fluid sphere]{Arbitrary gravastar case study: the fate of a negative pressure perfect fluid sphere} 
\label{case_study_2}\index{perfect fluid!sphere}
  
  In this section, we consider an arbitrary gravastar as defined in section (\ref{arbitrary_gravastar_def}), and we will show that if the gravastar configuration is perfect fluid, the same problems  as those encountered in the previous section (\ref{case_study1}) will arise in this more general case. 
  
  For any arbitrary gravastar that is static spherically symmetric, remember that the geometry is given by
  \begin{equation}
ds^2=-\exp  \left( 2  \int_r ^\infty g(\tilde{r})d\tilde{r}  \right) dt^2 +\frac{dr^2}{\left(    1-\frac{2m(r)}{r}  \right)} +r^2 d\Omega^2,
\end{equation}
where
  \begin{equation}
g(r)=\frac{m(r)+4 \pi p_r(r)r^3}{r^2 \left(    1-\frac{2m(r)}{r}\right)}.  
\end{equation}

\begin{theorem}
If an arbitrary gravastar is perfect fluid and finite, and if we permit the violation of the NEC\index{energy conditions!NEC}\index{NEC}, then there exists a pressure pole of order $1$ and it is the only way the TOV equation can be satisfied.
\end{theorem}

\begin{proof}
Using equation (\ref{eq_dm}) that relates the density $\rho$  to the mass $m(r)$, we can obtain an expression for the compactness $\chi = 2m(r)/r$ which satisfies:
\begin{equation}
\left[{2m(r)\over r}\right]' = 8\pi\; \rho \;r - {2m\over r^2} 
= {8\pi\;r\over3} \; [3\rho-\bar{\rho}] = {8\pi\;r\over3}\; [3(\rho+p) -(\bar{\rho}+3p)].
\end{equation}
But the first term on the RHS is non-negative by the NEC\index{energy conditions!NEC}\index{NEC}, while the second term is by definition negative on $[0,r_g)$, so the compactness $2m(r)/r$ is monotone increasing on the range $[0,r_g)$. This is somewhat unusual, while the compactness of a normal perfect fluid star tends to increase as one moves outwards, in a normal star it  also can be subject to oscillations that make the overall picture quite subtle~\cite{Harrison:1965, Visser:2002ww}.\\

Let's consider now a perfect fluid \index{perfect fluid!sphere}sphere with negative central pressure that satisfies the NEC\index{energy conditions!NEC}\index{NEC}. Since we have already seen that isotropy\index{isotropy} is violated on $(r_g,r_\mathrm{max})$, the only way we can maintain the perfect fluid nature of the sphere is if $r_g\to\infty$, which also implies that $r_0\to\infty$. But since we do not want a horizon\index{horizon} to form, the compactness must be bounded above by unity. But we have just shown the compactness is monotonic and therefore,
\begin{equation}
\lim_{r\to\infty} {2m(r)\over r} = \chi_*; \qquad \chi_*\in(0,1].
\end{equation}
So not only does a NEC-satisfying perfect-fluid  gravastar expand to infinite volume, it also has infinite mass.

To avoid the physical size of the gravastar blowing up to $r_0\to\infty$, our options now are rather limited: we could permit the development of a horizon\index{horizon} at finite $r$, which defeats the whole point of the exercise,  or we could permit something even worse.  

\subsubsection{First order pressure pole}
If we permit NEC\index{energy conditions!NEC}\index{NEC} violations then it is possible to arrange for the  development of a pressure pole at finite $r<r_0$. Let's see how this is possible by first considering  the following definition.
\begin{definition}
Set
\begin{equation}
p(r) \approx {\Gamma\over r-r_p},
\end{equation} in the vicinity of the pressure pole of order $1$ at $r_p$ where $\Gamma$ is a positive constant.
\end{definition}

This definition is compatible with the \emph{isotropic} TOV equation. First for $r<r_p$ the pressure is negative, and for $r>r_p$ the pressure is positive, which is appropriate for a modified gravastar model. Secondly
\begin{equation}
p' \approx {-\Gamma\over(r-r_p)^2} < 0.
\end{equation}
Third, assuming $\rho$ remains finite, close to the pole ($r\approx r_p$) we have
\begin{eqnarray}
 - {(\rho+p) (m+4\pi pr^3)\over r^2(1-2m/r)} & \approx &
 - {[\rho+\Gamma/(r-r_p)]\;[m+4\pi r^3\Gamma/(r-r_p)] \over r^2(1-2m(r)/r)}\\
& \approx & - {[\Gamma/(r-r_p)]\;[4\pi r^3\Gamma/(r-r_p)] \over r^2(1-2m(r)/r)}\\ 
& \approx & - {4\pi r_p \Gamma^2\over (1-2m(r_p)/r_p)} {1\over(r-r_p)^2} < 0.
\end{eqnarray}
So the TOV equation \emph{can} be satisfied in the vicinity of the pole provided 
\begin{equation}
 - {(\rho+p) (m+4\pi pr^3)\over r^2(1-2m/r)}   \approx p_r' \qquad \text{at}\; \;  r=r_g,
\end{equation}
that is provided we set
\begin{equation}
\Gamma = {1-2m(r_p)/r_p\over 4\pi r_p} > 0.
\end{equation}
That is, if the gravastar configuration is \index{perfect fluid}perfect fluid and finite in extent, then this pole is the only way the TOV equation can be satisfied. 
\end{proof}

\subsubsection{$n^{\text{th}}$ order pressure pole}
Now we can look at higher order pole and see if they also have the nice property of being compatible with the isotropic TOV equation.
\begin{definition}
Set
\begin{equation}
p(r) \approx {\Gamma\over \left(r-r_p \right)^n},
\end{equation} in the vicinity of the pressure pole of order $n>1$ at $r_p$ where $\Gamma$ is a positive constant.  \label{def_pole_n}
\end{definition}

This definition is compatible with the \emph{isotropic} TOV equation for \textit{odd value of $n$ only}. First for $r<r_p$ the pressure is negative if $n$ is odd, and for $r>r_p$ the pressure is positive if $n$ is odd. However, this definition is not compatible with the isotropic TOV equation for even values of $n$. Secondly
\begin{equation}
p' \approx {-n\Gamma\over(r-r_p)^{(n+1)}} < 0.
\end{equation}
Third, assuming $\rho$ remains finite, close to the pole ($r\approx r_p$) we have
\begin{eqnarray}
 - {(\rho+p) (m+4\pi pr^3)\over r^2(1-2m/r)} & \approx &
 - {[\rho+\Gamma/\left(r-r_p \right)^n]\;[m+4\pi r^3\Gamma/\left(r-r_p \right)^n] \over r^2(1-2m(r)/r)}\\
& \approx & - {[\Gamma/\left(r-r_p \right)^n]\;[4\pi r^3\Gamma/\left(r-r_p \right)^n] \over r^2(1-2m(r)/r)}\\ 
& \approx & - {4\pi r_p \Gamma^2\over (1-2m(r_p)/r_p)} {1\over \left(r-r_p \right)^{2n}} < 0.
\end{eqnarray}
So the TOV equation \emph{can} be satisfied in the vicinity of the pole provided 
\begin{equation}
 - {(\rho+p) (m+4\pi pr^3)\over r^2(1-2m/r)}   \approx p_r' \qquad \text{at}\; \;  r=r_g,
\end{equation}
that is provided we set
\begin{equation}
\Gamma = {1-2m(r_p)/r_p\over 4\pi r_p}  \; n \; \left(r-r_p \right)^{n-1} \geqslant 0. \label{gamma}
\end{equation}
First, equation (\ref{gamma}) is in total contradiction with the Definition \ref{def_pole_n}  of $\Gamma$ that is supposed to be a \emph{constant}. Secondly, it shows that the isotropic TOV equation \emph{cannot} hold at the pressure pole of order $n$.

 Therefore, higher-order poles do not even have this nice property of being compatible with the isotropic TOV equation.

\begin{theorem}
An arbitrary finite gravastar with a perfect fluid configuration that has a pressure pole leads to a curvature singularity.
\end{theorem}
\begin{proof}
For any isotropic static spherically symmetric spacetime the orthonormal components of the\index{Riemann!tensor} Riemann tensor are (see, for instance, \cite[p 110]{Visser:1995cc}):
\begin{eqnarray}
R_{\hat t \hat r \hat t \hat r}  &=& {4\pi\over3}\; [3(\rho+p)-2\bar\rho]; \\
R_{\hat t \hat \theta \hat t \hat \theta}& = &{4\pi\over3}\; [3p +\bar\rho]; \\
R_{\hat r \hat \theta \hat r \hat \theta} &= &{4\pi\over3}\; [3\rho-\bar\rho]; \\
R_{\hat \theta \hat \phi \hat \theta \hat \phi}& =& {4\pi\over3}\; [2\bar\rho]. 
\end{eqnarray}
It is clear that if the pressure has a pole of order $n \geqslant 1$, then, providing $\rho$ remains finite as per assumption, as $R_{\hat t \hat r \hat t \hat r}$ and $R_{\hat t \hat \theta \hat t \hat \theta}$ depend on $p$, they will also have a pole, and consequently, a naked \index{singularity!naked singularity}\index{naked singularity}singularity will arise.
\end{proof}

Even if the isotropic TOV equation holds with pressure pole in the gravastar configuration, this situation is physically inappropriate as the pressure pole introduces a naked\index{singularity!naked singularity} \index{naked singularity}singularity. The physically correct deduction form this analysis is that gravastar-like objects must violate pressure isotropy\index{isotropy}. 
 
 \subsubsection{Summary}
In this section, we have shown that a static spherically symmetric object with positive density, negative central pressure, and vanishing pressure at the surface, (which are the defining features of a gravastar),  cannot be supported by isotropic pressure alone without a pole in the pressure. Despite the fact that the presence of a simple pole in the pressure is compatible with the isotropic TOV equation, we must reiterate that such a pressure pole is unphysical because it is a naked\index{singularity!naked singularity}\index{naked singularity} singularity. Therefore, gravastar-like objects must violate pressure \index{isotropy}isotropy.

\section{Gravastars with anisotropy}\index{anisotropy}

In section (\ref{no_isotropy}), we have demonstrated that gravastars cannot have isotropic pressure alone, and that, as a consequence, anisotropic pressure must be part of the configuration or at least in some parts of it. In this section, we are going to see when the \index{anisotropy}anisotropy for the pressure is necessary for the TOV equation to hold, and what specific features the equation of state must have.

\subsection{Bounds on the anisotropy pressure}  \label{bounds_anisotropy}
\index{anisotropy}
Once we accept that perfect fluid spheres are not what we are looking for to model gravastars, one might wonder what happens to the Buchdahl--Bondi bound for isotropic fluid spheres. It has been shown that for $\rho' < 0$ and $p_t \leq p_r$ the $8/9$ bound still holds\footnote{See section (\ref{case_study1})}. However if the transverse stress is allowed to exceed the radial stress, (e.g., $p_t > p_r$), then the upper limit shifts to $2M/R < \kappa \leq 1$, where $\kappa$ depends on the magnitude of the maximal stress anisotropy~\cite{Guven:1999wm}\footnote{For more related work on anisotropic stars see~\cite{Herrera:2004xc, Herrera:2001vg, Herrera:2002bm}}.\\

Remember that in section \ref{inconsistencies} we defined the crust of the gravastar as the region $r_g<r<r_{\mathrm{max}}$, where the physics is still definitely ``unusual", and where the local acceleration due to gravity is inward, but the pressure still rises as one moves outward. The dimensionless anisotropy \index{anisotropy}parameter $\Delta$ is a function of the radial pressure, tangential pressure and positive density:
\begin{equation}
\Delta = \frac{p_t - p_r}{\rho} = \frac{r}{2} \left[
\frac{p_r'}{\rho} + \left(1+\frac{p_r}{\rho} \right)\, g \right]
\, .    \label{delta_again}
\end{equation}

\begin{theorem}
In the crust of a gravastar, the compactness is not limited to the Buchdahl--Bondi bound, but only by the magnitude of the maximal pressure anisotropy\index{anisotropy} and the regularity of the metric: 
\begin{equation}
\Delta \geq \frac{1}{4}\, \frac{2m/r}{1-2m/r} > 0  \qquad r_g<r<r_0   \label{delta_lower_bound}
\end{equation}
and

\begin{equation}
0 \leq \frac{r\,p_r'}{2\,\rho} < \Delta < \frac{r\,p_r'}{2\,\rho} 
+ \frac{1}{4}\, \frac{2m/r}{1-2m/r}  \qquad r_0<r<r_{\mathrm{max}}
\end{equation}

Furthermore, the tangential pressure is greater than the radial pressure $p_t>p_r$ for $r_g<r<r_{\mathrm{max}}$.
\end{theorem}

\begin{proof}
In section \ref{inconsistencies}, we have determined that anisotropy \index{anisotropy}is necessary in the crust\footnote{Isotropy\index{isotropy} possibly fails for an even larger region.}, that is, when $r_g<r<r_{\mathrm{max}}$. We can now find bounds on $\Delta$ in that interval by inserting the definition of the gravitational acceleration $g(r)$ into equation (\ref{delta_again}). We get,

\begin{equation}
\Delta = \frac{r}{2} \left[ \frac{p_r'}{\rho} +
\left(1+\frac{p_r}{\rho} \right)\, \frac{m+4\pi p_r\, r^3}{r^2
\left[ 1-\frac{2m}{r} \right] } \right] \, .
\end{equation}

Now we know that for the interval $r\in [r_0,r_\mathrm{max}]$, the radial pressure is positive $p_r \geq 0$ and increasing so $p_r' \geq 0$. Therefore, we can write the inequality,
\begin{eqnarray}
\Delta &  \geq &  \frac{r}{2}\,\left(1+\frac{p_r}{\rho} \right)\, \frac{m}{r^2[1-2m/r]} \; > \; 0  \\
\Delta &  \geq &  \frac{r}{2}\, \frac{m}{r^2[1-2m/r]}  \; > \; 0 ,
     \end{eqnarray}
     that leads to the simple lower bound:
\begin{equation}
\Delta \geq \frac{1}{4}\, \frac{2m/r}{1-2m/r} > 0. \label{delta_1}
\end{equation}

The inequality on $\Delta$ (\ref{delta_1}), for the interval $r\in [r_0,r_\mathrm{max}]$, shows that it is positive, consequently, in this region, we have $p_t>p_r>0$.

Now let's see what happens for the interval $r\in (r_g,r_0)$. We know that in this area, the radial pressure is negative $p_r<0$, but it is increasing and $\d p_r/ \d r>0$. From these conditions on the radial pressure and its derivative, we can write,
\begin{eqnarray}
\Delta &  \leqslant &  \frac{r}{2}\,\left(  \frac{ p_r' }{\rho} +1 \right)\,   \frac{m+4\pi p_r\, r^3}{r^2
\left( 1-2m/r \right) } \\
\Delta &  \leqslant &  \frac{r p_r'}{2\rho} + \frac{r}{2}\, \frac{m}{r^2[1-2m/r]}  \\
\Delta &  \leqslant &  \frac{r p_r'}{2\rho} +  \frac{1}{4}\, \frac{2m/r}{1-2m/r}.
     \end{eqnarray}
Remember that we assume that the NEC\index{energy conditions!NEC}\index{NEC} holds in the configuration of the gravastar. This leads to
\begin{eqnarray}
\rho + p_r & >& 0 \\
p_r & > &-\rho\\
\dfrac{p_r}{\rho} & >& -1\\
\left(   1+ \dfrac{p_r}{\rho} \right)  & >& 0. \label{ineq_NEC}
\end{eqnarray}
This last inequality (\ref{ineq_NEC}), allows us to write a lower bound on $\Delta$:
\begin{equation}
\left(  \frac{ p_r' }{\rho} +1 \right)\,   \frac{m+4\pi p_r\, r^3}{r^2 \left( 1-2m/r \right) }
\qquad \Longrightarrow \qquad \Delta \geqslant   \frac{r p_r'}{2\rho} \geqslant 0 .
\end{equation}
 Consequently, for the interval  $r\in (r_g,r_0)$, we find the weaker bounds
 \begin{equation}
0 \leq \frac{r\,p_r'}{2\,\rho} < \Delta < \frac{r\,p_r'}{2\,\rho}
+ \frac{1}{4}\, \frac{2m/r}{1-2m/r}
\end{equation}

\end{proof}

\begin{theorem}
In the interval $[r_0,r_\mathrm{max}]$, if the DEC\index{energy conditions!DEC}\index{DEC} holds, then 
\begin{equation}
\mathrm{DEC} \qquad \Longrightarrow \qquad \Delta \leqslant 1.
\end{equation}
\end{theorem}

\begin{proof}
For the first inequality on $\Delta$ (\ref{delta_1}), in the region $r\in [r_0,r_\mathrm{max}]$, the DEC\index{energy conditions!DEC}\index{DEC} is satisfied if
\begin{eqnarray}
\rho \geqslant 0, \\
- \rho \leqslant p_r \leqslant \rho, \\
- \rho \leqslant p_t \leqslant \rho .
\end{eqnarray}
This leads to the more useful inequalities,
\begin{eqnarray}
\left\| \dfrac{p_r}{\rho} \right\|  & \leqslant & 1, \\
 \left\| \dfrac{p_t}{\rho} \right\|  & \leqslant & 1 .
\end{eqnarray}
But because the radial pressure is positive for the interval $[r_0,r_\mathrm{max}]$, we can write,
\begin{equation}
\Delta =\dfrac{p_t-p_r}{\rho} \leqslant \dfrac{p_t}{\rho} \leqslant 1.
\end{equation}
And therefore, if the DEC\index{energy conditions!DEC}\index{DEC} is to be satisfied, it implies the following upper bound on $\Delta$:
\begin{equation}
\mathrm{DEC} \qquad \Longrightarrow \qquad \Delta \leqslant 1.
\end{equation}
\end{proof}

\begin{theorem}
Any gravastar that is sufficiently close to forming a horizon\index{horizon}, that is $2m/r>4/5$, will violate the DEC\index{DEC} in its ``crust''.
\end{theorem}

\begin{proof}
If the DEC\index{energy conditions!DEC}\index{DEC} is to be satisfied we must at the very least have $\Delta \leq 1$.  
Therefore, when the DEC \index{DEC}holds, we have
\begin{eqnarray}
1 & \geqslant & \Delta  \geq   \frac{1}{4}\, \frac{2m/r}{1-2m/r} > 0 \\
1   &\geqslant  &  \frac{1}{4}\, \frac{2m/r}{1-2m/r} > 0 \\
4  &\geqslant  & 5 \frac{2m}{r}\\
\frac{4}{5} & \geqslant &   \frac{2m}{r}
\end{eqnarray}
Hence, the DEC\index{energy conditions!DEC}\index{DEC} is guaranteed to be violated whenever $2m/r>4/5$. 
If the gravastar is sufficiently close to forming a horizon\index{horizon}, in the sense that $ 2m/r> 4/5$ somewhere in the range $[r_0,r_\mathrm{max}]$, then the DEC\index{DEC} must also be violated at this point. Even if we were to discard the entire positive-pressure region $(r_0,R)$, we can nevertheless still apply this bound at $r_0$ itself: if $2m(r_0)/r_0>4/5$ then the DEC\index{DEC} is violated at $r_0$. Consequently, any gravastar that is sufficiently close to forming a horizon\index{horizon} will violate the DEC\index{DEC} in its ``crust''.
\end{proof}

\subsection{Minimizing the anisotropy} \index{anisotropy}
We have shown that the anisotropy is necessary for gravastar-like objects in their ``crust", where the pressure is increasing as one moves outwards and the local force of gravity is inwards. In this section, we attempt to minimize the region over which anisotropy\index{anisotropy} is present.

Remember that he dimensionless anisotropy\index{anisotropy} parameter $\Delta$ is deduced from the anisotropy TOV equation (\ref{eq_full_TOV}), and gives,
\begin{equation}
\Delta = \frac{p_t - p_r}{\rho} = \frac{r}{2} \left[
\frac{p_r'}{\rho} + \left(1+\frac{p_r}{\rho} \right)\, g \right]
\, .   
\end{equation}

\subsubsection{What happens at $r_g$?   }
At the point $r_g$, we can write (since $g(r_g)=0$ by definition \ref{def_rg})
\begin{equation}
\Delta (r_g)= \frac{r_g p_r'(r_g)}{2 \rho} \geqslant 0.
\end{equation}
To confine the anisotropy \index{anisotropy}to the smallest interval possible, we have to set
\begin{equation}
p_r'(r_g)=0,
\end{equation}
which corresponds to an \index{singularity!extremality events!inflexion}inflexion point for the radial pressure.

\subsubsection{What happens at $r_0$?   }
At the point $r_0$, the radial pressure is zero, and we can write,
\begin{equation}
\Delta(r_0) = \frac{r_0\,p_r'}{2\,\rho} + \frac{1}{4}\,
\frac{2m/r_0}{1-2m/r_0} > 0 \, .
\end{equation}
Hence, at the point of zero radial pressure, the anisotropy\index{anisotropy} cannot vanish.

\subsubsection{What happens at $r_\mathrm{max}$?   }
At the point $r_\mathrm{max}$, the radial pressure reaches its maximum value. If we take the limit from below, we can write,
\begin{equation}
\Delta(r_\mathrm{max}^-) = \frac{1}{4}\, \frac{2m/r_\mathrm{max}}{1-2m/r_\mathrm{max}}
\left(1+\frac{p_r}{\rho} \right)\, \left(1+\frac{4\pi p_r\, r_\mathrm{max}^3}{m
} \right) \, > 0 \, .
\end{equation}
Remember that the anisotropy\index{anisotropy} is not necessary beyond the peak $\forall \; r > r_\mathrm{max}$ (in the atmosphere), and therefore, it is possible to arrange $\Delta=0$. To confine the anisotropy \index{anisotropy} to the smallest interval possible, we have to set 
\begin{equation}
\Delta(r \rightarrow r_\mathrm{max}^+) = 0.
\end{equation}
However, this leads to a discontinuity in the derivative of the radial pressure $p_r'$ and the \index{anisotropy}anisotropy parameter $\Delta$, as well as a ``kink" in the pressure profile $p_r$ at $r_\mathrm{max}$.

From the TOV equation (\ref{eq_full_TOV}), we can deduce the limit from above for the pressure derivative,
\begin{equation}
p_r'(r_\mathrm{max}^+) = - \left(\rho+p_r \right)\, \left( \frac{ m+4\pi p_r\, r_\mathrm{max}^3}{ r^2   \left( 1 -2m/r_\mathrm{max} \right) } \right).
\end{equation}
$\Delta$ at $r_\mathrm{max}^-$ can be rewritten as
\begin{equation}
\Delta(r_\mathrm{max}^-) = \frac{1}{2}\,  \left(1+\frac{p_r}{\rho} \right)\, \left( \frac{ m+4\pi p_r\, r_\mathrm{max}^3}{ r   \left( 1 -2m/r_\mathrm{max} \right) } \right).
\end{equation}
Therefore, we can still keep the radial pressure and the density continuous which will imply that,
\begin{equation}
p_r'(r_\mathrm{max}^+) = -
{2 \rho_\mathrm{max}\over r_\mathrm{max}} \; \Delta(r_\mathrm{max}^-).
\end{equation}

\begin{figure}[htb]
\begin{center}
\input{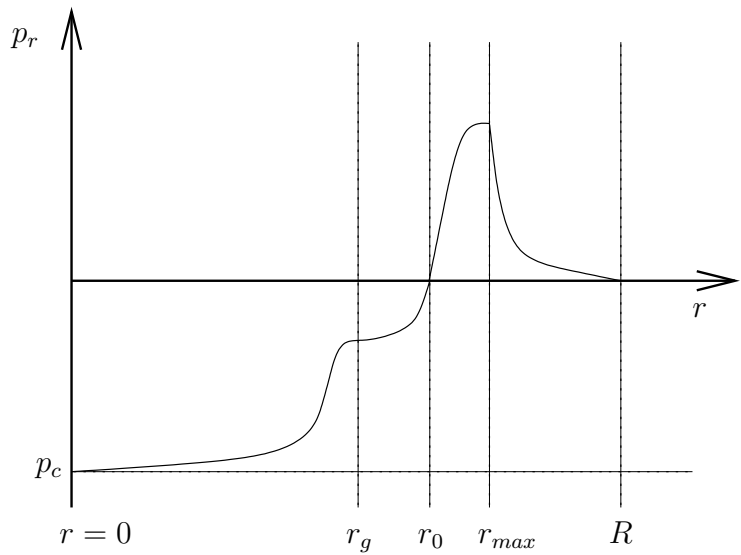}
\end{center}
\caption[Qualitative sketch of radial pressure for a ``perfect fluid'' gravastar.]{\label{fig4}
Qualitative sketch of radial pressure as a function of $r$ for a gravastar that is as near as possible to a perfect fluid. Note the\index{singularity!extremality events!inflexion} inflexion point at $r_g$ and the kink at $r_\mathrm{max}$.}
\end{figure}

The implications of confining the pressure \index{anisotropy}anisotropy to the smallest
interval possible are shown in figure \ref{fig4}, where the anisotropy is confined to the region $r\in(r_g,r_\mathrm{max}]$.

\subsubsection{How does this relate to other gravastar models?   }

The Mazur--Mottola model is recovered as the limiting case where
$r_g\to r_0\leftarrow r_\mathrm{max}$\footnote{See section (\ref{models}) for more details}. This results in the fact that all the \emph{important} anisotropy\index{anisotropy} is confined in their inner thin shell. The anisotropy
$\Delta\to\infty$, because $p'\to\infty$. Effectively $\Delta$ is replaced by 
choosing an appropriate finite (in
this case negative) surface tension $\vartheta$ and surface energy
density $\sigma$ which is given by the Israel--Lanczos--Sen 
junction conditions~\cite{ Barrabes:1991ng,Israel:1966rt, Lanczos, Sen, Visser:2003ge}. 

The second, outer thin shell which is present in the Mazur--Mottola model
is \emph{not} a physical necessity, but is merely a convenient
way to avoid an infinitely diffuse atmosphere that would
arise otherwise from the equation of state $p=\rho$. A finite
surface radius $R$ can be modelled by altering the equation of
state slightly to include a finite surface density $\rho_S$ which is
reached for vanishing pressure: $\rho(p)=\rho_S+p$. Then, the
outer thin shell can be omitted when joining the gravastar metric
onto the Schwarzschild\index{Schwarzschild!exterior solution} exterior metric.

\subsection{Features of the anisotropic equation of state} \label{features_eq_state}

In the case of a perfect\index{perfect fluid} fluid, the geometry is completely defined by the set of differential equations (\ref{eq_dm}), (\ref{TOV_iso}) plus initial conditions and an equation of state, which is commonly written as $\rho=\rho(p)$ or equivalently $p=p(\rho)$.

However, in this chapter we have pointed out that gravastars with perfect fluid sphere are a lost cause, and one needs to replace the isotropic TOV equation (\ref{TOV_iso}) with its anisotropic counterpart  (\ref{eq_full_TOV}). By doing so, we also need to introduce an additional free function $\Delta(r)$. In order to close the set of equations, it is necessary to define \textit{two} equations of state.

\subsubsection{Two equations of state}

Two equations of state must be chosen, but which criteria does one take into account?

\begin{itemize}
\item $p_r(\rho)$ and $p_t(\rho)$

This is a rather strong assumption that forces $p_r$ and $p_t$ to change in lock-step.

\item density profile $\rho(r)$ and a single equation of state / pressure profile $p_r(r)$ and a single equation of state

This is the strategy adopted, for example, in \cite{Dymnikova:1992ux, Dymnikova:2001fb, Dymnikova:2003vt, Dymnikova:2004qg, Dymnikova:1999cz}.

\item Specify any two profiles $\rho(r)$, $p_r(r)$, and $p_t(r)$ by hand, and use the anisotropic TOV to calculate the remaining profile.

This is rather unphysical.

\item finding an additional (differential) equation

This is another possibility to obtain a well defined solution, that might be motivated by some appropriate variational principle. An example would consists of minimizing the ``total  anisotropy"
\begin{equation}
  \frac{1}{R} \int_0^R (\rho\Delta)^2 \d r 
\quad\Rightarrow\quad 
\delta\left[\int_0^R (\rho\Delta)^2 \d r \right] = 0\, .
\end{equation}
No matter whether it is an equation of state for the transverse pressure or a variational principle for the total \index{anisotropy}anisotropy, the extra equation should (if the gravastar model is to be even qualitatively correct)  be responsible for maintaining stability of the gravastar over a wide range of total masses and central pressures. In other words, the closed set of equations must be self-regulating if it is to be physically interesting. The gravastar should shift to a new  stable configuration when the total mass changes. 

\end{itemize}

The fact that, there will be regions where $\Delta >0$ in the gravastar configuration suggests that it might be most efficient to choose the two distinct ``equations of state" as being  equations for $\rho$ and for $\Delta$.\\

What variables should these equations of state depend on? It seems obvious that they should depend on the radial pressure $p_r$, but in view of the lower bounds on $\Delta$ in equation (\ref{delta_lower_bound}), it is clear that to avoid the formation of an horizon\index{horizon} in the gravastar configuration, the equations of state should be sensitive to the compactness $2m/r$.

Therefore, we should posit equations of state of the form
 \begin{equation}
\rho= \rho(r,p_r,2m/r); \qquad \Delta= \Delta(r,p_r,2m/r);
\end{equation}
and solve the paired differential equations
\begin{eqnarray}
{\d p_r(r)\over \d r} &=& 
 -\frac{[\rho(r,p_r(r),2m(r)/r)+p_r(r)]\;[2m(r)/r+8\pi \;p_r(r) \;r^2 ]}{2\,r\;[1-2m(r)/r]} 
\nonumber \\
 &&
+{2\;\rho(r,p_r(r),2m(r)/r)\;\Delta(r,p_r(r),2m(r)/r)\over r};
\end{eqnarray}
\begin{equation}
{\d m(r)\over \d r}  = 4\pi \;\rho(r,p_r(r),2m(r)/r) \;r^2.
\end{equation}

The success of this gravastar model depends on whether or not one can find physically realistic equations of state that take into account the effect of the compactness $2m/r$ for large ranges of central negative pressure $p_c$ and total mass $M$.

\subsubsection{Potential issue:   }  

Some traditional relativists might argue that there is an important issue of principle, by stating that the variable $2m/r$ is not detectable by local physics. There might also argue that an equation of state that depends on $2m/r$ somehow violates the Einstein \index{Einstein!equivalence principle}equivalence principle\index{equivalence principle}\footnote{See section \ref{intro} for the equivalence principle}. 

But this is incorrect because $2m/r$ is certainly measurable by \emph{quasi-local effects} in small but \emph{finite size regions}. For any static spherically symmetric spacetime the orthonormal components of the Riemann \index{Riemann!tensor}tensor are (see, for instance, \cite[p 110]{Visser:1995cc}):
\begin{eqnarray}
R_{\hat t \hat r \hat t \hat r}  &=& {4\pi\over3}\; [3(\rho-p_r+2p_t)-2\bar\rho]; \\
R_{\hat t \hat \theta \hat t \hat \theta}& = &{4\pi\over3}\; [3p_r +\bar\rho]; \\
R_{\hat r \hat \theta \hat r \hat \theta} &= &{4\pi\over3}\; [3\rho-\bar\rho]; \\
R_{\hat \theta \hat \phi \hat \theta \hat \phi}& =& {4\pi\over3}\; [2\bar\rho]. 
\end{eqnarray}

The Riemann\index{Riemann!tensor} tensor is certainly measurable in finite-sized regions, so in particular $\bar\rho$ is measurable. Likewise $r$ is measurable in finite-sized regions, and therefore $2m/r= (8\pi/3)\;\bar\rho\;r^2$ is measurable.

 Consequently the compactness $2m/r$ is a \emph{quasi-local measurable quantity} and it can meaningfully be put into the equation of state without violating the equivalence \index{equivalence principle}principle. 
 
 Note that in either the Mazur--Mottola scenario~\cite{Mazur:2001fv, Mazur:2004ku, Mazur:2004fk} or the Laughlin \emph{et al.} scenario~\cite{Chapline:2000en,Chapline:2002} the gravastar material is assumed to be a quantum condensate, and therefore sensitive to non-local physics. The point is that we do not need to appeal to quantum non-locality to get $2m/r$ into the equation of state. \\

 However, note that while this argument demonstrates that possible equations of state are at least conceivable, that is not the same as explicitly demonstrating that such equations of state actually exist. 

\section{Results and discussion}

This chapter has hopefully served as a brief introduction to the idea of gravastars
and how they are configured, as well as some of their properties.\\

 We have discussed the qualitative features one would expect from a gravastar configuration that has finite continuous radial pressure profile everywhere and that does not have delta function transition layers (\ref{arbitrary_gravastar_def}). More importantly, we posit that
\begin{itemize}
\item Inside the gravastar, $ r < R$, the density is everywhere positive and finite.

\item The  central pressure is negative, $p_c<0$, and in fact $p_c = - \rho_c$, and the spacetime in the centre is regular.

\item The spacetime is assumed to \emph{not} possess an event horizon\index{horizon}. 

\item We choose to enforce the WEC\index{energy conditions!WEC}\index{WEC} throughout the gravastar but impose no restrictions on other energy conditions\index{energy conditions}.

\item Our model describes a radial pressure that starts negative at the centre, then increases till $p_r=0$, becomes positive, and starts decreasing after a maximum $p_r=p_\mathrm{max}$ at $r=r_\mathrm{max}$ eventually reaching a second zero at the surface.

\end{itemize}

Adopting $(t,r,\theta, \phi)$ coordinates, the interior geometry for gravastars can be written in the form,
\begin{equation}
ds^2=-\exp  \left( 2  \int_r ^\infty g(\tilde{r})d\tilde{r}  \right) dt^2 +\frac{dr^2}{\left(    1-2m(r)/r  \right)} +r^2 d\Omega^2
\end{equation}
where $d\Omega^2=d\theta^2+\sin^2 \theta d\phi^2$, and $m(r)$ represents the total mass-energy inside the radius $r$ integrated from the centre $r=0$ until arbitrary $r$. The gravitational acceleration $g(r)$ is written as:
\begin{equation}
g(r)=\frac{4 \pi r}{3}\frac{\bar{\rho} +3p_r(r)}{1-2m(r)/r}.
\end{equation}

We have used the isotropy\index{isotropy} TOV equation where the tangential pressure is equal to the radial pressure $p_r=p_t$,
  \begin{equation}
\frac{d p}{d r} = -(\rho+p) g,
\end{equation}
  and the more general anisotropy \index{anisotropy}TOV equation,
  \begin{equation}
\frac{d p_r}{d r} = -\frac{4\pi r}{3} \frac{(\rho+p_r) (\bar{\rho}+3p_r)}{1-2m(r)/r}+\frac{2 \rho \Delta}{r},
\end{equation}
where we have defined the anisotropy dimensionless parameter
\begin{equation}
\Delta =\frac{p_t-p_r}{\rho}.
\end{equation}

From theses specific features and the\index{isotropy} isotropy TOV equation, we have analyzed and deduced that gravastars with perfect\index{perfect fluid!sphere} fluid sphere do not exist, that is, gravastar-like objects must have anisotropic pressure. \\

We also have extracted some generic information concerning the features that the equations of state of a gravastar should have. More specifically, any gravastar-like object, that has negative central pressure $p_c$, must have anisotropic pressure in their ``crust"; the ``crust" being the region where the pressure is increasing as one moves outward and the local force of gravity is inward (\ref{inconsistencies}). Since the compactness is monotonic, we have, assuming the NEC\index{energy conditions!NEC}\index{NEC},
 \begin{equation}
\lim_{r\to\infty} {2m(r)\over r} = \chi_*; \qquad \chi_*\in(0,1].
\end{equation}
 Not only do perfect fluid gravastars lead to inconsistencies when the\index{isotropy} isotropy TOV equation holds for finite continuous pressure, but, when the pressure is allowed to be discontinuous for the TOV equation to hold, they lead to an infinite-size infinite-mass object, or a naked\index{singularity!naked singularity} \index{naked singularity}singularity as the pressure exhibits a pole (\ref{case_study1} and \ref{case_study_2}).\\

 If we assume the DEC\index{energy conditions!DEC}\index{DEC}, we can explicitly bound the magnitude of the anisotropy\index{anisotropy}, required in the crust, in terms of the local compactness $2m(r)/r$. 
 We have shown that,
 \begin{equation}
\Delta \geq \frac{1}{4}\, \frac{2m/r}{1-2m/r} > 0  \qquad r_g<r<r_0    \label{ineq1}
\end{equation}
and
\begin{equation}
0 \leq \frac{r\,p_r'}{2\,\rho} < \Delta < \frac{r\,p_r'}{2\,\rho} 
+ \frac{1}{4}\, \frac{2m/r}{1-2m/r}  \qquad r_0<r<r_{\mathrm{max}}.   \label{ineq2}
\end{equation}
The second inequality (\ref{ineq2}) is weaker than the first one (\ref{ineq1}), and it might still be possible to find better bounds.

 The magnitude of the anisotropy\index{anisotropy} becomes arbitrary large for gravastars that are sufficiently close to forming a horizon\index{horizon}, that is,
 \begin{equation}
\dfrac{2m}{r}>\dfrac{4}{5},
\end{equation}
  in which case the DEC\index{energy conditions!DEC}\index{DEC} must be violated (\ref{bounds_anisotropy}). 
 
 The compactness bounds are harder to state as well: if one is prepared to ignore the
 DEC\index{energy conditions!DEC}\index{DEC} then it seems that the compactness can be arbitrarily close to one.
 
 When trying to minimize the anisotropy\index{anisotropy} region in the gravastar model, one has to pay the price of a discontinuity in $p_r'$ and $\Delta$, and a ``kink" in the radial pressure profile $p_r$.\\

 We have also deduced, that for configurations that are \index{horizon}horizon-avoiding for any total mass and negative central pressure, the equations of state of gravastar matter must depend on the local compactness $2m(r)/r$. We posit that the equations of state are of the form
  \begin{equation}
\rho= \rho(r,p_r,2m/r); \qquad \Delta= \Delta(r,p_r,2m/r);
\end{equation}
and that one should solve the paired differential equations
\begin{eqnarray}
{\d p_r(r)\over \d r} &=& 
 -\frac{[\rho(r,p_r(r),2m(r)/r)+p_r(r)]\;[2m(r)/r+8\pi \;p_r(r) \;r^2 ]}{2\,r\;[1-2m(r)/r]} 
\nonumber \\
 &&
+{2\;\rho(r,p_r(r),2m(r)/r)\;\Delta(r,p_r(r),2m(r)/r)\over r};
\end{eqnarray}
\begin{equation}
{\d m(r)\over \d r}  = 4\pi \;\rho(r,p_r(r),2m(r)/r) \;r^2.
\end{equation}
\\

The success of this gravastar model depends on whether or not one can find physically realistic equations of state that take into account the effect of the compactness $2m/r$. However, we have not explicitly demonstrated that such equations of state actually exists.
 Note that, as discussed in (\ref{features_eq_state}), this unusual equation of state does not violate the equivalence principle\index{equivalence principle}.\\

 Even though we are personally agnostic to the existence or non-existence of gravastars, the study and understanding of their properties are important to comprehend what observational data regarding astrophysical black\index{black hole} holes are telling us.


\chapter{Conclusion}
Two main problems in general relativity have been addressed in this thesis:
\begin{itemize}
\item The first problem is relevant to cosmology and singularities.
\item The second looks at the gravastar model developed by Mazur and Mottola, a possible alternative to black holes.
\end{itemize}

Main results have already been summarized to a large extent at the end of each respective chapter. However, the main important results are highlighted below:
\begin{itemize}
\item Cosmology:\\
Three issues have been explored concerning this topic:
\begin{itemize}
\item We have developed a complete catalogue of the various Òcosmological milestonesÓ
in terms of a generalized power series expansion of the FRW scale factor. This power series expansion is sufficiently general to accommodate all commonly occurring models considered in the literature.
\item With the notion of a generalized power series in hand, it is possible at
a purely kinematic level to address the question of when a Òcosmological milestoneÓ
corresponds to a curvature singularity, and what type of singularity is
implied. An important result is that there are only a few cases of cosmological milestones that are not polynomial curvature singularities.
And the only two situations in which a cosmological milestone is not a derivative curvature singularity correspond to :
\begin{itemize}
\item  an extremality event;
\item  a FRW geometry that smoothly asymptotes near the cosmological milestone to the Riemann-flat  Milne universe. 
\end{itemize}

\item Finally, this definition of cosmological milestones in terms of generalized power series enables us to perform a complete model-independent check on the validity or otherwise of the classical energy conditions.
In particular we provide a complete catalogue of those bangs/crunches, sudden
singularities and extremality events for which the NEC, the WEC, the SEC and
the DEC are satisfied.
Depending on one's attitude towards the energy conditions, one could use
this catalogue as a guide towards deciding on potentially interesting scenarios to
investigate. In particular, the DEC is satisfied all the way to the singularity for a few range of sudden singularities $(\eta_0=0)$ that are sufficiently ÒgentleÓ. 
\end{itemize}

\item The gravastar model:
\begin{itemize}
\item We have discussed the qualitative features one would expect from a gravastar configuration
that has finite continuous radial pressure profile everywhere and that does
not have delta function transition layers.
\item From these specific features and the isotropic TOV equation, we have analyzed and
deduced that gravastars with perfect fluid sphere do not exist, that is, gravastar-like
objects must have anisotropic pressure.
\item We also have extracted some generic information concerning the features that the
equations of state of a gravastar should have: for configurations that are horizon-avoiding for any
total mass and negative central pressure, the equations of state of gravastar matter
must depend on the local compactness $2m(r)/r$. The success of this gravastar model depends on whether or not one can find physically realistic equations of state that take into account the effect of the compactness $2m(r)/r$.
\end{itemize}

\end{itemize}

To conclude, I would like to point out several open problems that have not yet been analyzed.\\

The results posited in cosmology assume an isotropic and homogeneous FRW universe. However, it would be interesting to look at what happens when assuming an inhomogeneous universe. Calculations would become much harder and not as simple as in the homogeneous case, but will we still have some energy conditions satisfied all the way through the cosmological milestones?\\

Concerning the gravastar model, its success depends on whether or not one can find physically
realistic equations of state that take into account the effect of the compactness. However, we have not explicitly demonstrated that such equations of state actually exist. This is a challenge that we leave for the future.

\appendix
 \chapter{Main spacetime metrics considered}
This appendix assembles all the main metrics that have been mentioned in this thesis.

\section{Static spacetime}
 For a \emph{static} spacetime, in the chosen coordinate system with arbitrary $\{ x^a \}$, the metric components are of the form: 
\begin{equation}
ds^2= -V^2 (x^1, x^2, x^3)dt^2 +\sum_{a,b=1}^{3}g_{ab}(x^1,x^2,x^3)dx^a dx^b,  
\end{equation}
where $V^2=-\xi_a \xi^a$ and $\xi^a$ is a Killing vector field. \\
This metric is first mentioned in section \ref{Schwarzschild_solution}.

\section{Schwarzschild exterior solution}
 The Schwarzschild solution, describing static spherical symmetric vacuum spacetimes, is in standard coordinates $(t,r,\theta,\phi)$:
\begin{equation}
ds^2=- \left(1-\frac{2GM}{r}   \right)dt^2+\frac{1}{ \left(1-\frac{2GM}{r}   \right)} dr^2 +r^2d\Omega^2     \label{A_exterior}
\end{equation}
where $d\Omega^2=d\theta^2+\sin^2 \theta d\phi^2$. This solution is also known as the ``Schwarzschild exterior solution".
This is true for any spherically symmetric vacuum solution to Einstein's equations.
 $M$ is a parameter that can be interpreted as the conventional Newtonian mass that  would be measured by studying orbits at large distances from the gravitating source. \\
This metric is first mentioned in section \ref{Schwarzschild_solution}.

\section{Friedmann-Robertson-Walker geometry} 
In cosmology, the homogeneous and isotropic cosmological model called the Friedmann-Robertson-Walker geometry, (FLRW Friedmann-Lema\^itre-Robertson-Walker geometry) is a good description of our universe and is derived by symmetry considerations without using the Einstein equations of general relativity. The metric is given by:
\begin{equation}
ds^2=-dt^2+a(t)^2 \left\{   \frac{dr^2}{1-kr^2} +r^2 \left[  d\theta^2+\sin^2 \theta  \;  d\phi^2 \right]   \right\}   
\end{equation}
where, $a(t)$ is the scale factor of the universe.
There are only three values of interest for the parameter $k$:
\begin{itemize}
\item $k=-1$, this corresponds to a negative curvature (for the hyperboloid)
\item $k=0$, this corresponds to no curvature (flat space)
\item $k=+1$, this corresponds to a positive curvature (for the $3$-sphere)
\end{itemize}
The assumptions of homogeneity and isotropy alone determine the spacetime metric up to three discrete possibilities of spatial geometry $k$ and the arbitrary positive function of the scale factor $a(t)$.\\
This metric is first mentioned in section \ref{cosmology}.

\section{Static spherically symmetric spacetime}
 The most general static spherically symmetric line element in four dimensions can be written in the canonical form~\cite{Wald:1984rg}:
\begin{equation}
ds^2=-f(r) dt^2+h(r) dr^2+r^2 d\Omega,
\end{equation}
where $d\Omega^2=d\theta^2+\sin^2 \theta d\phi^2$. \\
This metric is first mentioned in section \ref{arbitrary_gravastar_def}.

\section{Interior of a gravastar}
 Adopting $(t,r,\theta, \phi)$ coordinates we write any static spherically symmetric geometry, and in particular for the \emph{interior of a gravastar}, in the form:
\begin{equation}
ds^2=-\exp  \left( 2  \int_r ^\infty g(\tilde{r})d\tilde{r}  \right) dt^2 +\frac{dr^2}{\left(    1-\frac{2m(r)}{r}  \right)} +r^2 d\Omega^2,
\end{equation}
where the gravitational acceleration $g(r)$ is:
\begin{equation}
g(r)=\frac{4 \pi r}{3}\frac{\bar{\rho} +3p_r(r)}{1-\frac{2m(r)}{r}}.
\end{equation}
This general solution is also know as an ``interior solution". Of course, the metric of the exterior of the gravastar is obviously the Schwarzschild exterior solution (\ref{A_exterior}).\\
This metric is first mentioned in section \ref{arbitrary_gravastar_def}.

\section{General Schwarzschild interior solution}
 The Schwarzschild exterior solution in standard coordinates $(t,r,\theta,\phi)$ is given by equation (\ref{A_exterior}). A different solution is the general Schwarzschild interior solution for values of $r$ smaller than the Schwarzschild radius $R_{\mathrm{Schwarzschild}}$ is given by:
\begin{equation}
ds^2=-  \left(  A +B \sqrt{1-\frac{2m_*(r)}{r}} \right)^2 dt^2 + \left(  \frac{1}{1-\frac{2m_*(r)}{r}} \right) dr^2+r^2 d\Omega^2   \label{general_interior}
\end{equation}
where $d\Omega^2=d\theta^2+\sin^2 \theta d\phi^2$ and $m_*(r) = M (r/R)^3$.
This general solution describes static spherically symmetric stars with constant density independent of $r$ from the centre to the surface of the star.\\
This metric is first mentioned in section \ref{case_study1}.

\section[Schwarzschild interior solution as a function of the mass and radius]{Schwarzschild interior solution as a function of the total mass $M$ and the surface radius $R$}
 If one wishes to determine the metric of the Schwarzschild interior solution entirely as a function of the total mass $M$ and the surface radius $R$ instead of $A$ and $B$, it can be found that:
\begin{equation}
\left\{\begin{array}{ccc}     
 A &=&   \sqrt{1-\dfrac{2M}{R}}  \\ 
B  &= &    -\dfrac{1}{3}    .
     \end{array}\right.
\end{equation}

In this case the metric becomes:
\begin{equation}
ds^2=-  \left(  \sqrt{1-\frac{2M}{R}}-\frac{1}{3} \sqrt{1-\frac{2m_*(r)}{r}} \right)^2 dt^2 + \left(  \frac{1}{1-\frac{2m_*(r)}{r}} \right) dr^2+r^2 d\Omega^2     
\end{equation}
where $m_*(r) = M (r/R)^3$.\\
This metric is first mentioned in section \ref{case_study1}.


\chapter[Necessary and sufficient conditions for big bangs, bounces, rips...]
{Necessary and sufficient conditions for big bangs, bounces, crunches, rips, 
sudden singularities, extremality events, and more... }%
\centerline{\bf
C\'eline Catto\"en and Matt Visser}
\vskip 0.2 cm
\centerline{ {\footnotesize School of Mathematics, Statistics, and Computer Science,} }
\centerline{{\footnotesize Victoria University of Wellington, }}
\centerline{{\footnotesize P.O.Box 600, Wellington, New Zealand} }

\vskip 0.2 cm
\noindent
{\small Until recently, the physically relevant  singularities occurring in FRW cosmologies had traditionally been thought to be limited to the ``big bang'', and possibly a ``big crunch''. However, over the last few years, the zoo of cosmological singularities considered in the literature has become  considerably more extensive, with ``big rips'' and ``sudden singularities'' added to the mix, as well as renewed interest in non-singular cosmological events such as ``bounces'' and ``turnarounds''. 
In this article we present an extensive catalogue of such cosmological milestones, both at the kinematical and dynamical level. First, using generalized power series, purely kinematical definitions of these cosmological events are provided in terms of the behaviour of the scale factor $a(t)$. The notion of a ``scale-factor singularity'' is defined, and its relation to curvature singularities (polynomial and differential) is explored. Second, dynamical information is extracted by using the Friedmann equations (without assuming even the existence of any equation of state)  to place constraints on whether or not the classical energy conditions are satisfied at the cosmological milestones.  We use these considerations to derive necessary and sufficient conditions for the existence of cosmological milestones such as bangs, bounces, crunches, rips, sudden singularities, and extremality events. Since the classification is extremely general,  and modulo certain technical assumptions complete, the corresponding results are to a high degree model-independent: In particular, we provide a characterization of the class of bangs, crunches, and sudden singularities for which the dominant energy condition is satisfied.}

\vskip 0.2 cm
\noindent

\centerline{{\small gr-qc/0508045; Accepted for publication in {\bf Classical and Quantum Gravity}.}}
\centerline{ {\footnotesize  \texttt{
celine.cattoen@mcs.vuw.ac.nz,
 matt.visser@mcs.vuw.ac.nz}}}

\vskip 1 cm
Note that the full version of this paper can be found in Class.Quant.Grav.22:4913-4930, 2005 or there is an e-Print Archive version: gr-qc/0508045.
  

\newcommand{\arxiv}[1]{[arXiv:#1]}
\newcommand{\sign}{\mathrm{sign}\:}

\chapter{Gravastars must have anisotropic pressures}
\centerline{\bf
C\'eline Catto\"en,
Tristan Faber, and Matt Visser}
\centerline{ {\footnotesize School of Mathematics, Statistics, and Computer Science,} }
\centerline{{\footnotesize Victoria University of Wellington, }}
\centerline{{\footnotesize P.O.Box 600, Wellington, New Zealand}} 
\vskip 0.2 cm
{\small One of the very small number of serious alternatives to the usual concept of an astrophysical
black hole is the ``gravastar'' model developed by Mazur and Mottola; and a related phase-transition model due to Laughlin \emph{et al}. We consider a generalized class of similar models that exhibit continuous pressure --- without the presence of infinitesimally thin shells. By considering the usual TOV equation for static solutions with negative central pressure, we find that gravastars cannot be perfect fluids --- anisotropic pressures in the ``crust'' of a gravastar-like object are unavoidable. The anisotropic TOV equation can then be used to bound the pressure anisotropy.
The transverse stresses that support a gravastar permit a higher compactness than is given by the Buchdahl--Bondi bound for perfect fluid stars. Finally we comment on the qualitative features of the equation of state that gravastar material must have if it is to do the desired job of preventing horizon formation.}
\vskip 0.2 cm
\noindent
\centerline{{\small gr-qc/0505137; } 
\hskip 0.2 cm
{\small Published as {\bf Classical and Quantum Gravity 22 (2005) 4189-4202}}}
\centerline{ {\footnotesize  \texttt{
celine.cattoen@mcs.vuw.ac.nz,
tristan.faber@mcs.vuw.ac.nz, matt.visser@mcs.vuw.ac.nz}}}



\chapter{Effective refractive index tensor for weak-field gravity}

\centerline{\bf
 Petarpa Boonserm,
C\'eline Catto\"en,
Tristan Faber, }
\centerline{\bf Matt Visser,
and
Silke Weinfurtner}

\vskip 0.5 cm
\centerline{ {\footnotesize School of Mathematics, Statistics, and Computer Science,}} 
\centerline{{\footnotesize Victoria University of Wellington, }}
\centerline{{\footnotesize P.O.Box 600, Wellington, New Zealand}} 



\vskip 0.2 cm

\noindent
  {\small Gravitational lensing in a weak but otherwise arbitrary
  gravitational field can be described in terms of a $3\times3$
  tensor, the ``effective refractive index''. If the sources
  generating the gravitational field all have small internal fluxes,
  stresses, and pressures, then this tensor is automatically isotropic
  and the ``effective refractive index'' is simply a scalar that can
  be determined in terms of a classic result involving the Newtonian
  gravitational potential. In contrast if anisotropic stresses are
  ever important then the gravitational field acts similarly to an
  anisotropic crystal.  We derive simple formulae for the refractive
  index tensor, and indicate some situations in which this will be
  important.}

\vskip 0.2 cm
\noindent
{{\small gr-qc/0411034; }
\hskip 0.25 cm
{\small Published as {\bf Classical and Quantum Gravity 22 (2005) 1905-1915}}}
\centerline{ {\footnotesize  \texttt{
petarpa.boonserm@mcs.vuw.ac.nz,
celine.cattoen@mcs.vuw.ac.nz,
tristan.faber@mcs.vuw.ac.nz,}}} 
\centerline{{\footnotesize  \texttt{matt.visser@mcs.vuw.ac.nz,
silke.weinfurtner@mcs.vuw.ac.nz}}}

\backmatter

\newpage
\newpage
\addcontentsline{toc}{chapter}{Bibliography}

\begin{thebibliography}{10}

\bibitem{Abramowicz:2002vt}
Marek~A. Abramowicz, Wlodek Kluzniak, and Jean-Pierre Lasota.
\newblock ``{N}o observational proof of the black-hole event-horizon".
\newblock {\em Astron. Astrophys.}, 396:L31--L34, 2002.
\newblock [arXiv:astro-ph/0207270].

\bibitem{Allen:2005xh}
L.~E. Allen.
\newblock ``{C}osmological perturbations through a simple bounce".
\newblock {\em AIP Conf. Proc.}, 736:182--187, 2005.

\bibitem{Barcelo:2002bv}
Carlos Barcelo and Matt Visser.
\newblock ``{T}wilight for the energy conditions?".
\newblock {\em Int. J. Mod. Phys.}, D11:1553--1560, 2002.
\newblock [arXiv:gr-qc/0205066].

\bibitem{Barrabes:1991ng}
C.~Barrabes and W.~Israel.
\newblock ``{T}hin shells in general relativity and cosmology: The lightlike
  limit".
\newblock {\em Phys. Rev.}, D43:1129--1142, 1991.

\bibitem{Barrow:1988xi}
J.~D. Barrow.
\newblock ``{T}he premature recollapse problem in closed inflationary
  universes".
\newblock {\em Nucl. Phys.}, B296:697--709, 1988.

\bibitem{Barrow:2004hk}
John~D. Barrow.
\newblock ``{M}ore general sudden singularities".
\newblock {\em Class. Quant. Grav.}, 21:5619--5622, 2004.
\newblock [arXiv:gr-qc/0409062].

\bibitem{Barrow:2004xh}
John~D. Barrow.
\newblock ``{S}udden future singularities".
\newblock {\em Class. Quant. Grav.}, 21:L79--L82, 2004.
\newblock [arXiv:gr-qc/0403084].

\bibitem{Barrow:2004he}
John~D. Barrow and Christos~G. Tsagas.
\newblock ``{N}ew isotropic and anisotropic sudden singularities".
\newblock {\em Class. Quant. Grav.}, 22:1563--1571, 2005.
\newblock [arXiv:gr-qc/0411045].

\bibitem{Bekenstein:1975ww}
J.~D. Bekenstein.
\newblock ``{N}onsingular general relativistic cosmologies".
\newblock {\em Phys. Rev.}, D11:2072--2075, 1975.

\bibitem{Bilic:2005sn}
Neven Bilic, Gary~B. Tupper, and Raoul~D. Viollier.
\newblock ``{B}orn-{I}nfeld phantom gravastars".
\newblock 2005.
\newblock [arXiv:astro-ph/0503427].

\bibitem{Boonserm:2004wp}
Petarpa Boonserm, Celine Cattoen, Tristan Faber, Matt Visser, and Silke
  Weinfurtner.
\newblock ``{E}ffective refractive index tensor for weak field gravity".
\newblock {\em Class. Quant. Grav.}, 22:1905--1916, 2005.
\newblock [arXiv:gr-qc/0411034].

\bibitem{Bouhmadi-Lopez:2004me}
Mariam Bouhmadi-Lopez and Jose~A. Jimenez~Madrid.
\newblock ``{E}scaping the big rip?".
\newblock {\em JCAP}, 0505:005, 2005.
\newblock [arXiv:astro-ph/0404540].

\bibitem{Brustein:2001xh}
Ram Brustein, Stefano Foffa, and Avraham~E. Mayo.
\newblock ``{C}ausal entropy bound for non-singular cosmologies".
\newblock {\em Phys. Rev.}, D65:024004, 2002.
\newblock [arXiv:hep-th/0108098].

\bibitem{Burnett:1993bn}
Gregory~A. Burnett.
\newblock ``{R}ecollapse of the closed tolman space-times".
\newblock {\em Phys. Rev.}, D48:5688--5696, 1993.
\newblock [arXiv:gr-qc/9308003].

\bibitem{Calderon:2004bi}
Hector Calderon and William~A. Hiscock.
\newblock ``{Q}uantum fields and ``big rip" expansion singularities".
\newblock {\em Class. Quant. Grav.}, 22:L23--L26, 2005.
\newblock [arXiv:gr-qc/0411134].

\bibitem{Caldwell:1999ew}
R.~R. Caldwell.
\newblock ``{A} phantom menace?".
\newblock {\em Phys. Lett.}, B545:23--29, 2002.
\newblock [arXiv:astro-ph/9908168].

\bibitem{Carroll:2004st}
Sean~M. Carroll.
\newblock ``{S}pacetime and geometry: An introduction to general relativity".
\newblock San Francisco, USA: Addison-Wesley (2004) 513 p.

\bibitem{Cattoen:2005he}
Celine Cattoen, Tristan Faber, and Matt Visser.
\newblock ``{G}ravastars must have anisotropic pressures".
\newblock 2005.
\newblock [arXiv:gr-qc/0505137].

\bibitem{Cattoen:2005dx}
Celine Cattoen and Matt Visser.
\newblock ``{N}ecessary and sufficient conditions for big bangs, bounces,
  crunches, rips, sudden singularities, and extremality events".
\newblock 2005.
\newblock [arXiv:gr-qc/0508045].

\bibitem{Chapline:2000en}
G.~Chapline, E.~Hohlfeld, R.~B. Laughlin, and D.~I. Santiago.
\newblock ``{Q}uantum phase transitions and the breakdown of classical general
  relativity".
\newblock {\em Int. J. Mod. Phys.}, A18:3587--3590, 2003.
\newblock [arXiv:gr-qc/0012094].

\bibitem{Chapline:2002}
G.~Chapline, R.~B. Laughlin, and D.~I. Santiago.
\newblock ``{E}mergent relativity and the physics of black hole interiors".

\bibitem{Chimento:2004ps}
Luis~P. Chimento and Ruth Lazkoz.
\newblock ``{O}n big rip singularities".
\newblock {\em Mod. Phys. Lett.}, A19:2479--2484, 2004.
\newblock [arXiv:gr-qc/0405020].

\bibitem{Curbelo:2005dh}
Ruben Curbelo, Tame Gonzalez, and Israel Quiros.
\newblock ``{I}nteracting phantom energy and avoidance of the big rip
  singularity".
\newblock 2005.
\newblock [arXiv:astro-ph/0502141].

\bibitem{Dymnikova:1992ux}
I.~Dymnikova.
\newblock ``{V}acuum nonsingular black hole".
\newblock {\em Gen. Rel. Grav.}, 24:235--242, 1992.

\bibitem{Dymnikova:2001fb}
Irina Dymnikova.
\newblock ``{C}osmological term as a source of mass".
\newblock {\em Class. Quant. Grav.}, 19:725--740, 2002.
\newblock [arXiv:gr-qc/0112052].

\bibitem{Dymnikova:2003vt}
Irina Dymnikova.
\newblock ``{S}pherically symmetric space-time with the regular de {S}itter
  center".
\newblock {\em Int. J. Mod. Phys.}, D12:1015--1034, 2003.
\newblock [arXiv:gr-qc/0304110].

\bibitem{Dymnikova:2004qg}
Irina Dymnikova and Evgeny Galaktionov.
\newblock ``{S}tability of a vacuum nonsingular black hole".
\newblock {\em Class. Quant. Grav.}, 22:2331--2358, 2005.
\newblock [arXiv:gr-qc/0409049].

\bibitem{Dymnikova:1999cz}
Irina~G. Dymnikova.
\newblock ``{T}he algebraic structure of a cosmological term in spherically
  symmetric solutions".
\newblock {\em Phys. Lett.}, B472:33--38, 2000.
\newblock [arXiv:gr-qc/9912116].

\bibitem{PDBook}
S.~{Eidelman}, K.G. {Hayes}, K.A. {Olive}, M.~{Aguilar-Benitez}, C.~{Amsler},
  D.~{Asner}, K.S. {Babu}, R.M. {Barnett}, J.~{Beringer}, P.R. {Burchat}, C.D.
  {Carone}, C.~{Caso}, G.~{Conforto}, O.~{Dahl}, G.~{D'Ambrosio}, M.~{Doser},
  J.L. {Feng}, T.~{Gherghetta}, L.~{Gibbons}, M.~{Goodman}, C.~{Grab}, D.E.
  {Groom}, A.~{Gurtu}, K.~{Hagiwara}, J.J. {Hern\'andez-Rey}, K.~{Hikasa},
  K.~{Honscheid}, H.~{Jawahery}, C.~{Kolda}, {Kwon} Y., M.L. {Mangano}, A.V.
  {Manohar}, J.~{March-Russell}, A.~{Masoni}, R.~{Miquel}, K.~{M\"onig},
  H.~{Murayama}, K.~{Nakamura}, S.~{Navas}, L.~{Pape}, C.~{Patrignani},
  A.~{Piepke}, G.~{Raffelt}, M.~{Roos}, M.~{Tanabashi}, J.~{Terning}, N.A.
  {T\"ornqvist}, T.G. {Trippe}, P.~{Vogel}, C.G. {Wohl}, R.L. {Workman}, W.-M.
  {Yao}, P.A. {Zyla}, B.~{Armstrong}, P.S. {Gee}, G.~{Harper}, K.S. {Lugovsky},
  S.B. {Lugovsky}, V.S. {Lugovsky}, A.~{Rom}, M.~{Artuso}, E.~{Barberio},
  M.~{Battaglia}, H.~{Bichsel}, O.~{Biebel}, P.~{Bloch}, R.N. {Cahn},
  D.~{Casper}, A.~{Cattai}, R.S. {Chivukula}, G.~{Cowan}, T.~{Damour},
  K.~{Desler}, M.A. {Dobbs}, M.~{Drees}, A.~{Edwards}, D.A. {Edwards}, V.D.
  {Elvira}, J.~{Erler}, V.V. {Ezhela}, W.~{Fetscher}, B.D. {Fields},
  B.~{Foster}, D.~{Froidevaux}, M.~{Fukugita}, T.K. {Gaisser}, L.~{Garren},
  H.-J. {Gerber}, G.~{Gerbier}, F.J. {Gilman}, H.E. {Haber}, C.~{Hagmann},
  J.~{Hewett}, I.~{Hinchliffe}, C.J. {Hogan}, G.~{H\"ohler}, P.~{Igo-Kemenes},
  J.D. {Jackson}, K.F. {Johnson}, D.~{Karlen}, B.~{Kayser}, D.~{Kirkby}, S.R.
  {Klein}, K.~{Kleinknecht}, I.G. {Knowles}, P.~{Kreitz}, Yu.V. {Kuyanov},
  O.~{Lahav}, P.~{Langacker}, A.~{Liddle}, L.~{Littenberg}, D.M. {Manley}, A.D.
  {Martin}, M.~{Narain}, P.~{Nason}, Y.~{Nir}, J.A. {Peacock}, H.R. {Quinn},
  S.~{Raby}, B.N. {Ratcliff}, E.A. {Razuvaev}, B.~{Renk}, G.~{Rolandi}, M.T.
  {Ronan}, L.J. {Rosenberg}, C.T. {Sachrajda}, Y.~{Sakai}, A.I. {Sanda},
  S.~{Sarkar}, M.~{Schmitt}, O.~{Schneider}, D.~{Scott}, W.G. {Seligman}, M.H.
  {Shaevitz}, T.~{Sj\"ostrand}, G.F. {Smoot}, S.~{Spanier}, H.~{Spieler},
  N.J.C. {Spooner}, M.~{Srednicki}, A.~{Stahl}, T.~{Stanev}, M.~{Suzuki}, N.P.
  {Tkachenko}, G.H. {Trilling}, G.~{Valencia}, K.~{van Bibber}, M.G. {Vincter},
  D.~{Ward}, B.R. {Webber}, M.~{Whalley}, L.~{Wolfenstein}, J.~{Womersley},
  C.L. {Woody}, O.V. {Zenin}, and R.-Y. {Zhu}.
\newblock ``{\bf review of particle physics}".
\newblock {\em {Physics Letters B}}, 592:1+, 2004.
\newblock url http://pdg.lbl.gov.

\bibitem{Ellis:2002we}
George F.~R. Ellis and Roy Maartens.
\newblock ``{T}he emergent universe: Inflationary cosmology with no
  singularity".
\newblock {\em Class. Quant. Grav.}, 21:223--232, 2004.
\newblock [arXiv:gr-qc/0211082].

\bibitem{Ellis:2003qz}
George F.~R. Ellis, Jeff Murugan, and Christos~G. Tsagas.
\newblock ``{T}he emergent universe: An explicit construction".
\newblock {\em Class. Quant. Grav.}, 21:233--250, 2004.
\newblock [arXiv:gr-qc/0307112].

\bibitem{Fernandez-Jambrina:2004yy}
L.~Fernandez-Jambrina and Ruth Lazkoz.
\newblock ``{G}eodesic behaviour of sudden future singularities".
\newblock {\em Phys. Rev.}, D70:121503, 2004.
\newblock [arXiv:gr-qc/0410124].

\bibitem{Ford:1985qt}
L.~H. Ford.
\newblock ``{U}nstable fields and the recollapse of an open universe".
\newblock {\em Phys. Lett.}, A110:21, 1985.

\bibitem{Frampton:2004tn}
Paul~H. Frampton.
\newblock ``{E}nhanced big rip without conventional dark energy".
\newblock 2004.
\newblock [arXiv:astro-ph/0407353].

\bibitem{Frampton:2004xn}
Paul~H. Frampton and Tomo Takahashi.
\newblock ``{B}igger rip with no dark energy".
\newblock {\em Astropart. Phys.}, 22:307--312, 2004.
\newblock [arXiv:astro-ph/0405333].

\bibitem{Gasperini:2004ss}
M.~Gasperini, Massimo Giovannini, and G.~Veneziano.
\newblock ``{C}osmological perturbations across a curvature bounce".
\newblock {\em Nucl. Phys.}, B694:206--238, 2004.
\newblock [arXiv:hep-th/0401112].

\bibitem{Gliner:1966}
E.~B. Gliner.
\newblock {\em Sov. Phys.}, 1966.
\newblock JETP 22 378.

\bibitem{Gliner:2002yh}
E.~B. Gliner.
\newblock ``{I}nflationary universe and the vacuumlike state of physical
  medium".
\newblock {\em Phys. Usp.}, 45:213--220, 2002.

\bibitem{Gordon:2002jw}
Christopher Gordon and Neil Turok.
\newblock ``{C}osmological perturbations through a general relativistic
  bounce".
\newblock {\em Phys. Rev.}, D67:123508, 2003.
\newblock [arXiv:hep-th/0206138].

\bibitem{Gott:2003pf}
J.~Richard~III Gott, M.~Juri\'c, D.~Schegel, F.~Hoyle, M.~Vogeley, M.~Tegmark,
  N.~Bahcall, and J.~Brinkmann.
\newblock ``{A} map of the universe".
\newblock {\em Astrophys. J.}, 624:463, 2005.
\newblock [arXiv:astro-ph/0310571].

\bibitem{Guven:1999wm}
Jemal Guven and Niall O'Murchadha.
\newblock ``{B}ounds on $2m/{R}$ for static spherical objects".
\newblock {\em Phys. Rev.}, D60:084020, 1999.
\newblock [arXiv:gr-qc/9903067].

\bibitem{Harrison:1965}
B.K. Harrison, K.S. Thorne, M.~Wakano, and J.A. Wheeler.
\newblock ``{G}ravitation theory and gravitational collapse".
\newblock 1965.
\newblock University of Chicago Press, Chicago.

\bibitem{Hartle:2003yu}
J.~B. Hartle.
\newblock ``{A}n introduction to {E}instein's general relativity".
\newblock 1985.
\newblock San Francisco, USA: Addison-Wesley (2003) 582 p.

\bibitem{Herrera:2001vg}
L.~Herrera, A.~Di~Prisco, J.~Ospino, and E.~Fuenmayor.
\newblock ``{C}onformally flat anisotropic spheres in general relativity".
\newblock {\em J. Math. Phys.}, 42:2129--2143, 2001.
\newblock [arXiv:gr-qc/0102058].

\bibitem{Herrera:2004xc}
L.~Herrera et~al.
\newblock ``{S}pherically symmetric dissipative anisotropic fluids: A general
  study".
\newblock {\em Phys. Rev.}, D69:084026, 2004.
\newblock [arXiv:gr-qc/0403006].

\bibitem{Herrera:2002bm}
L.~Herrera, J.~Martin, and J.~Ospino.
\newblock ``{A}nisotropic geodesic fluid spheres in general relativity".
\newblock {\em J. Math. Phys.}, 43:4889--4897, 2002.
\newblock [arXiv:gr-qc/0207040].

\bibitem{Hochberg:1998vm}
David Hochberg, Carmen Molina-Paris, and Matt Visser.
\newblock ``{T}olman wormholes violate the strong energy condition".
\newblock {\em Phys. Rev.}, D59:044011, 1999.
\newblock [arXiv:gr-qc/9810029].

\bibitem{Hoyle:2001kn}
Fiona Hoyle and Michael~S. Vogeley.
\newblock ``{V}oids in the {PSC}z survey and the updated {Z}wicky catalog".
\newblock {\em Astrophys. J.}, 566:641, 2002.
\newblock [arXiv:astro-ph/0109357].

\bibitem{Hoyle:2003hc}
Fiona Hoyle and Michael~S. Vogeley.
\newblock ``{V}oids in the 2d{F} galaxy redshift survey".
\newblock {\em Astrophys. J.}, 607:751--764, 2004.
\newblock [arXiv:astro-ph/0312533].

\bibitem{Hwang:2001zt}
Jai-chan Hwang and Hyerim Noh.
\newblock ``{N}on-singular big-bounces and evolution of linear fluctuations".
\newblock {\em Phys. Rev.}, D65:124010, 2002.
\newblock [arXiv:astro-ph/0112079].

\bibitem{Israel:1966rt}
W.~Israel.
\newblock ``{S}ingular hypersurfaces and thin shells in general relativity".
\newblock {\em Nuovo Cim.}, B44S10:1, 1966.

\bibitem{Lake:2004fu}
Kayll Lake.
\newblock ``{S}udden future singularities in {FLRW} cosmologies".
\newblock {\em Class. Quant. Grav.}, 21:L129, 2004.
\newblock [arXiv:gr-qc/0407107].

\bibitem{Lanczos}
K~Lanczos.
\newblock ``{F}lachenhafte verteilung der materie in der einsteinschen
  gravitationstheorie".
\newblock {\em Ann. Phy (Leipzig)}, 74:518--540, 1924.

\bibitem{Liapunov}
S.~Lefschetz.
\newblock ``{D}ifferential equations: Geometric theory".
\newblock Dover, 1977, New York.

\bibitem{Matzner:1986us}
R.~A. Matzner and A.~Mezzacappa.
\newblock ``{A} 3-dimensional closed universes without collapse in the
  5-dimensional {K}aluza-{K}lein theory".
\newblock {\em Phys. Rev.}, D32:3114--3117, 1985.

\bibitem{Matzner:1986nt}
R.~A. Matzner and A.~Mezzacappa.
\newblock ``{P}rofessor {W}heeler and the crack of doom: Closed cosmologies in
  the 5-{D} {K}aluza-{K}lein theory".
\newblock {\em Found. Phys.}, 16:227--248, 1986.

\bibitem{Mazur:2001fv}
Pawel~O. Mazur and Emil Mottola.
\newblock ``{G}ravitational condensate stars".
\newblock 2001.
\newblock [arXiv:gr-qc/0109035].

\bibitem{Mazur:2004ku}
Pawel~O. Mazur and Emil Mottola.
\newblock ``{D}ark energy and condensate stars: Casimir energy in the large".
\newblock 2004.
\newblock [arXiv:gr-qc/0405111].

\bibitem{Mazur:2004fk}
Pawel~O. Mazur and Emil Mottola.
\newblock ``{G}ravitational vacuum condensate stars".
\newblock {\em Proc. Nat. Acad. Sci.}, 111:9545--9550, 2004.
\newblock [arXiv:gr-qc/0407075].

\bibitem{Medved:2003sk}
A.~J.~M. Medved.
\newblock ``{A}natomy of a bounce".
\newblock {\em Class. Quant. Grav.}, 21:2749--2760, 2004.
\newblock [arXiv:hep-th/0307258].

\bibitem{Miritzis:2005hg}
John Miritzis.
\newblock ``{T}he recollapse problem of closed frw models in higher-order
  gravity theories".
\newblock 2005.
\newblock [arXiv:gr-qc/0505139].

\bibitem{Molina-Paris:1998tx}
Carmen Molina-Paris and Matt Visser.
\newblock ``{M}inimal conditions for the creation of a friedman-
  robertson-walker universe from a ``bounce"".
\newblock {\em Phys. Lett.}, B455:90--95, 1999.
\newblock [arXiv:gr-qc/9810023].

\bibitem{Nojiri:2004ip}
Shin'ichi Nojiri and Sergei~D. Odintsov.
\newblock ``{Q}uantum escape of sudden future singularity".
\newblock {\em Phys. Lett.}, B595:1--8, 2004.
\newblock [arXiv:hep-th/0405078].

\bibitem{Parker:1990mk}
Leonard Parker and Yi~Wang.
\newblock ``{A}voidance of singularities in relativity through two-body
  interactions".
\newblock {\em Phys. Rev.}, D42:1877--1883, 1990.

\bibitem{D'inverno}
R.D'Inverno.
\newblock {\em Introducing Einstein's Relativity}.
\newblock (Clarendon Press, Oxford\begin{scriptsize}\end{scriptsize}), 2002.

\bibitem{Sahni:1991ks}
Varun Sahni, Hume Feldman, and Albert Stebbins.
\newblock ``{L}oitering universe".
\newblock {\em Astrophys. J.}, 385:1--8, 1992.

\bibitem{Sahni:2004fb}
Varun Sahni and Yuri Shtanov.
\newblock ``{D}id the universe loiter at high redshifts ?".
\newblock {\em Phys. Rev.}, D71:084018, 2005.
\newblock [arXiv:astro-ph/0410221].

\bibitem{Sen}
N~Sen.
\newblock ``{U}ber die grenzbedingungen des schwerefelds an unstetigen
  kreisflachen".
\newblock {\em Ann. Phy (Leipzig)}, 73:365--396, 1924.

\bibitem{Stefancic:2004kb}
Hrvoje Stefancic.
\newblock `` ``{E}xpansion" around the vacuum equation of state: Sudden future
  singularities and asymptotic behavior".
\newblock {\em Phys. Rev.}, D71:084024, 2005.
\newblock [arXiv:astro-ph/0411630].

\bibitem{Gravitation}
C.~W. Misner K.~S. Thorne and J.~A. Wheeler.
\newblock {\em ``{G}ravitation"}.
\newblock (Freeman, San Fransisco\begin{scriptsize}\end{scriptsize}), 1972.

\bibitem{Tippett:2004xj}
Benjamin~K. Tippett and Kayll Lake.
\newblock ``{E}nergy conditions and a bounce in flrw cosmologies".
\newblock 2004.
\newblock [arXiv:gr-qc/0409088].

\bibitem{Veneziano:2003sz}
G.~Veneziano.
\newblock ``{A} model for the big bounce".
\newblock {\em JCAP}, 0403:004, 2004.
\newblock [arXiv:hep-th/0312182].

\bibitem{Virey:2005ih}
J.~M. Virey et~al.
\newblock ``{O}n the determination of the deceleration parameter from
  supernovae data".
\newblock {\em Phys. Rev.}, D72:061302, 2005.
\newblock [arXiv:astro-ph/0502163].

\bibitem{Visser:1995cc}
Matt Visser.
\newblock ``{L}orentzian wormholes: From {E}instein to {H}awking".
\newblock Woodbury, USA: AIP (1995) 412 p.

\bibitem{Visser:2004bf}
Matt Visser.
\newblock ``{C}osmography: Cosmology without the {E}instein equations".
\newblock 2004.
\newblock [arXiv:gr-qc/0411131].

\bibitem{Visser:2003vq}
Matt Visser.
\newblock ``{J}erk and the cosmological equation of state".
\newblock {\em Class. Quant. Grav.}, 21:2603--2616, 2004.
\newblock [arXiv:gr-qc/0309109].

\bibitem{Visser:2003ge}
Matt Visser and David~L. Wiltshire.
\newblock ``{S}table gravastars - an alternative to black holes?".
\newblock {\em Class. Quant. Grav.}, 21:1135--1152, 2004.
\newblock [arXiv:gr-qc/0310107].

\bibitem{Visser:2002ww}
Matt Visser and Nicolas Yunes.
\newblock ``{P}ower laws, scale invariance, and generalized frobenius series:
  Applications to newtonian and {TOV} stars near criticality".
\newblock {\em Int. J. Mod. Phys.}, A18:3433--3468, 2003.
\newblock [arXiv:gr-qc/0211001].

\bibitem{Wald:1984rg}
R.~M. Wald.
\newblock ``{G}eneral relativity".
\newblock Chicago, Usa: Univ. Pr. ( 1984) 491p.

\end{thebibliography}

\newpage
\newpage
\addcontentsline{toc}{chapter}{Index}
\printindex

\end{document}